\documentclass{article}

\usepackage{microtype}
\usepackage{kbordermatrix}
\usepackage{amsmath,amssymb,amsthm}
\usepackage{enumitem}
\usepackage{mathtools}
\usepackage{relsize}
\usepackage{euscript}

\usepackage{scalerel}
\usepackage{subcaption}

\newtheorem{theorem}{Theorem}[section]
\newtheorem{lemma}[theorem]{Lemma}
\newtheorem{definition}[theorem]{Definition}
\newtheorem{proposition}[theorem]{Proposition}
\newtheorem*{proposition-no}{Proposition}

\newtheorem{corollary}[theorem]{Corollary}
\newtheorem{conjecture}[theorem]{Conjecture}

\newtheorem{remark}[theorem]{Remark}

\newtheorem{example}[theorem]{Example}

\let\al=\alpha
\let\gm=\gamma
\let\Gm=\Gamma

\let\vf=\varphi

\let\sg=\sigma

\let\th=\theta

\let\dg=\dagger
\def\rel{R}
\def\relo{Q}

\def\cG{\mathcal G}

\def\cH{\mathcal H}
\def\cI{\mathcal I}
\def\cJ{\mathcal J}
\def\cK{\mathcal K}
\def\cP{\mathcal P}
\def\cS{\mathcal S}

\def\cC{\mathcal{C}}
\def\RT{{\mathfrak{R}}}

\def\ba{\mathbf{a}}
\def\bb{\mathbf{b}}
\def\bc{\mathbf{c}}

\def\bs{\mathbf{s}}
\def\bx{\mathbf{x}}

\def\sH{\mathsf{H}}
\def\sG{\mathsf{G}}
\def\sK{\mathsf{K}}

\def\sS{\mathsf{S}}

\def\sZ{\mathsf{Z}}
\def\const{\mathsf{c}}

\def\wJ{{\widetilde \cJ}}
\def\wM{{\widetilde M}}
\def\zz{{\underline{0}}}
\def\zo{{\underline{1}}}

\def\BIS{\mathsf{BIS}}
\def\IS{\mathsf{IS}}
\def\CSP{\mathsf{CSP}}
\def\SAT{\mathsf{SAT}}

\def\NCSP{\mathsf{\#CSP}}
\def\NpCSP{\#_p\mathsf{CSP}}
\def\Aut{\mathsf{Aut}}
\def\Fix{\mathsf{Fix}}
\def\Stab{\mathsf{Stab}}

\def\Part{\mathsf{Part}}

\def\ghom#1{\#\mathsf{Hom}(#1)}
\def\ghomk#1#2{\#_{#1}\mathsf{Hom}(#2)}

\def\Hom{\mathsf{Hom}}

\def\CSP{\mathsf{CSP}}

\def\Aut{\mathsf{Aut}}
\def\Fix{\mathsf{Fix}}

\def\ext{\mathsf{\# ext}}

\def\hom{\mathsf{hom}}
\def\inj{\mathsf{inj}}

\def\bip{{\mathsf{bip}}}
\def\nbip{{\mathsf{nonbip}}}
\def\graph{{\mathsf{graph}}}
\def\LZ{{LZ}}
\def\RZ{{RZ}}
\def\WT#1{\widetilde #1}

\let\eps=\emptyset

\def\ang#1{\langle#1\rangle}
\let\sse=\subseteq
\def \carts{\; \small\Box \;}
\let\tm=\times
\def\zd{,\dots,}
\def\vv#1{\mathbf{#1}}
\def\vc#1#2{#1_1,\dots,#1_{#2}}
\def\modp{$\#_p$P}
\def\modk{$\#_k$P}

\newcommand{\FUNC}[3]{#1 : #2 \rightarrow #3}

\newcommand{\edge}[2]{\left\langle #1,#2 \right\rangle}

\newcommand{\nedge}[2]{(#1,#2)}
\newcommand{\constCSP}[2]{\langle #1,#2 \rangle}

\usepackage{scalerel}
\usepackage{stackengine}
\stackMath

\newcommand{\Rside}{{\breve{R}}}
\newcommand{\Lside}{{\breve{L}}}
\newcommand{\vvec}[2]{
    \begin{pmatrix}
    {#1}_1 \\
    \vdots \\
    {#1}_{#2}
    \end{pmatrix}
}
\newcommand{\nvedge}[2]{
    \begin{pmatrix}
    {#1} \\
    {#2}
    \end{pmatrix}
}
\newcommand{\vvecad}[2]{
    \begin{pmatrix}
    {#1} \\
    \vdots \\
    {#2}
    \end{pmatrix}
}

\newtheorem{innercustomthm}{Theorem}
\newenvironment{custom_num_theorem}[1]
  {\renewcommand\theinnercustomthm{#1}\innercustomthm}
{\endinnercustomthm}

\begin{document}
\date{}

\title{Complexity classification of counting graph homomorphisms modulo a prime number}
\author{Andrei A.Bulatov\thanks{abulatov@sfu.ca} 
\and Amirhossein Kazeminia\thanks{akazemin@sfu.ca} 
}
\maketitle

\begin{abstract}
Counting graph homomorphisms and its generalizations such as the Counting Constraint Satisfaction Problem (CSP),  variations of the Counting CSP, and counting problems in general have been intensively studied since the pioneering work of Valiant. While the complexity of exact counting of graph homomorphisms (Dyer and Greenhill, 2000) and the Counting CSP (Bulatov, 2013, and Dyer and Richerby, 2013) is well understood, counting modulo some natural number has attracted considerable interest as well. In their 2015 paper Faben and Jerrum suggested a conjecture stating that counting homomorphisms to a fixed graph $H$ modulo a prime number is hard whenever it is hard to count exactly, unless $H$ has automorphisms of certain kind. In this paper we confirm this conjecture. As a part of this investigation we develop techniques that widen the spectrum of reductions available for modular counting and apply to the general CSP rather than being limited to graph homomorphisms.
\end{abstract}


\section{Introduction}


In this paper we tackle the problem of counting graph homomorphisms modulo a prime number. Counting problems in general have been intensively studied since the pioneering work by Valiant \cite{Valiant79:complexity,Valiant79:computing}. For a problem $A$ from NP the corresponding counting problem asks about the number of accepting paths of a nondeterministic Turing machine that solves the problem $A$. In many cases rather than counting accepting paths we may need to compute a more tangible number. One such case is the Constraint Satisfaction Problem (CSP), in which the question is to decide the existence of an assignment of values to variables subject to a given collection of constraints. Thus, in the Counting CSP the objective is to find the number of such assignments. The counting CSP also allows for generalizations such as partition functions \cite{Barvinok16:combinatorics,Bulatov05:partition} that yield connections with areas such as statistical physics, see, e.g.\ \cite{Jerrum93:polynomial,Lieb81:general}. While the complexity of exact counting solutions of a CSP is now well understood \cite{Dyer00:complexity,Bulatov13:counting,Dyer13:effective,Dalmau04:complexity}, modular counting such as finding the parity of the number of solutions remains widely open. Although the focus of this paper is on graph homomorphisms that are a special kind of the CSP, we intensively use the general CSP as a technical tool, apply the techniques developed for the general CSP, and also try to state our results in a form as general as possible.

\paragraph{Homomorphisms and the Constraint Satisfaction Problem}
A \emph{relational signature} $\sg$ is a collection of \emph{relational symbols} each of which is assigned a positive integer, the \emph{arity} of the symbol. A \emph{relational structure} $\cH$ with signature $\sg$ is a set $H$ and an \emph{interpretation} $\rel^\cH$ of each $\rel\in\sg$, where $\rel^\cH$ is a relation or a predicate on $H$ whose arity equals that of $\rel$. The structure $\cH$ is finite if both $H$ and $\sg$ are finite. All the structures in this paper are finite. The set $H$ is said to be the \emph{base set} or the \emph{universe} of $\cH$. We will use for the base set the same letter as for the structure, only in the regular font. A structure with signature $\sg$ is often called a \emph{$\sg$-structure}. Structures with the same signature are called \emph{similar}.

Let $\cG,\cH$ be similar structures with signature $\sg$. A \emph{homomorphism} from $\cG$ to $\cH$ is a mapping $\vf:G\to H$ such that for any $\rel\in\sg$, say, of arity $r$, if $\rel^\cG(\vc ar)$ is true for $\vc ar\in G$, then $\rel^\cH(\vf(a_1)\zd\vf(a_r))$ is true as well. The set of all homomorphisms from $\cG$ to $\cH$ is denoted $\Hom(\cG,\cH)$. The cardinality of $\Hom(\cG,\cH)$ is denoted by $\hom(\cG,\cH)$. A homomorphism $\vf$ is an \emph{isomorphism} if it is bijective and the inverse mapping $\vf^{-1}$ is a homomorphism from $\cH$ to $\cG$. If $\cH$ and $\cG$ are isomorphic, we write $\cH\cong\cG$. A homomorphism of a structure to itself is called an \emph{endomorphism}, and an isomorphism to itself is called an \emph{automorphism}.

The Constraint Satisfaction Problem can be defined in multiple ways; one of the standard ones actually involves constraints imposed on sets of variables. However, for our purposes the definition promoted by Feder and Vardi \cite{Feder98:computational} is the most convenient one. It is also equivalent to the other definitions. \emph{The CSP} asks, given similar relational structures $\cG$ and $\cH$, to decide whether there is a homomorphism from $\cG$ to $\cH$. 

While the general CSP is of course NP-complete, certain restrictions of the problem, first, allow one to model specific combinatorial problems, and, second, give rise to problems of lower complexity. The most widely studied way to restrict the CSP, and the one we use here, is to fix the target structure $\cH$. By $\CSP(\cH)$ we denote the problem, in which given a structure $\cG$ similar to $\cH$, the goal is to decide whether there is a homomorphism from $\cG$ to $\cH$. CSPs of this form are sometimes called nonuniform CSPs. In the literature such restrictions are also often given through constraint languages.

\paragraph{Counting CSP and graph homomorphisms}
In the (exact) Counting CSP the goal is to find the number $\hom(\cG,\cH)$ of homomorphisms from a relational structure $\cG$ to a relational structure $\cH$. Restricted versions of the Counting CSP can be introduced in the same way as for the decision one. In the counting version of $\CSP(\cH)$ denoted $\NCSP(\cH)$ the goal is to find $\hom(\cG,\cH)$ for a given structure $\cG$.
A graph can be viewed as a relational structure with one symmetric binary relation, however, for historical reasons, we use slightly different notation for graphs. In particular, we use $H$ rather than $\cH$ and $\mathsf{Hom}(H),\ghom H$ rather than $\CSP(H),\NCSP(H)$. The problem $\mathsf{Hom}(H)$ is often called the $H$-Coloring problem in the graph theory literature, see, e.g.,  \cite{Hell04:homomorphism}. Its counting variant is called the $\#H$-Coloring problem.

The complexity class the Counting CSP belongs to is \#P, the class of problems of counting accepting paths of polynomial time nondeterministic Turing machines. There are several ways to define reductions between counting problems, but the most widely used ones are parsimonious reductions and Turing reductions. A \emph{parsimonious reduction} from a counting problem $\#A$ to a counting problem $\#B$ is an algorithm that, given an instance $I$ of $\#A$, produces (in polynomial time) an instance $I'$ of $\#B$ such that the answers to $I$ and $I'$ are equal. A \emph{Turing reduction} is a polynomial time algorithm that solves $\#A$ using $\#B$ as an oracle. Completeness in \#P is then defined in the standard way. This paper and all the papers we cite predominantly use Turing reductions.

The complexity of the $\#H$-Coloring problem was characterized by Dyer and Greenhill \cite{Dyer00:complexity}. It turns out that this problem can be solved in polynomial time if and only if every connected component of $H$ is either an isolated vertex, or a complete graph with all loops present, or a complete bipartite graph. Otherwise $\#H$-Coloring is \#P-complete. This theorem was generalized through a sequence of intermediate results \cite{Dyer07:counting,Creignou96:complexity,ref:BULATOV_TowardDichotomy,Bulatov05:partition} to a complete complexity classification of $\NCSP(\cH)$ for arbitrary finite relational structures $\cH$ by Bulatov \cite{Bulatov13:counting} and Dyer and Richerby \cite{Dyer13:effective}. 

\paragraph{Modular counting}
In this paper we study the problem of counting solutions to a CSP modulo a prime number $p$. If a relational structure $\cH$ is fixed, this problem is denoted by $\NpCSP(\cH)$. More precisely, in $\NpCSP(\cH)$ the objective is, given a relational structure $\cG$, to find the number of homomorphisms from $\cG$ to $\cH$ modulo $p$. If the relational structure $\cH$ is a graph, this problem is also denoted by $\#_p\Hom(H)$. There are several complexity classes related to modular counting. The more established type of classes is $\mathsf{Mod}_k$P, the class of problems deciding whether the number of accepting paths of a polynomial time nondeterministic Turing machine is \emph{not} divisible by $k$, \cite{Cai89:power,Hertrampf90:relations}. In particular, if $k=2$ then $\mathsf{Mod}_k$P is the class $\oplus$P. However, problems of counting accepting paths naturally belong to classes of the form \modk,   introduced in \cite{faben2008complexity} that contain problems of counting accepting paths modulo $k$. The standard notions of reduction are again parsimonious reduction (only now the equality of answers is understood modulo $k$) and Turing reduction. Faben in \cite{faben2008complexity} studied basic properties of such classes, in particular, he identified several \modk-complete problems. 

A priori, the relationship of the complexity of such problems with that of regular counting problems is unclear, except modular counting cannot be harder than exact counting. Later we will see examples when modular counting problems are much easier than their exact counterparts. The problem $\#_k CSP(\cH)$ of counting solutions of CSPs modulo a number $k$, which is not necessarily prime, is, of course, also very natural. As is easily seen, solving $\#_k CSP(\cH)$ reduces to solving several problems of the form $\#_{p^\ell} CSP(\cH)$, where $p$ is prime. However, problems of the form $\#_{p^\ell} CSP(\cH)$ for $\ell>1$ have not been studied and are beyond the scope of this paper.

In the case of the CSP, the research has mostly been focused on graph homomorphisms. The only exceptions we are aware of are a result of Faben \cite{faben2008complexity} who characterized the complexity of counting the solutions of a Generalized Satisfiability problem modulo an integer, a generalization of \cite{faben2008complexity} to problems with weights by  Guo et al.\ \cite{guo_et_al:LIPIcs:2011:3015}, and a study of the complexity of a certain class of a more general Holant problem by Guo et al.\ \cite{Guo13:symmetric}. The study of modular counting of graph homomorphisms has been much more vibrant. 

Before discussing the results of this study we need to mention the automorphism group of a graph or, more generally, of a relational structure. The automorphisms of $\cH$ form a group with respect to composition denoted $\Aut(\cH)$. The order of an automorphism $\pi\in\Aut(\cH)$ is the smallest number $k$ such that $\pi^k$ is the identity permutation. An element $a\in H$ is a fixed point of $\pi\in\Aut(\cH)$ if $\pi(a)=a$. The set of all fixed points of $\pi$ is denoted by $\Fix(\pi)$.

A systematic study of counting graph homomorphisms modulo a prime number $p$ was initiated by Faben \cite{faben2008complexity}, and later continued  by Faben and Jerrum \cite{ref:CountingMod2Ini}. One of the first observations they made concerns the cases where exact and modular counting clearly deviate from each other. As Faben and Jerrum observed, the automorphism group $\Aut(H)$ of a graph $H$ plays a very important role in solving the $\#_p\Hom(H)$ problem. Let  $\vf$ be a homomorphism from a graph $G$ to $H$. Then composing $\vf$ with an element from $\Aut(H)$ we again obtain a homomorphism from $G$ to $H$. Thus, $\Aut(H)$ acts on the set $\Hom(G,H)$ of all homomorphisms from $G$ to $H$. If $\Aut(H)$ contains an automorphism $\pi$ of order $p$ (or \emph{$p$-automorphisms}), the cardinality of the orbit of $\vf$ is divisible by $p$, unless $\pi\circ\vf=\vf$, that is, the range of $\vf$ is within the set of fixed points $\Fix(\pi)$ of $\pi$. Therefore this orbit contributes 0 modulo $p$ into the total homomorphism count from $G$ to $H$. This motivates the following construction. Let $H^\pi$ denote the subgraph of $H$ induced by $\Fix(\pi)$. We write $H\rightarrow_p H'$ if there is $\pi\in\Aut(H)$ such that $H'$ is isomorphic to $H^\pi$. We also write $H\rightarrow^*_p H'$ if there are graphs $\vc H k$ such that $H$ is isomorphic to $H_1$, $H'$ is isomorphic to $H_k$, and $H_1\rightarrow_p H_2\rightarrow_p\dots\rightarrow_p H_k$. 

\begin{lemma}[\cite{ref:CountingMod2Ini}]\label{lem:aut-reduction}
Let $H$ be a graph and $p$ a prime number. Up to an isomorphism there is a unique smallest (in terms of the number of vertices) graph $H^{*p}$ such that $H\rightarrow_p^* H^{*p}$, and for any graph $G$ it holds 
\[
\hom(G,H)\equiv\hom(G,H^{*p})\pmod p.
\]
Moreover, $H^{*p}$ does not have automorphisms of order $p$.
\end{lemma}

Often Lemma~\ref{lem:aut-reduction} helps to reduce the complexity of modular counting.

\begin{example}\label{exa:3-coloring}
Consider the 3-Coloring problem. Since permuting colors in a proper coloring produces another proper coloring, the number of 3-colorings of a graph is always $0\pmod3$, that is, counting 3-colorings modulo 3 is trivial. From a more formal perspective, the $\#_33$-Coloring problem is equivalent to $\#_3\Hom(\sK_3)$, where $\sK_3$ is a complete graph with 3 vertices. Any permutation of the vertices of $\sK_3$ is an automorphism, and the cyclic permutation of all vertices has order $3$. Therefore by Lemma~\ref{lem:aut-reduction} $\#_3\Hom(\sK_3)$ is equivalent to counting homomorphisms to an empty graph.
\end{example}

Graphs in Lemma~\ref{lem:aut-reduction} can be replaced with relational structures without changing the result, see Lemma~\ref{lem:aut-reduction-structures} in Section~\ref{sec:factors}. We will call relational structures $p$-automorphisms \emph{$p$-rigid}. By Lemmas~\ref{lem:aut-reduction},~\ref{lem:aut-reduction-structures}, and~\ref{lem:reducedForm} it suffices to determine the complexity of $\NpCSP(\cH)$ for $p$-rigid structures $\cH$. 

Faben and Jerrum \cite{ref:CountingMod2Ini} and then G\"obel, Lagodzinski, and Seidel \cite{ref:CountingModPToTrees_gbel_et_al_LIPIcs} conjectured that $p$-automorphisms are the only reason why counting homomorphisms modulo $p$ can be easier than exact counting. 

\begin{conjecture}[\cite{ref:CountingMod2Ini,ref:CountingModPToTrees_gbel_et_al_LIPIcs}]\label{con:conjecture}
For a $p$-rigid graph $H$ the problem $\#_p\Hom(H)$ is \modp-complete (solvable in polynomial time) if and only if $\#\Hom(H)$ is \#P-complete (solvable in polynomial time).
\end{conjecture}

\paragraph{The existing results}
The research in modular counting has mostly been aimed at verifying Conjecture~\ref{con:conjecture}. In \cite{ref:CountingMod2Ini} Faben and Jerrum proved their conjecture in the case when $H$ is a tree and $p=2$. This result has been extended  by G\"{o}bel et al.\ first to the class of cactus graphs \cite{ref:CountingMod2ToCactus} and then to the class of square-free graphs \cite{ref:CountingMod2ToSquarefree} (a graph is square-free if it does not contain a 4-cycle), still for $p=2$. Next, G\"obel, Lagodzinski, and Seidel confirmed Conjecture~\ref{con:conjecture} for trees and arbitrary prime $p$ \cite{ref:CountingModPToTrees_gbel_et_al_LIPIcs}. Kazeminia and Bulatov \cite{ref:CountingModPToSquarefree} confirmed the conjecture in the case of square-free graphs and arbitrary prime $p$. Focke et al.\ \cite{ref:CountingMod2ToKMinor,Focke21:counting} used some techniques from \cite{ref:CountingModPToSquarefree} to prove the conjecture for $K_4$-minor-free graphs and $p=2$. Finally, Lagodzinski et al.\ \cite{ref:CountingModPToK33free} considered quantum graphs and quantum homomorphisms, where quantum graphs are simply formal sums of graphs, and quantum homomorphisms are homomorphism-like constructions for quantum graphs defined in an appropriate way. They proved that $\ghomk pH$ for a quantum graph is \modp-complete whenever it is hard for any of its component graphs. Also, they showed that $\ghomk pH$ is polynomial time interreducible with $\ghomk p{H'}$, where $H'$ is a certain bipartite graph. Finally, they confirmed Conjecture~\ref{con:conjecture} for bipartite graphs that do not contain $K_{3,3}$ without an edge or a \emph{domino} as an induced subgraph (a domino is a bipartite graph obtained from $K_{3,3}$ by removing two non-incident edges). The last two results use intricate structural properties of graphs and a massive case analysis.

In this paper we confirm Conjecture~\ref{con:conjecture} for arbitrary graphs.

\begin{theorem}\label{the:main-intro}
For any prime $p$ and any graph $H$ the problem $\ghomk pH$ is solvable in polynomial time if and only if $\#\Hom(H^{*p})$ is solvable in polynomial time. Otherwise it is \modp-complete. 
\end{theorem}

By the result of Dyer and Greenhill \cite{Dyer00:complexity} $\#\Hom(H^{*p})$ (and therefore $\ghomk pH$) is solvable in polynomial time if and only if every connected component of $H^{*p}$ is a complete graph with all loops present or a complete bipartite graph. 

Our main technical contributions into the study of modular counting graph homomorphisms are twofold. First, we expand the signature of graphs with new relational symbols so that, on one hand, this does not change the complexity of the counting problem, and on the other hand, the resulting relational structure is richer and allows for more straightforward and concise arguments. Second, we do not design any particular gadgets for our reductions, but rather prove that the existence of such gadgets follows from the $p$-rigidity of a graph or a structure. 

In the next section we outline the main steps of the proof.


 


\section{Outline of the proof}\label{sec:outline}

In this section we give a detailed outline of the proof of  Theorem~\ref{the:main-intro}. The easiness part of Theorem~\ref{the:main-intro}, that is, if $\ghom{H^{*p}}$ is polynomial-time solvable then $\ghomk pH$ is polynomial-time solvable, follows from Lemma~\ref{lem:aut-reduction}. Indeed, for any input graph $G$, Lemma~\ref{lem:aut-reduction} guarantees that $\hom(G,H)\equiv\hom(G,H^{*p})\pmod p$, and the latter value can be computed in polynomial time. Therefore we focus on the proof of hardness. 

We will need some additional definitions and notation.

\smallskip

\textbf{Graphs and their expansions.}
Although the main result of the paper is about graphs, we use relational structures with richer signatures. Let $\sg,\sg'$ be signatures such that $\sg\sse\sg'$. A $\sg'$-structure $\cH'$ is said to be an \emph{expansion} of a $\sg$-structure $\cH$ if it has the same base set, and for every $\rel\in\sg$ the interpretations $\rel^\cH$ and $\rel^{\cH'}$ are equal. If $\sg'=\sg\cup\{\relo\}$, and $\relo$ has a certain interpretation $S$ on $H$, we often abuse the terminology and write $\cH+S$ instead of $\cH'$. 

The majority of structures we deal with are expansions of graphs. We therefore reserve $E$ to denote the graph edge relation. To deal with bipartite graphs properly it will be convenient to introduce a special kind of expansions of graphs that we call \emph{$\graph$-structures}. In order to avoid confusion we use a different font for such structures, say, $\sH$, rather than $H$. In the case of a nonbipartite graph $H$ the corresponding $\graph$-structure $\sH$ is just the graph itself. If $H$ is a bipartite graph, then $\sH$ includes the binary edge relation $E$ and two unary predicates $\Lside$ and $\Rside$ interpreted as the two parts of the bipartition (the choice of parts is arbitrary, but fixed). If we need to distinguish whether we are dealing with bipartite or nonbipartite case, we refer to $\sH$ as a \emph{$\bip$-} or \emph{$\nbip$-structure}, respectively. Every expansion of a $\graph$- ($\bip$-, $\nbip$-)structure is called a \emph{graph expansion} (\emph{bipartite graph expansion}, \emph{nonbipartite graph expansion}). We again use a different font in this case, say, $\cH$. A bipartite graph expansion $\cH$ has symbols $E,\Lside,\Rside$ in its signature, where $E$ is an edge relation of a bipartite graph, and $\Lside,\Rside$ are the two parts of the bipartition. By $\sH$ we denote the underlying $\graph$-structure with the base set $H$ and signature $\{E,\Lside,\Rside\}$ if $H$ is a bipartite graph, and signature $\{E\}$ otherwise. 
\smallskip

The proof of hardness is a chain of reductions, so here we explain each of these steps, the main constructions and the techniques used. Then we fill in the remaining details and proofs in the subsequent sections. The idea is to reduce one of three known $\#_p$P-hard problems depending on whether the graph $H$ is bipartite or not. In the case of bipartite graphs the problem is $\#_p \BIS(\alpha, \beta)$ of counting weighted independent sets in a bipartite graph to $\#_p \Hom(H)$. In the case of nonbipartite graphs, for simple graphs we reduce the $\#_p3\SAT$ problem of counting the number of satisfying assignments to a 3-CNF. Otherwise we reduce either $\#_p \BIS(\alpha, \beta)$ or $\#_p\IS$, the problem of counting independent sets in general graphs. This is done in five steps as shown in Figure~\ref{eq:chain_of_reductions}. We define all the intermediate problems as we go, and give a brief explanation on how the reduction works in each step. We start with a graph $H$ that satisfies the hardness conditions of Theorem~\ref{the:main-intro}. The first three steps (right to left) transform the graph $H$ into a relational structure $\cH^\const$ that is a graph expansion. The last two steps depend on whether or not $H$ is bipartite and in the nonbipartite case, whether or not the graph is simple. In the bipartite case we identify a further expansion $\cH^\sZ$ of $\cH^\const$ of a particular kind, and then show that $\#_p \BIS(\alpha, \beta)$ is reducible to the CSP over this expansion. In the nonbipartite case if $H$ has no loops, we find an expansion $\cH^{3\SAT}$ of $\cH^\const$ such that $\#_p3\SAT$ is reducible to $\#_p\CSP(\cH^{3\SAT})$. Otherwise we find an expansion $\cH^\dg$ of $\cH^\const$ such that either $\#_p \BIS(\alpha, \beta)$ or $\#_p\IS$ is reducible to $\#_p\CSP(\cH^\dg)$.
\begin{figure}[t]
    \centering
\begin{align*}
    &\#_p\BIS(\al, \beta)\leq_T 
    \#_p \CSP(\cH^\sZ) 
    &\leq_T& \\
    &\text{$\#_p3\SAT$}
    \leq_T \#_p\CSP(\cH^{3\SAT}) 
    &\leq_T& \ \  \#_p\CSP(\cH^\const)
    \leq_T \ghomk{p}{H^{*p}}
    \leq_T  \ghomk pH\\
    &\#_p\BIS [\#_p\IS]\leq_T 
    \#_p \CSP(\cH^\dg) 
    &\leq_T&
\end{align*}
    \caption{The chain of reductions}
    \label{eq:chain_of_reductions}
\end{figure}

\subsection{$\ghomk{p}{H^{*p}} \leq_T \ghomk pH$} 

Recall that $H^{*p}$ denotes the $p$-rigid graph obtained from $H$ by means of $p$-automorphisms in Lemma~\ref{lem:aut-reduction}. In this step of the proof, we use Lemma~\ref{lem:aut-reduction} to obtain a parsimonious (modulo $p$) reduction from $\ghomk{p}{H^{*p}}$ to $\ghomk pH$. 

\subsection{$\#_p\CSP(\cH^\const)\leq_T \ghomk{p}{H^{*p}}$} 
\label{sec:to-constants} 

One of the standard techniques in studying homomorphisms is \emph{pinning}. Let $\cG,\cH$ be two similar relational structures. To pin a vertex $v\in G$ to a vertex $a\in H$ means to require that we only consider homomorphisms of $\cG$ to $\cH$ that map $v$ to $a$. Clearly, if we allow pinning in a problem such as $\NCSP(\cH)$, it only becomes harder. However, often under certain conditions the problem with pinning can be reduced to the one without. In order to make this connection precise we again translate it into the terminology of relational structures.

Let $\cH$ be a $\sg$-structure and $a\in H$. The \emph{constant relation} $C_a=\{(a)\}$ is a unary relation that contains just a single element. Consider the expansion $\sg^\const$ of the signature $\sg$, given by $\sg^\const=\sg\cup\{C_{H,a}\mid a\in H\}$. Here we assume that $C_{H,a}\not\in \sg$ for any $a\in H$. For the structure $\cH$ by $\cH^\const$ we denote the $\sg^\const$-structure with the same base set as $\cH$ and such that for every $a\in H$ the predicate symbol $C_{H,a}\in\sg^\const$ is interpreted as $C_a$, and for every symbol $R\in\sg$ the interpretations $R^\cH$ and $R^{\cH^\const}$ are the same. It is easy to see that $\CSP(\cH^\const),\NCSP(\cH^\const)$ or $\NpCSP(\cH^\const)$ are the same as $\CSP(\cH),\NCSP(\cH),\NpCSP(\cH)$ with pinning allowed. 

Thus the claim that the problem with pinning is reducible to the one without is equivalent to saying that the problem over $\cH^\const$ is reducible to the one over $\cH$. Such statements were proved in \cite{Bulatov05:classifying} for the decision CSP, i.e., that  $\CSP(\cH^\const)$ is polynomial time reducible to $\CSP(\cH)$, provided $\cH$ is a core, that is, it does not have endomorphisms that are not automorphisms. It is proved in \cite{ref:BULATOV_TowardDichotomy} that $\NCSP(\cH^\const)$ is polynomial time reducible to $\NCSP(\cH)$ for any $\cH$. In the case of $\ghomk pH$ it was proved in \cite{ref:CountingMod2ToSquarefree} for $p=2$ and in \cite{ref:CountingModPToTrees_gbel_et_al_LIPIcs} for arbitrary prime $p$ that if $H$ is $p$-rigid then $\NpCSP(H^\const)$ is polynomial time reducible to $\ghomk pH$. In Section~\ref{sec:adding-constants} we prove a generalization of this result for arbitrary relational structures.

\begin{theorem}\label{the:adding-constants-intro}
Let $\cH$ be a $p$-rigid $\sg$-structure. 
Then $\NpCSP(\cH^\const)$ is polynomial time reducible to $\NpCSP(\cH)$.
\end{theorem}

\begin{remark}\label{rem:constants}
For any relational structure $\cH$ the structure $\cH^\const$ does not have any nontrivial automorphisms, in particular, it is $p$-rigid, and so is any of its induced substructures. We will assume this from this point on, and this is of course one of the major advantages of working with structures of this kind. However, several times we will need to work with direct powers of the structure $\cH^\const$, and such powers of $\cH^\const$ may have nontrivial automorphisms. The reason is that formally speaking if $\cH$ is a $\sg$-structure then $\cH^\const$ has the signature $\sg\cup\{C_{H,a}\mid a\in H\}$. Therefore, so does, say $(\cH^\const)^2$. For the latter structure $(b,c)\in C_{H,a}^{(\cH^\const)^2}$ if and only if $b,c\in C_{H,a}^{\cH^\const}$, that is, if and only if $b=c=a$. This means that the involution, that is, the mapping  $(a,b)\mapsto(b,a)$, is an automorphism of $(\cH^\const)^2$.
\end{remark}

Further reductions are split into three cases: bipartite graphs without loops, non-bipartite graphs without loops, and graphs with loops. 
\subsection{$\#_p \CSP(\cH^\sZ)\leq_T\#_p\CSP(\cH^\const)$}

We begin with the bipartite case without loops. In this step we first identify a substructure, a \emph{thick Z-graph}, of $\cH^\const$ of a special kind. We denote the structure $\cH^\const$ augmented with this new substructure as a new predicate by $\cH^\sZ$. Then it turns out that a thick $\sZ$-graph can be defined in such a way that $\NpCSP(\cH^\sZ)$ is polynomial time reducible to $\NpCSP(\cH^\const)$, thus completing this step of reduction. 

Before we consider specifically bipartite graphs, we explain a technique that is used in both, bipartite and nonbipartite cases. 

\subsubsection{Primitive positive definitions and polynomial time reductions}\label{sec:pp-defs}

Primitive positive definitions (pp-definitions for short) have played a major role in the study of the CSP. It has been proved in multiple circumstances that expanding a relational structures with pp-definable relations does not change the complexity of the corresponding CSP. This has been proved for the decision CSP in \cite{Jeavons:algebraic,Bulatov05:classifying} and the exact Counting CSP \cite{ref:BULATOV_TowardDichotomy}. The reader is referred to \cite{ref:POlymorphismAndUsethem_barto2017polymorphisms} for details about primitive positive definitions and their use in the study of the CSP.

Let $=_H$ denote the equality relation on the set $H$.
Let $\cH$ be a relational structure with the base set $H$. A \emph{primitive positive} formula in $\cH$ is a first-order formula 
\[
\exists \vc ys\Phi(\vc xk,\vc ys),
\] 
where $\Phi$ is a conjunction of atomic formulas of the form $z_1=_Hz_2$ or $\rel(\vc z\ell)$, $\vc z\ell\in\linebreak\{\vc xk,\vc ys\}$, and $\rel$ is a predicate of $\cH$. We say that $\cH$ \emph{pp-defines} a predicate $\rel$ if there exists a pp-formula $\exists \vc ys\Phi(\vc xk,\vc ys)$ such that 
\[\rel(\vc xk)=\exists \vc ys\Phi(\vc xk,\vc ys),\]
that is, $(\vc ak)\in\rel$ if and only if for some values $\vc bs\in H$ the formula $\Phi(\vc ak,\vc bs)$ is true.

Let $\cH$ be a relational structure and $\rel$ a relation on $H$. By $\cH+\rel$ we denote the expansion of $\cH$ with the relation $\rel$. Jeavons et al.\ \cite{Jeavons:algebraic} and Bulatov and Dalmau \cite{ref:BULATOV_TowardDichotomy} proved that if $\cH$ pp-defines $\rel$ then $\CSP(\cH+\rel)$ (respectively, $\NCSP(\cH+\rel)$) is polynomial time reducible to $\CSP(\cH)$ (respectively, to $\NCSP(\cH)$). 

The following theorem shows that expanding a relational structure $\cH$ with a pp-definable predicate does not change the complexity of modular counting provided $\cH$ is a graph expansion and is $p$-rigid.

\begin{theorem}\label{the:pp-intro}
Let $p$ be a prime and $\cH$ a $p$-rigid relational structure that is a graph expansion. If $\rel$ is pp-definable in $\cH$ then $\NpCSP(\cH+\rel)$ is polynomial time reducible to $\NpCSP(\cH)$.
\end{theorem}

A unary relation that is pp-definable in a relational structure is said to be a \emph{subalgebra} of the structure.

While the analogous reduction in \cite{Jeavons:algebraic,Bulatov05:classifying} is straightforward and that in \cite{ref:BULATOV_TowardDichotomy} uses interpolation techniques, the main tool in proving Theorem~\ref{the:pp-intro} is careful counting homomorphisms modulo $p$. A full proof for Theorem~\ref{the:pp-intro} is given in Section~\ref{sec:pp-def}. We show using the $p$-rigidity of $\cH$ and the M\"obius inversion formula that $\rel$ has a pp-definition of a special kind that allows for a polynomial time reduction. However, as we will see repeatedly in Section~\ref{sec:pp-def}, the existence of such a pp-definition depends on the structure of direct powers of $\cH$, or more precisely, on the structure of automorphisms of direct powers of $\cH$. This requires an in-depth analysis of the automorphism group of $\cH^\ell$ in Section~\ref{sec:auto}. Note that the required properties of $\Aut(\cH^\ell)$ are only known for nonbipartite graphs and we prove them for graph expansions, including bipartite graph extensions, and this restricts the scope of Theorem~\ref{the:pp-intro} to structures of this kind.

\subsubsection{Thick Z-Graphs and $\cH^\sZ$}
We now focus on bipartite graph expansions.

\begin{figure}[h]
    \centering
    \includegraphics[height=4cm]{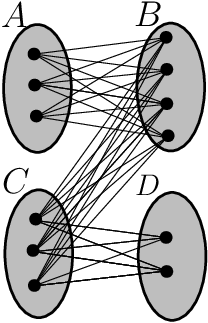}
    \caption{Thick Z-graph}
    \label{fig:generalizedN-intro}
\end{figure}

The special type of bipartite graphs we need in this step is shown in Figure~\ref{fig:generalizedN-intro}. More precisely, a $\bip$-structure $\sZ=(\Lside^\sZ \cup \Rside^\sZ,E,\Lside,\Rside)$ is said to be a \emph{thick Z-graph} if its vertices can be partitioned into four nonempty sets $A,B,C,D$, where $A,C\sse \Lside^\sZ, B,D\sse \Rside^\sZ$ in such a way that the sets $A\cup B,C\cup B,C\cup D$ induce complete bipartite subgraphs of $\sZ$, and there are no edges between vertices from $A\cup D$. The relational structure $\cH^\const$ we deal with is a bipartite graph expansion. We say that the thick Z-graph $\sZ$ is pp-definable in $\cH^\const$ if the sets $\Lside^\sZ\sse \Lside^{\cH^\const}$ and $\Rside^\sZ\sse \Rside^{\cH^\const}$ are pp-definable in $\cH$, and $\Lside^\sZ \cup \Rside^\sZ$ induces $\sZ$. 

The structure $\cH^\sZ$ is defined as the expansion of $\cH$ by two new unary predicates $\LZ$ and $\RZ$ interpreted on $H$ as $\Lside^\sZ$ and $\Rside^\sZ$, respectively. These new predicates can be used in an input structure $\cG$ of $\NpCSP(\cH^\sZ)$ to force some vertices of $\cG$ to be mapped into $\sZ$. 

\begin{lemma}\label{lem:getting-Z-graph-intro}
Let $\cH$ be a bipartite graph expansion. If the underlying bipartite graph of $\cH$ is not a complete bipartite graph, then $\cH^\const$ contains a pp-definable thick Z-graph.
\end{lemma}

\subsection{$\#_p\BIS(\alpha, \beta )\leq_T \#_p \CSP{(\cH^\sZ})$} 

In the final step for the bipartite case we reduce $\#_p\BIS(\al,\beta)$ to $\NpCSP{(\cH^\sZ})$. The problem $\#_p\BIS(\alpha,\beta )$ was introduced by G{\"o}bel, Lagodzinski and Seidel \cite{ref:CountingModPToTrees_gbel_et_al_LIPIcs}. Let $\al,\beta\not\equiv0\pmod p$. Then $\#_p\BIS(\al,\beta)$ is the problem defined as follows: given a bipartite graph, or, actually, a $\bip$-structure, $\sG$ with bipartition $\Lside^\sG,\Rside^\sG$, find the value
\[
\mathcal{Z}_{\alpha,\beta}(\sG)=\sum_{I\in\IS(\sG)}\al^{|I\cap \Lside^\sG|}\cdot\beta^{|I\cap \Rside^\sG|}
\]
modulo $p$, where $\IS(\sG)$ denotes the set of independent sets of $\sG$. In the same paper $\#_p\BIS(\al,\beta)$ was also shown to be \modp-complete for any $\al,\beta\not\equiv0\pmod p$. 

Before explaining the argument we need to introduce one more concept.

We sometimes need to restrict the set of homomorphisms between relational structures $\cG$ and $\cH$ to those that map specific elements of $\cG$ to specific sets of elements of $\cH$. This can be achieved in two ways. First, the signature of $\cG,\cH$ can be expanded by introducing unary predicates similar to constant relations as in Section~\ref{sec:to-constants}. However, a more flexible way is to use structures with distinguished vertices.

A pair $(\cH,\vv a)$, where $\vv a=(\vc ak)\in H^k$ for some $k$, will be called a \emph{relational structure $\cH$ with distinguished vertices $\vv a$}. For structures with distinguished vertices $(\cG,\vv a),(\cH,\vv b)$, $\vv b=(\vc bk)\in H^k$, such that $\cG,\cH$ are similar, a homomorphism from $(\cG,\vv a)$ to $(\cH,\vv b)$ is a homomorphism $\vf$ from $\cG$ to $\cH$ that maps $a_i$ to $b_i$, $i\in[k]$. The set of all homomorphisms from $(\cG,\vv a)$ to $(\cH,\vv b)$ is denoted by $\Hom((\cG,\vv a),(\cH,\vv b))$. Also, $\hom((\cG,\vv a),(\cH,\vv b))=|\Hom((\cG,\vv a),(\cH,\vv b))|$.

This notion can be slightly generalized replacing $\vv b$ with a sequence $\vc Bk\sse H$:
\begin{align*}
\Hom((\cG,\vv a),(\cH,\vc Bk))&=\bigcup_{b_i\in B_i, i\in[k]} \Hom((\cG,\vv a),(\cH,\vc bk))\qquad\text{and}\\
\hom((\cG,\vv a),(\cH,\vc Bk))&=\sum_{b_i\in B_i, i\in[k]} \hom((\cG,\vv a),(\cH,\vc bk))
\end{align*}

Observe that $\#_p\BIS(\al,\beta)$ can also be viewed as $\NpCSP(\sZ)$, where $\sZ$ is a thick Z-graph with parts $A,B,C,D$, $A\cup C= \Lside^\sZ, B\cup D=\Rside^\sZ$, such that $\al=\frac{|A|}{|C|}$ and $\beta=\frac{|D|}{|B|}$. Suppose that $F$ is an instance of $\#_p\BIS(\al,\beta)$, that is, a bipartite graph. The observation above means that, as $\cH^\sZ$ contains a pp-definable thick Z-graph $\sZ$ we can consider homomorphisms from $F$ to $\cH^\sZ$ and force all the vertices of $F$ to be mapped into $\sZ$ thus simulating $\#_p\BIS(\al,\beta)$. If none of $|A|,|B|,|C|,|D|$ is $0\pmod p$, this gives us the \modp-completeness of $\NpCSP{(\cH^\sZ})$. However, there is no way to guarantee this and we need to make one more step to overcome this hurdle. 

\begin{lemma}\label{lem:z-gadgets-intro}
Let $\sg^\sZ$ be the signature of $\cH^\sZ$. There are  $\sg^\sZ$-structures $(\cK_L,x),(\cK_R,x)$ with distinguished vertices such that 
\begin{align*}
    \begin{split}
        \hom((\cK_L,x),(\cH^\sZ, A)) \equiv \alpha_1 &\not \equiv 0 \pmod p,\\
        \hom((\cK_L,x),(\cH^\sZ, C)) \equiv \alpha_2 &\not \equiv 0 \pmod p,\\
        \hom((\cK_R,x),(\cH^\sZ, D)) \equiv \beta_1  &\not \equiv 0 \pmod p,\\
        \hom((\cK_R,x),(\cH^\sZ, B)) \equiv \beta_2  &\not \equiv 0 \pmod p,
    \end{split}
\end{align*}
and 
\begin{align*}
    \begin{split}
       \hom((\cK_L,x),(\cH^\sZ,v)) & =0, \qquad \text{for $v\in  \Lside^\cH-(A\cup C)$}, \\
       \hom((\cK_R,x),(\cH^\sZ,v) & =0    \qquad \text{for $v\in \Rside^\cH - (B\cup D)$}.
    \end{split}
\end{align*}
\end{lemma}

We need a similar result for other cases, so we prove a more general result in Section~\ref{sec:nested}.

\begin{figure}[h]
    \centering
    \includegraphics[height=3.5cm]{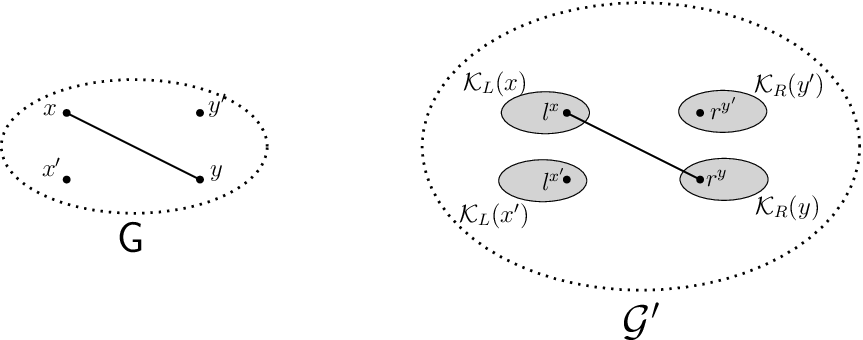}
    \caption{Reduction of BIS}
    \label{fig:z-reduction}
\end{figure}

Let $\sG$ be an instance of $\#_p\BIS(\al_1\al_2^{-1},\beta_1\beta_2^{-1})$. For every vertex $v$ of $\sG$ we attach a copy of $(\cK_L,x)$ if $v\in \Lside^\sG$, and a copy of $(\cK_R,x)$ if $v\in \Rside^\sG$, identifying the distinguished vertex $x$ with $v$ as shown in Figure~\ref{fig:z-reduction}. We then show that
\[
\hom(\cG',\cH^\sZ)\equiv\sum_{I\in\IS(G)}(\al_1\al_2^{-1})^{|I\cap \Lside^G|}\cdot(\beta_1\beta_2^{-1})^{|I\cap \Rside^G|}\pmod p,
\]
where $\cG'$ is the structure obtained by applications of $(\cK_L,x),(\cK_R,x)$ to $G$. The result follows.

\subsection{$\#_p\CSP(\cH^{3\SAT})\le_T\#_p\CSP(\cH^\const)$}\label{sec:H3SAT} 
In this step, we assume that the graph $H$ is non-bipartite and contains no loops. We need a special type of structure that somehow simulates the $3\SAT$ problem. A relational structure $\sS=(O\cup I,R_{ijk},i,j,k\in\{O,I\})$ is said to be a \emph{$3\SAT$ structure} if $O,I\ne\eps$ and 
\[
R_{ijk}=(O\cup I)^3-\{(a,b,c)\mid a\in i, b\in j, c\in k\},
\]
for every $i,j,k\in\{O,I\}$. 
We say that the $3\SAT$ structure $\sS$ is pp-definable in $\cH^\const$ if the sets $O,I$, and $O\cup I$, the equivalence relation on $O\cup I$ with equivalence classes $O$ and $I$, as well as the relations $R_{ijk},i,j,k\in\{O,I\}$ are pp-definable in $\cH$. 

If the $3\SAT$ structure $\sS$ is pp-definable in $\cH^\const$, then $\cH^{3\SAT}$ is defined as the expansion of $\cH^\const$ by a new unary predicate $O\cup I$ and the relations $R_{ijk},i,j,k\in\{O,I\}$. The new unary predicate can be used in an input structure $\cG$ of $\NpCSP(\cH^{3\SAT})$ to force some vertices of $\cG$ to be mapped into $\sS$. 

The following lemma is straightforward from the results of \cite{ref:BULATOV_HColoring}.

\begin{lemma}\label{lem:thick-3SAT}
Let $\cH$ be a nonbipartite graph expansion (of a graph without loops) that contains all the constant relations. If the underlying graph of $\cH$ is not a complete graph or a single vertex, then $\cH^\const$ contains a pp-definable $3\SAT$ structure.
\end{lemma}

\subsection{$\#_p3\SAT\le_T\#_p\CSP(\cH^{3\SAT})$} 

In this step we need another construction that has proved to be useful in the CSP research. Let $\cH$ be a $\sg$-structure and $\th$ an equivalence relation on $H$. By $\cH/_\th$ we denote the \emph{factor structure} defined as follows. 
\begin{itemize}\label{def:factor-structure}
    \item
    $\cH/_\th$ is a $\sg$-structure.
    \item 
    The base set of $\cH/_\th$ is $H/_\th=\{a/_\th\mid a\in H\}$, where $a/_\th$ denotes the $\th$-class containing $a$.
    \item
    For any $\rel\in\sg$, say, $k$-ary, $\rel^{\cH/_\th}=\{(a_1/_\th\zd a_k/_\th)\mid (\vc ak)\in\rel^\cH\}$.
\end{itemize}  
Now, let $G\sse H$ be a subalgebra of $\cH$, that is, a unary relation pp-definable in $\cH$, $\cG$ the substructure of $\cH$ induced by $G$, and $\th$, an equivalence relation on $G,$ a congruence of $\cH$. Then $\cG/_\th$ is said to be a quotient structure of $\cH$. Note that in this definition $\th$ is pp-definable in $\cH$ rather than in $\cG$.

If $\cG$ is a quotient structure of $\cH$, then $\CSP(\cG)$ (the decision problem) is polynomial time reducible to $\CSP(\cH)$, \cite{Bulatov05:classifying}, and $\NCSP(\cG)$ (the exact counting problem) is polynomial time reducible to $\NCSP(\cH)$, \cite{ref:BULATOV_TowardDichotomy}. In Section~\ref{sec:factor} we prove that a similar reducibility holds for modular counting as well, provided $\cH$ is a graph expansion.

\begin{theorem}\label{the:quotient-intro}
Let $p$ be a prime and $\cH$ a $p$-rigid relational structure that is a graph expansion. If $\cG$ is a quotient structure of $\cH$ then $\NpCSP(\cG)$ is polynomial time reducible to $\NpCSP(\cH)$.
\end{theorem}

If $H$ is a simple nonbipartite graph, then by Lemma~\ref{lem:thick-3SAT} there is a subalgebra $G=O\cup I$ of $\cH^{3\SAT}$ and a congruence $\th$ on $G$ with classes $O$ and $I$ such that the quotient structure $\cG/_\th$ of $\cH^{3\SAT}$ contains the relations $R^\sS_{ijk}$, $i,j,k\in\{0,1\}$, where we treat $0=O/_\th, 1=I/_\th$. These relations are given by 3-clauses on $\{0,1\}$, and therefore implement $3\SAT$. By Theorem~\ref{the:quotient-intro} $\#_p3\SAT$ is polynomial time reducible to $\NpCSP(\cH^{3\SAT})$.

\subsection{$\#_p\CSP(\cH^\dg)\le_T\#_p\CSP(\cH^\const)$}

If the graph $H$ has loops, we construct an expansion of $\cH^c$ with one of the two kinds of subalgebras. A reflexive graph $G=(V,E)$ is called a \emph{thick star} if there is a partition of $V$ into $V_0,V_1\zd V_k$, $k\ge2$, such that the subgraph $G|_{V_i}$ of $G$ induced by $V_i$ is a clique with all loops present, every vertex from $V_0$ is connected with an edge with every other vertex, and there are no other edges. A graph $G=(V,E)$ is said to be an \emph{independent 3-path} if there is a partition of $V$ into $V_0,V_1,V_2$ such that $G|_{V_2}$ is an independent set and contains no loops, $G|_{V_1}$ is a disjoint union of complete graphs $V_1^1\zd V_1^k$ with all loops present, $G|_{V_0\cup V_1^i}$ is a complete graph with all loops present for $i\in[k]$, every vertex of $V_2$ is connected with all the vertices of $V_0$, but to no vertex of $V_1$.

\begin{figure}[h]
    \centering
    \begin{subfigure}[t]{0.45\textwidth}
    \centering
    \includegraphics[height=4.5cm]{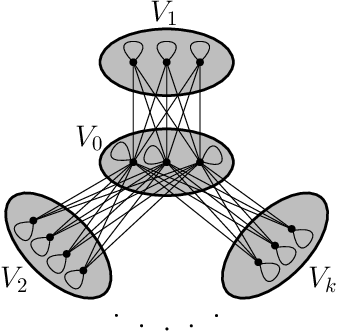} 
    \caption{}
    \end{subfigure}
    ~
    \begin{subfigure}[t]{0.45\textwidth}
    \centering
    \includegraphics[height=4.5cm]{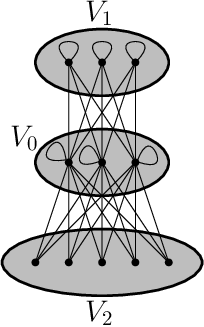}
    \caption{}
    \end{subfigure}
    \caption{The structure of (a) thick star and (b) an independent 3-path.}
    \label{fig:thick-star}
\end{figure}

\begin{lemma}\label{lem:reflexive-reduction-intro}
Let $H$ be a connected graph that is not a clique and contains a loop. Then there exists a subalgebra $W$ of $\cH^\const$ such that $H|_W$ is a thick star (with $k\ge2$) or an independent 3-path.
\end{lemma}

Finally, let $\cH^\dg$ denote the expansion of $\cH^\const$ with the unary predicate $W$.

\subsection{$\#_p\BIS, \ \#_p\IS \le_T\#_p\CSP(\cH^\dg)$}

The reduction from $\#_p\BIS$ (or from $\#_p\IS$) is less intuitive than that in the case of bipartite graphs, because neither a thick star nor independent 3-path are bipartite graphs. A simpler case is when $V_1=\emptyset$ in an independent 3-path. If this is the case, then the independent 3-path is very close to the homomorphism analogue of $\#\IS$, which is $\Hom(K_2^*)$, where $K_2^*$ is a 2-element edge, say $01$, with a loop attached to $1$. Indeed, for any graph $G$ and any homomorphism $\vf: G\to K_2^*$, the preimage of 0 is an independent set. Vice versa, for any independent set $I$ of $G$ the mapping that takes the vertices from $I$ to 0 and the rest to 1 is a homomorphism. In the case of independent 3-paths with $V_1=\emptyset$ the reduction works the same way except we also need to account for the sizes of $V_0,V_2$. This can be done in a manner similar to finding a non-degenerate Z-graph.

In the remaining cases it will be convenient to uniformize the notation for both thick stars and independent 3-paths. Let $A=V_0$ and $B=V_1$ in both cases. Then let $C=V_2\cup\dots\cup V_k$ for thick stars and $C=V_2$ in the case of independent 3-paths. Then, as is easily seen $A$ and $A\cup B$ are subalgebras of $\cH^\const$. Indeed, $A\cup B=W\cap N_H(b)$ where $W$ is the subalgebra identified in Lemma~\ref{lem:reflexive-reduction-intro} and $b$ is an arbitrary vertex from $B$. Then $A=(A\cup B)\cap N_H(c)$, where $c$ is an arbitrary vertex from $C$. Similar to Z-graphs, we are interested in non-degenerate thick stars ind independent 3-paths.

\begin{lemma}\label{lem:nondegenerate-star-intro}
Let $W$ be a subalgebra of $\cH^\const$ such that $H|_W$ is a thick star or an independent 3-path. There are  structures $(\cK'_L,x),(\cK'_R,x)$ with distinguished vertices such that 
\begin{align*}
    \begin{split}
        \hom((\cK'_L,x),(\cH^\dg, A)) &\equiv 1  \pmod p,\\
        \hom((\cK'_L,x),(\cH^\dg, B)) &\equiv 1  \pmod p,\\
        \hom((\cK'_R,x),(\cH^\dg, A\cup B)) &\equiv 1  \pmod p,\\
        \hom((\cK'_R,x),(\cH^\dg, C)) &\equiv 1  \pmod p,
    \end{split}
\end{align*}
and 
\begin{align*}
    \begin{split}
       \hom((\cK'_L,x),(\cH^\dg,v)) & =0 \qquad \text{for $v\not\in A\cup B$}, \\
       \hom((\cK'_R,x),(\cH^\dg,v) & =0    \qquad \text{for $v\not\in W$}.
    \end{split}
\end{align*}
\end{lemma}

Similar to the bipartite case, given an instance $\sG$ of an instance of $\#_p\BIS(1,1)$, for every vertex $v$ of $\sG$ we attach a copy of $(\cK'_L,x)$ if $v\in \Lside^\sG$, and a copy of $(\cK'_R,x)$ if $v\in \Rside^\sG$, identifying the distinguished vertex $x$ with $v$. We then show that
\[
\hom(\cG',\cH^\dg)\equiv|\IS(G)|\pmod p,
\]
where $\cG'$ is the structure obtained by applications of $(\cK'_L,x),(\cK'_R,x)$ to $G$. The key to the equality above is that for any homomorphism $\vf:\cG'\to\cH^\dg$ the vertices from $\Lside^\sG$ mapped to $B$ and the vertices from $\Rside^\sG$ mapped to $C$ form an independent set of $\sG$.


\section{Properties of Relational Structures and M\"{o}bius inversion}\label{sec:gadgets}

In this section, first we prove several results about relational structures that will be used later. Then we claim the existence of certain well-behaved pp-definitions and homomorphisms for $p$-rigid relational structures.

\subsection{Factors, products and homomorphisms}\label{sec:factors}

For a relational structure $\cH$, an \emph{automorphism} is an injective homomorphism into itself. The automorphisms of $\cH$ form a group with respect to composition denoted $\Aut(\cH)$. The set of all automorphisms with $a\in H$ as a fixed point is the \emph{stabilizer} of $a$ denoted $\Stab_\cH(a)$. It is always a subgroup of $\Aut(\cH)$.

We will use two basic facts from group theory. First, for any prime $p$, if $G$ is a group and $|G|\equiv0\pmod p$, then $G$ contains an element of order $p$. In particular, if $|\Aut(\cH)|\equiv0\pmod p$ for a relational structure $\cH$, then $\cH$ has an automorphism of order $p$. We call such automorphisms \emph{$p$-automorphisms}.

Second, for a subgroup $H$ of a group $G$, a \emph{coset} of $G$ modulo $H$ is a set of the form $aH=\{ah\mid h\in H\}$ for an element $a\in G$. Then $G:H$ denotes the set of cosets of $G$ modulo $H$. In this case Lagrange's theorem holds claiming $|G|=|H|\cdot|G:H|$.

We will also need several concepts related to homomorphisms. In particular we use expansions of relational structures and structures with distinguished vertices and their homomorphisms introduced in Section~\ref{sec:outline}. 
Note that a relational structure $(\cG,\vv a)$ with distinguished vertices can be viewed as an expansion of $\cG$ with $k$ additional unary symbols, one for each distinguished vertex. In such an interpretation a homomorphism of structures with distinguished vertices is just a homomorphism between the corresponding expansions.

Next we introduce two types of products of structures. The \emph{direct product} of $\sg$-structures $\cH,\cG$, denoted $\cH\tm\cG$ is the $\sg$-structure with the base set $H\tm G$ and such that the interpretation of $\rel\in\sg$ is given by $\rel^{\cH\tm\cG}((a_1,b_1)\zd(a_k,b_k))=1$ if and only if $\rel^\cH(\vc ak)=1$ and $\rel^\cG(\vc bk)=1$. By $\cH^\ell$ we will denote the \emph{$\ell$th power} of $\cH$, that is, the direct product of $\ell$ copies of $\cH$. The direct product $(\cG,\vv x)\tm(\cH,\vv y)$ of structures $(\cG,\vv x),(\cH,\vv y)$ is defined to be $(\cH\tm\cG,(x_1,y_1)\zd(x_r,y_r))$. 

Let $[n]$ denote the set $\{1\zd n\}$. Two $r$-tuples $\vv{x} $ and $\vv{y}$ have the same \emph{equality type} if $x_i = x_j$ if and only if $y_i = y_j$ for $i,j\in[r]$. Let $(\cG,\vv x)$ and $(\cH,\vv y)$ be structures with $r$ distinguished vertices and such that $\vv x$ and $\vv y$ have the same equality type. Then $(\cG,\vv x)\odot(\cH,\vv y)$ denotes the structure that is obtained by taking the disjoint union of $\cG$ and $\cH$ and identifying every $x_i$ with $y_i$, $i\in[r]$. The distinguished vertices of the new structure are $\vc xr$.

The following statement is straightforward.

\begin{proposition}[(5.28),(5.30), p.61 of \cite{lovasz2012large}]\label{pro:ProductHom}
Let $(\cG,\vv x),(\cH,\vv y),(\cK,\vv z)$ be similar relational structures with $r$ distinguished vertices. Then
\begin{align*}
\hom((\cG,\vv x)\odot(\cH,\vv y),(\cK,\vv z)) &= \hom((\cG,\vv x),(\cK,\vv z))\cdot\hom((\cH,\vv y),(\cK,\vv z)); \\  
\hom((\cK,\vv z),((\cG,\vv x)\times(\cH,\vv y)) &= \hom((\cK,\vv z),(\cG,\vv x))\cdot\hom((\cK,\vv z),(\cH,\vv y)).   
\end{align*}
Moreover, $\vf\in\Hom((\cK,\vv z),((\cG,\vv x)\times(\cH,\vv y))$ if and only if there exist a homomorphism $\vf_1 \in \Hom((\cK,\vv z),(\cG,\vv x))$ and $\vf_2\in\Hom((\cK,\vv z),(\cH,\vv y))$ such that $\vf(v)=(\vf_1(v),\vf_2(v))$ for $v\in K$.
\end{proposition}

We will also need another simple observation. By $\inj((\cG,\vv{x}),(\cH,\vv{y}))$ we denote the number of injective homomorphisms from $(\cG,\vv x)$ to $(\cH,\vv y)$.

\begin{lemma}[\cite{ref:CountingMod2ToSquarefree}]
Let $(\cG,\vv{y})$ and $(\cH,\vv{y})$ be relational structures with $r$ distinguished vertices. If $\vv{x},\vv{y}$ do not have the same equality type, then $\inj((\cG,\vv{x}),(\cH,\vv{y}))=0$.
\end{lemma}

Factor structures (see page~\pageref{def:factor-structure}) often appear in relation with homomorphisms. If $\vf$ is a homomorphism from a structure $\cG$ to a structure $\cH$, then the \emph{kernel} $\th$ of $\vf$, denoted $\ker(\vf)$, is the equivalence relation on $G$ given by 
\[
(a,b) \in\th \text{ if and only if } \vf(a) =\vf(b).
\]
For an equivalence relation $\th$ on $G$ by $\hom_\th(\cG,\cH)$ and $\Hom_\th((\cG,\vv x),(\cH,\vv y))$ we denote the number of homomorphisms from $\cG$ to $\cH$ (from $(\cG,\vv x)$ to $(\cH,\vv y)$) with kernel $\th$.
The homomorphism $\vf$ gives rise to a homomorphism $\vf/_\th$ from $\cG/_\th$ to $\cH$, where $\vf/_\th(a/_\th)=\vf(a)$, $a\in G$. The homomorphism $\vf/_\th$ is always injective. 

We also define factor structures for structures with distinguished vertices as follows. Let $(\cH,\vv a)$ be a structure with $k$ distinguished vertices and $\th$ an equivalence relation on $H$. Then $(\cH,\vv a)/_\th=(\cH/_\th, (a_1/_\th\zd a_k/_\th))$. 

The following lemma is straightforward.

\begin{lemma}\label{lem:hom-for-factors}
Let $(\cH,\vc ak)$ be a $\sg$-structure with distinguished vertices and $\theta$ an equivalence relation on $H$. Then the mapping $\vf\colon H\to H/_\th$, where $\vf(x)=x/_\theta$ is a homomorphism from $(\cH,\vc ak)$ to $(\cH/_\th,(a_1/_\th\zd a_k/_\th))$.
\end{lemma}

\subsection{$p$-Reduced form of $\cH$}\label{sec:p-rigid}

We observed in the introduction that a major complication when studying the complexity of modular counting $\NpCSP(\cH)$ are $p$-automorphisms of $\cH$. We call a structure $\cH$ \emph{$p$-rigid} if it does not have $p$-automorphisms. 

As was mentioned, $p$-automorphisms allow for a reduction of $\NpCSP(\cH)$ as follows. Let $\pi$ be a $p$-automorphism of $\cH$. By $\pi^\ell$ we denote
 the composition $\pi\circ\dots\circ\pi$ of $\ell$ applications of $\pi$. For an instance $\cG$ and any homomorphism $\vf$ from $\cG$ to $\cH$ the mappings $\pi^\ell\circ\vf$, $\ell\in[p]$, are also homomorphisms. This means that the homomorphism $\vf$ makes a contribution into $\hom(\cG,\cH)$ other than $0\pmod p$ only if $\pi\circ\vf=\vf$. This happens only if $\vf$ maps $\cG$ to the set of \emph{fixed points} $\Fix(\pi)$ of $\pi$, that is, $\Fix(\pi)=\{a\in H\mid \pi(a)=a\}$. By $\cH^\pi$ we denote the relational structures obtained by restricting $\cH$ to $\Fix(\pi)$.

\begin{lemma}\label{lem:aut-reduction-structures} 
If $\cH$ is a relational structure, and $\pi$ an $p$-automorphism of $\cH$, then for any structure $\cG$
\begin{equation*}
    \hom(\cG,\cH)\equiv \hom(\cG, \cH^{\pi}) \pmod{p}
\end{equation*}
\end{lemma}

\begin{proof}
Let $H$ and $H^\pi$ denote the universes of $\cH$ and $\cH^{\pi}$ respectively. For a similar structure $\cG$ with universe $G$, we show that the number of homomoprhisms which use at least one element of $H- H^{\pi}$ is $0\pmod p$. 

Given any homomorphism $\FUNC{\vf}{\cG}{\cH}$, consider the  homomorphism  $\pi \circ \vf$. This is still a homomorphism  which is different from $\vf$ as there is some element $v \in G$ such that $\vf(v) \in H- H^{\pi}$, and so $\pi(\vf(v)) \neq \vf(v)$. On the other hand $\pi^p \circ \vf$ is just $\vf$, as $\pi$ is a $p$-automorphism. So $\pi$ acts as a permutation of order $p$ on the set $\Hom(\cG, \cH)$. Moreover, the orbit of $\pi$ containing a homomorphism that has at least one element from $H- H^\pi$ in its range has size $p$. 
\end{proof}

The binary relation $\rightarrow_p$ on relational structures is defined as follows. For relational structures $\cH$ and $\cK$, we have $\cH \rightarrow_p \cK$ if and only if there exists an automorphism $\pi$ of $\cH$, of order $p$, such that $\cH^{\pi}=\cK$. If there exists a sequence of structures $\cH_1,\cH_2\zd \cH_\ell$ such that $\cH = \cH_1 \rightarrow_p \cH_2 \rightarrow_p\dots \rightarrow_p \cH_\ell = \cK$, we write $\cH \rightarrow_p^{*} \cK$ and say that $\cH$ \emph{$p$-reduces} to $\cK$. If $\cK$ is $p$-rigid, it is called a \emph{$p$-reduced form} associated with $\cH$. The following lemma is proved in Section~\ref{sec:INdandISO} by a light modification of the proof in \cite{ref:CountingMod2Ini}.

\begin{lemma}[\cite{ref:CountingMod2Ini}]\label{lem:reducedForm}
For a relational structure $\cH$ there is (up to an isomorphism) exactly one $p$-rigid structure $\cH^{* p}$ such that  $\cH \rightarrow_p \cH^{* p}$.
\end{lemma}


\subsection{Indistinguishability and isomorphism}\label{sec:INdandISO}

In this section we prove a variation of Lovasz's theorem about homomorphism counts and graph isomorphisms \cite{ref:CountingMod2ToSquarefree,ref:CountingModPToTrees_gbel_et_al_LIPIcs}. 

For a set $H$, let $\Part(H)$ denote the poset of partitions of $H$, where $\zo$ denotes the single class partition, $\zz$ the partition into 1-element classes, and $\eta\le\th$ means that $\eta$ is the finer partition of the two. 

\begin{lemma}\label{lem:lovasz-structure}
Let $(\cG, \vv{x})$ and $(\cH,\vv y)$ be $p$-rigid relational structures with $r$ distinguished vertices. Then, $(\cG, \vv{x}) \cong (\cH,\vv y)$ if and only if 
\begin{equation}
\hom((\cK, \vv{z}) , (\cG, \vv{x})) \equiv \hom((\cK, \vv{z}),(\cH, \vv{y})) \pmod{p} \label{equ:lovasz}
\end{equation} 
for all relational structures $(\cK, \vv z)$ with $r$ distinguished vertices.
\end{lemma}

\begin{proof}
The proof goes along the same lines as that in \cite{ref:CountingModPToTrees_gbel_et_al_LIPIcs}. If $(\cG, \vv{x})$ and $(\cH,\vv y)$ are isomorphic, then \eqref{equ:lovasz} obviously holds for all relational structures $(\cK, \vv{z})$. 

For the other direction, suppose that \eqref{equ:lovasz} is true for all $(\cK, \vv z)$.

First, we claim that this implies that $\vv{x}$ and $\vv{y}$ have the same equality type. Indeed, if they do not, then without loss of generality there are $i,j\in[r]$ such that $x_i = x_j$ but $y_i\ne y_j$. 
Let $\cK$ be the relational structure with the base set $\{\vc xr\}$  with empty predicates, and $(\vc xr)$ as distinguished vertices. Then $\hom((\cK, \vv{x}) , (\cG, \vv{x}))=1 \ne 0=  \hom((\cK, \vv{x}),(\cH, \vv{y}))$, a contradiction with the assumption that \eqref{equ:lovasz} holds for all $(\cK,\vv{x})$.

We show by induction on the number of vertices in $\cK$ that 
\begin{equation}\label{lavasEquInj}
\inj((\cK, \vv{z}),(\cG, \vv{x})) \equiv \inj((\cK, \vv{z}),(\cH, \vv{y})) \pmod{p}.  
\end{equation}
for all $(\cK, \vv{z})$. Let $n_0 = |\{\vc xr \}| = |\{\vc yr\}|$ be the number of distinct elements in $\vv x,\vv{y}$. For the base case of the induction, consider a relational structure $(\cK, \vv{z})$  such that $|K| \leq n_0$. If $\vv{z}$ does not have the same equality type as $\vv x,\vv{y}$, then $\inj((\cK, \vv{z}),(\cG, \vv{x})) = \inj((\cK, \vv{z}),(\cH, \vv{y})) = 0$.
If $\vv{x}$ has the same equality type as $\vv x,\vv{y}$, the only homomorphisms from $(\cK,\vv z)$ to $(\cG,\vv x),(\cH,\vv y)$ are the ones that map $z_i$ to $x_i,y_i$, respectively. Therefore, $\inj((\cK, \vv{z}),(\cG,\vv{x}))=\inj((\cK, \vv{z}) , (\cH, \vv{y}))$.

For the inductive step, let $n > n_0$ and assume that (\ref{lavasEquInj}) holds for all $(\cK, \vv{z})$ with $|K| < n$. Let $(\cK,\vv{z})$ be a relational structure with $|K|=n$, and let $\th$ be an equivalence relation on $K$. Then, as is easily seen 
\begin{align*}
&\hom_\th((\cK,\vv z),(\cG,\vv x))=\inj((\cK/_\th,\vv z/_\th),(\cG,\vv x))\\
\text{ and }&\quad \\
&\hom_\th((\cK,\vv z),(\cH,\vv y))=\inj((\cK/_\th,\vv z/_\th),(\cH,\vv y)).
\end{align*}
Then
\begin{align*}
\hom((\cK,\vv{z}),(\cG,\vv{x})) &= \inj((\cK,\vv{z}),(\cG,\vv{x})) + \sum_{\th\in\Part(K)-\{\zz\}} \inj((\cK,\vv{z})/_\th,(\cG,\vv{x}))\\
\hom((\cK,\vv{z}),(\cH,\vv{y})) &= \inj((\cK,\vv{z}),(\cH,\vv{y})) + \sum_{\th\in\Part(K)-\{\zz\}} \inj((\cK,\vv{z})/_\th,(\cH,\vv{y}))
\end{align*}
Since $\hom((\cK,\vv{z}),(\cG,\vv{x}))\equiv\hom((\cK,\vv{z}),(\cH,\vv{y}))\pmod p$ and $\inj((\cK,\vv{z})/_\th,(\cG,\vv{x}))\equiv\inj((\cK,\vv{z})/_\th,(\cH,\vv{y}))\pmod p$ for all $\th\in\Part(K)-\{\zz\}$ we get $\inj((\cK,\vv{z}),(\cG,\vv{x}))\equiv\inj((\cK,\vv{z}),(\cH,\vv{y}))\pmod p$.

Finally, we prove that \eqref{lavasEquInj} for $(\cK, \vv{z})=(\cG,\vv x)$ implies $(\cG, \vv{x}) \cong (\cH, \vv y)$. An injective homomorphism from a relational structure to itself is an automorphism. Since $(\cG, \vv{x})$ is $p$-rigid, $|\Aut(\cG,\vv x)|=\inj((\cG,\vv x),(\cG,\vv x))\not\equiv0\pmod p$. Therefore  $\inj(\cG,\vv{x}),(\cH,\vv y))\not\equiv0\pmod p$, meaning there is an injective homomorphism from $(\cG,\vv x)$ to $(\cH,\vv y)$. In a similar way, there is an injective homomorphism from $(\cH,\vv y)$ to $(\cG,\vv x)$. Thus, the two structures are isomorphic.
\end{proof}

We give a proof of Lemma~\ref{lem:reducedForm} here.

\begin{proof}[Proof of Lemma~\ref{lem:reducedForm}] 
Suppose the contrary, that $\cH_1$ and $\cH_2$ are two different non-isomorphic $p$-reduced forms of $\cH$. By Lemma~\ref{lem:aut-reduction-structures} for any structure $\cG$
\[
    \hom(\cG, \cH_1) \equiv \hom(\cG,\cH)\equiv \hom(\cG, \cH_2) \pmod{p}
\]
By Lemma~\ref{lem:lovasz-structure} we can conclude $\cH_1 \cong \cH_2$. The result follows.
\end{proof}

\subsection{M\"obius inversion}\label{sec:mobius-inversion}

We will use the \emph{M\"obius inversion formula}. Let $H$ be a set and let $\FUNC{M,N}{\Part(H)}{\mathbb{Z}}$ be some functions satisfying 
\[
M(\th)=\sum_{\eta\ge\th} N(\eta).
\]
Then 
\[
N(\zz)=\sum_{\th\in\Part(H)}w(\th) M(\th),
\]
where the function $\FUNC{w}{\Part(H)}{\mathbb{Z}}$ is given by:
\begin{itemize}
  \item $w(\zo)=1$,
  \item for any partition $\th<\zo$, $w(\theta) = -\mathlarger{\sum}\limits_{\gm >\theta} w(\gm)$.
\end{itemize}

The M\"obius inversion formula will mainly be used as the following lemmas indicate.

\begin{lemma}\label{lem:mobius}
Let $p$ be prime and let $\cH$ be a relational structure such that, for every $\cG$, $\hom(\cG,\cH)\equiv c\pmod p$, where $c$ does not depend on $\cG$. Then $\cH$ has a $p$-automorphism.
\end{lemma}

\begin{proof}
We use the following parameters in the M\"obius inversion formula: the set $H$ is the base set of $\cH$, $M(\th)=\hom(\cH/_\th,\cH)$, $N(\th)=\inj(\cH/_\th,\cH)$. Then clearly $N(\zz)=\Aut(\cH)$. By the formula we have
\[
N(\zz)=\sum_{\th\in\Part(H)}w(\th) M(\th)=\sum_{\th\in\Part(H)}w(\th)\hom(\cH/_\th,\cH)\equiv c\cdot\sum_{\th\in\Part(H)}w(\th)\equiv0\pmod p.
\]
The second last equality is due to the assumption that $\hom(\cG,\cH)\equiv c\pmod p$ for every $\cG$. The last equality follows from the definition of $w(\th)$, namely, that $w(\zz) = -\mathlarger{\sum}\limits_{\gm >\zz} w(\gm)$, where $\gm>\zz$ holds for all $\gm\in\Part(H)-\{\zz\}$. Therefore $|\Aut(\cH)|\equiv0\pmod p$ and $\cH$ has a $p$-automorphism.
\end{proof}

We call a subset $A\sse\cH^r$ \emph{automorphism-stable} if there is $\ba\in A$ such that the set $\Stab(\ba,A)=\{\pi\in\Aut(\cH^r)\mid \pi(\ba)\in A\}$ is a subgroup of $\Aut(\cH^r)$. Note that $\Stab(\ba,A)$ is always nonempty, as it contains the identity mapping. Also, by $\Stab(\vc\ba k)$ we denote the subset of $\Aut(\cH^r)$ that contains all the automorphisms for which each of $\vc\ba k$ is a fixed point.

\begin{lemma}\label{lem:mobius-point}
Let $p$ be prime, $\cH$ a relational structure. 
\begin{itemize}
    \item[(1)]
    Let $A\sse H^r$ be an automorphism-stable set. If for every $\cG$ and $\vv x\in\cG$, $\hom((\cG,\vv x),(\cH,A))\equiv c\pmod p$, where $c$ does not depend on $\cG$ and $\vv x$, then the structure $\cH^r$ has a $p$-automorphism $\pi\in\Stab(\ba,A)$ for some $\ba\in A$.
    \item[(2)]
    Let $\vc\ba k\in\cH^r$. If for every $\cG$ and $\vc xk\in\cG$, 
    \[
    \hom((\cG,\vc xk),(\cH,\vc\ba k))\equiv c\pmod p,
    \]
    where $c$ does not depend on $\cG$ and $\vc xk$, then the structure $\cH^r$ has a $p$-automorphism $\pi\in\Stab(\vc\ba k)$.
\end{itemize}
\end{lemma}

\begin{proof}
(1) Similar to Lemma~\ref{lem:mobius} we use the M\"obius inversion formula on $\Part(H^r)$. Let $\vv a\in A$ be the element witnessing that $A$ is automorphism-stable and set $M(\th)=\hom((\cH^r/_\th,\ba/_\th),(\cH^r,A))$, $N(\th)=\inj((\cH^r/_\th,\ba/_\th),(\cH^r,A))$. Then $N(\zz)\ne0$, as it includes the identity mapping, and as before we have
\begin{align*}
N(\zz) &=\sum_{\th\in\Part(H^r)}w(\th) M(\th)\\ &=\sum_{\th\in\Part(H^r)}w(\th)\hom((\cH^r/_\th,x/_\th),(\cH^r,A))\equiv c\cdot\sum_{\th\in\Part(H^r)}w(\th)\equiv0\pmod p.
\end{align*}
Note that $N(\zz)=\Stab(\ba,A)$ and therefore is a subgroup of $\Aut(\cH^r)$. As it has order that is a multiple of $p$, it contains a $p$-automorphism.
 
(2) In this case the argument is essentially the same. 
\end{proof}

The way we will apply Lemmas~\ref{lem:mobius} and ~\ref{lem:mobius-point} is as follows. If $\cH^r$ is $p$-rigid and $A$ is an automorphism-stable set, then by Lemma~\ref{lem:mobius-point}(1), there is a gadget $(\cG, \vv x)$ such that $\hom((\cG,\vv x),(\cH,A)) \not \equiv 0\pmod p$. This means, however, that we can apply Lemma~\ref{lem:mobius-point} only when $\cH^r$ is $p$-rigid. It is clear that, the $p$-rigidness of $\cH$ does not imply that $\cH^r$ is $p$-rigid. Hence, for applying Lemma~\ref{lem:mobius-point} we need to study the group $\Aut(\cH^r)$.


\section{Automorphisms of Relational Structures}\label{sec:auto}


The goal of this section is to study the automorphism group of $\sH^\ell$ where $\sH$ is a $\graph$-structure. In general, for arbitrary $\sg$-structures $\cH_1$ and $\cH_2$, it is known that $\Aut(\cH_1)\times \Aut(\cH_2)$ is a subgroup of $\Aut(\cH_1 \times \cH_2)$. For the reverse inclusion, 
it is shown in \cite{ref:ProductOfGraphs} that for \emph{$\RT$-thin} non-bipartite graphs $G_1$ and $G_2$, $\Aut(G_1)\times \Aut(G_2)=\Aut(G_1\times G_2)$. 
The relation $\RT_G$ on $V(G)$ for a graph $G$ is defined as follows: vertices $x$ and $x'$ are in $\RT_G$, if and only if  $N_G(x) = N_G(x')$, where $N_G(x) = \{y | 
\nedge{x}{y} \in E(G) \}$. 
Clearly, $\RT_G$ is an equivalence relation. A graph $G$ is called $\RT$-thin (aka \emph{graph without false twins}) if all of its $\RT_G$-classes have size one. If the graph $G$ is clear from the context we will use $\RT$ rather than $\RT_G$.

We denote the quotient of $G$ modulo $\RT$ by $G/_\RT$. Given $x \in V(G)$, let $[x] = \{x'\in V (G) | N_G(x') = N_G(x)\}$ denote the $\RT$-class containing $x$. Then $V(G/_\RT)=\{[x]\mid x\in V(G)\}$ and $E(G/_\RT)=\{\nedge{[x]}{[y]}\mid \nedge xy\in E(G)\}$. Since $\RT$ is defined entirely in terms of $E(G)$, it is easy to see that for an isomorphism $\vf : G \rightarrow H$, we have $x\RT y$ if and only if $\vf(x)\RT\vf(y)$. Thus, $\vf$ maps equivalence classes of $\RT_G$ to equivalence classes of $\RT_H$, and can be defined to act on $V(G/_\RT)$ by $\vf([x]) = [\vf(x)]$. This observation implies that for a graph $G$ and $\vf\in\Aut(G)$, if for some $x\in V(G)$ it holds that $\vf(x)\in[x]$, then $\vf([x])\sse[x]$. In particular, the following statement holds. An automorphism of a graph $G$ is called \emph{local} if it is the identity mapping on $G/_\RT$.

\begin{lemma}\label{lem:R-factor-auto}
Let $G$ be a graph and $\vf\in\Aut(G)$. Then the order of $\vf$ as a permutation of $G/_\RT$ divides the order of $\vf$ on $G$. If $\vf$ is a $p$-automorphism, then either $\vf$ is local, or it gives rise to a $p$-automorphism of $G/_\RT$.
\end{lemma}

The definition of $\RT$ can be extended to a $\graph$-structure $\sH$ by applying it on its underlying graph $H$. Hence, the $\RT$-classes of $\sH$ are the $\RT$-classes of $H$. Note that the definition of local automorphism for $\graph$-structures is the same as that for graphs.

Another structural feature of $\graph$-structures we use is prime factorization. A \emph{trivial} $\bip$-structure $\sK_{1,1}$ is just an edge:  $K_{1,1}=\{ a,b\}$, $E^{\sK_{1,1}}=\{ (a,b),(b,a) \}$, $\Lside^{\sK_{1,1}}=\{ a \}$, and $\Rside^{\sK_{1,1}}=\{ b \}$, and a trivial $\nbip$-structure is just a vertex with a loop attached to it.
A $\graph$-structure is \emph{prime} with respect to direct product if it is nontrivial and cannot be represented as the product of two nontrivial $\graph$-structures. We call $\sH=\sH_1 \times\dots \times \sH_r$ a \emph{prime factorization} of $\sH$ if all the $\sH_i$'s are prime.

In Section~\ref{sec:bip-auto} we state Theorem~\ref{the:splitting_automorphisms} about the automorphism group of $\graph$-structures, which generalizes the result of \cite{ref:ProductOfGraphs} to also include bipartite graphs. Later on, we use this theorem to study the reduced form of $\sH^\ell$.

\subsection{Automorphisms of $\bip$-structures}\label{sec:bip-auto}

We start this section with two results, Theorem~\ref{the:splitting_automorphisms} and Lemma~\ref{lem:factoring_R-thin}, that help us to study the automorphism group of $\cH^\ell$ later in the section. These results are known for nonbipartite graphs and therefore for $\nbip$-structures, see \cite{ref:ProductOfGraphs}. We omit proofs for $\bip$-structures here and move them to Section~\ref{sec:appendix}.

\begin{lemma}\label{lem:factoring_R-thin}
If $\sG$ and $\sH$ are $\graph$-structures and have no isolated vertices, then $(\cG \times \cH)/_\RT \cong \cG/_\RT \times \cH/_\RT$, and $\FUNC{\psi}{(\cG \times \cH)/_\RT}{ G/_\RT \times H/_\RT}$ with $\psi ([(x,y)])=([x],[y])$ is an isomorphism. Also $[(x, y)] = [x] \times [y]$.
\end{lemma}

For an automorphism $\vf$ of a $\graph$-structure $\sG$, let $\FUNC{\psi^\RT}{\sG/_\RT}{ \sG/_\RT}$ be the automorphism of $\sG/_\RT$ induced by $\vf$.

We will often need to represent mappings on direct powers $\sH^\ell$ of structures. Let $\vf$ be such a mapping. Every element of $\sH^\ell$ is a tuple, so we denote by $\vf(a_1\zd a_\ell)$ the action of $\vf$ on such a tuple. However, if we also have a prime factorization $\sH=\sH_1\tm\dots\tm\sH_r$, every element of $\cH$ is again a tuple. In this case we denote the action of $\vf$ as 
\[
\vf\left(\vvecad{a_{1,1}}{a_{1,r}}\zd \vvecad{a_{\ell, 1}}{a_{\ell, r}}\right).
\]

\begin{theorem}\label{the:splitting_automorphisms}
 Let $\sG=\sG_1 \times \dots \times\sG_r$ be a prime factorization of a $\graph$-structure $\sG$, where for $v\in G$, we have $v=(\vc{v}{r})$. Then for any automorphism $\psi$ of $\sG^{\ell}/_\RT$, there is a permutation $\pi$ of $[\ell]\times [r]$ such that $\psi$ can be split into $r\ell$ automorphisms:
\begin{align*}
    \psi([v_1]\zd [v_\ell])= \left( \vvecad{\psi_{1,1}([v_{\pi(1,1)}])}{\psi_{1,r}([v_{\pi(1,r)}])},...,\vvecad{\psi_{\ell,1}([v_{\pi(\ell,1)}])}{\psi_{\ell,r}([v_{\pi(\ell,r)}])} \right),
\end{align*}
where for $(i',j')=\pi(i,j)$, $\psi_{i,j}$ is an isomorphism from the $i'$th copy of $G_{j'}/_\RT$ to the $i$th copy of $G_j/_\RT$. 
\end{theorem}


Although Theorem~\ref{the:splitting_automorphisms} is only stated for $\graph$-structures (graphs), it has implications for more general relational structures, as well. In particular, if $\cH$ is a graph expansion, then every automorphism of $\cH^\ell$ is also an automorphism of $\sH^\ell$, where $\sH$ is the underlying $\graph$-structure of $\cH$, and therefore satisfies the conditions of Theorem~\ref{the:splitting_automorphisms}. Often additional predicates impose even stronger restrictions on the structure of automorphisms. For example, if $\cH$ has all the constant relations, we obtain the following result. Recall that $(\cH^\const)^\ell$ denotes the $\ell$th power of $\cH^\const$, that is, the structure $\cH$ equipped with all the constant relations, and $(\cH^\const)^\ell/_\RT$ is the factor-structure of $(\cH^\const)^\ell$ modulo the relation $\RT$. Also observe that $\cH^\const/_\RT$ has all the constant relations, as $C_a=\{([a])\}$ on $\cH^\const/_\RT$ for any $a\in\cH$.

\begin{proposition}\label{pro:power-rigid}
Let $\cH$ be a graph expansion, $\sH$ its underlying $\graph$-structure, and let $\sH=\sH_1\times ... \times \sH_r$ be a prime factorization  of $\sH$. Then for every automorphism $\vf$ of $(\cH^\const)^\ell/_\RT$ there are permutations $\vc\pi r$ of $[\ell]$ such that
\[
\vf\left(\vvecad{a_{1,1}}{a_{1,r}}\zd \vvecad{a_{\ell, 1}}{a_{\ell, r}}\right) =\left(\vvecad{a_{\pi_1(1),1}}{a_{\pi_r(1),r}}\zd \vvecad{a_{\pi_1(\ell),1}}{a_{\pi_r(\ell),r}}  \right).
\]
where $(a_{i,1}\zd a_{i,r})$ is an element of the $i$th copy of $\cH^\const/_\RT$.
\end{proposition}

\begin{proof}
By Theorem~\ref{the:splitting_automorphisms} there are isomorphisms $\vf_{1,1}\zd\vf_{\ell,r}$ of factors $\sH_1,...,\sH_r$ and a permutation $\pi$ of $[\ell]\tm[r]$ such that 
\begin{align*}
    \vf\left(\vvecad{a_{1,1}}{a_{1,r}}\zd \vvecad{a_{\ell, 1}}{a_{\ell, r}}\right)= \left( \vvecad{\vf_{1,1}(a_{\pi(1,1)})}{\vf_{1,k}(a_{\vf(1,r)})},...,\vvecad{\vf_{\ell,1}(a_{\pi(\ell,1)})}{\vf_{\ell,r}(a_{\pi(\ell,r)})} \right)
\end{align*}
Now observe that as $\cH^\const/_\RT$ has all the constant relations, for any $a \in H/_\RT$, where $a=(\vc{a}{r})$, we have
\begin{equation}\label{equ:diagonal}
\vf\left(\vvec ar \zd \vvec ar\right)=\vf(a\zd a)=(a\zd a)=\left(\vvec ar \zd \vvec ar \right)    
\end{equation}
Firstly, this implies that every $\vf_{i,j}$ is the identity mapping. This means that $\vf$ is basically the permutation $\pi$ of coordinates. Moreover, if for some $i\in[\ell],j\in[r]$, $\pi(i,j)=(i',j')$ and $j'\ne j$ then choosing any $(\vc ak)$ such that $a_i\ne a_j$ we get a contradiction. The result follows.
\end{proof}

\begin{corollary}\label{cor:bipartite-p-auto}
If $\cH$ is a graph expansion and $\ell<p$ then every $p$-automorphism of $(\cH^\const)^\ell$ is local. 
\end{corollary}

\begin{proof}
Let $\vf$ be a non-local $p$-automorphism of $(\cH^\const)^\ell$ and $\vf^\RT$ the corresponding automorphism of $(\cH^\const)^\ell/_\RT$. By Lemma~\ref{lem:R-factor-auto} $\vf^\RT$ has order $p$ and by Proposition~\ref{pro:power-rigid} there are permutations $\vc\pi r$ of $[\ell]$ such that
\[
\vf^\RT\left(\vvecad{a_{1,1}}{a_{1,r}}\zd \vvecad{a_{\ell, 1}}{a_{\ell, r}}\right) =\left(\vvecad{a_{\pi_1(1),1}}{a_{\pi_r(1),r}}\zd \vvecad{a_{\pi_1(\ell),1}}{a_{\pi_r(\ell),r}}  \right),
\]
for any $((a_{1,1}\zd a_{1,r})\zd (a_{\ell, 1}\zd a_{\ell, r}))\in(\cH^\const)^\ell$. As $\vf$ is a $p$-automorphism, so is $\vf^\RT$. Therefore every $\pi_i$ is either the identity mapping or has order $p$. The latter is impossible as $\ell<p$, and $\vf^\RT$ has to be the identity mapping.
\end{proof}

\subsection{Reduced form of $\cH^{\ell}$}
\def\cHClT{\widetilde{(\cH^\const)^{ \ell}}}
\def\sHClT{\widetilde{(\sH^\const)^{ \ell}}}
As Faben and Jerum showed in \cite{ref:CountingMod2Ini}, the $p$-reduced form of $(\cH^\const)^{\ell}$ is obtained by removing from a graph or a relational structure vertices that are not fixed under a $p$-automorphism. We will often use the $p$-reduced form of $(\cH^\const)^{\ell}$, and need to make sure that some vertices remain in that $p$-reduced form. In this section our goal is to show that if $\cH$ is $p$-rigid and equipped with all the constants, then we can guarantee that at least some vertices from certain specified sets are not eliminated when constructing the $p$-reduced form. For elements $a,b\in H$ we write $a\sim b$ if there is $\pi\in\Aut(\cH)$ such that $\pi(a)=b$.

Following Remark~\ref{rem:constants} observe that tuples from $(\sH^c)^\ell$ that are elements of a constant relation $C_{H,a}^{(\sH^\const)^\ell}$ are \emph{constant} tuples, that is, of the form $(a\zd a)$, where $a$ is the same as in the relation $C_{H,a}$. Such tuples are fixed points of any automorphism of  $(\sH^c)^\ell$. 

\begin{lemma}\label{lem:permutation-of-R-class}
For the structure $(\sH^c)^\ell$, $a\in (\sH^c)^\ell$, any permutation $\pi$ of the $\RT$-class $[a]$ such that every constant tuple from $[a]$ is a fixed point of $\pi$, can be extended to an automorphism of $(\sH^c)^\ell$. Therefore for  $a,b\in (\sH^c)^\ell$, if $a\RT b$ and none of $a,b$ is a constant tuple, then $a\sim b$. 
\end{lemma}

\begin{proof}
Let $\pi$ be a permutation of $[a]$ such that every constant tuple from $[a]$ is its fixed point. As is easily seen, the following mapping is an automorphism of $(\sH^\const)^\ell$.
\begin{equation*}
    \vf(v) = 
     \left\{ \begin{array}{cl}
   \pi(v)  & v \in [a], \\
    v & v \not \in [a]. 
\end{array}\right. 
\end{equation*}
Indeed, by the assumption on $\pi$ the mapping $\vf$ preserves all the constant relations. Moreover, as $\vf$ preserves the $\RT$-classes, it also preserves the binary edge relation of $\sH^\ell$.
Observe that if $a\RT b$ and none of $a,b$ is a constant tuple, there is a permutation $\pi$ of $[a]$ satisfying the conditions of the lemma and such that $\pi(a)=b$, hence there is an automorphism $\vf$ such that $\vf(a)=b$.
\end{proof}

Let $\widetilde{(\sH^\const)^{\ell}}$ be a substructure (subgraph) of $(\sH^\const)^{ \ell}$ obtained by reducing $(\sH^\const)^{ \ell}$ by all the local $p$-automorphisms. In other words $(\sH^\const)^{ \ell}\to_p^*\widetilde{(\sH^\const)^{\ell}}$, every step in this reduction is through a local $p$-automorphism, and $\widetilde{(\sH^\const)^{\ell}}$ has no local $p$-automorphisms.  Note that $\sHClT$ is NOT the reduced form of $(\sH^\const)^{\ell}$, as some non-local $p$-automorphisms may remain. More formally, for every $\RT$-class $[a]$ of $(\sH^\const)^{ \ell}$ let $C^a$ be the set of all the constant tuples from $[a]$. We select a subset $[a]^*\sse[a]-C^a$ such that $|[a]^*|<p$ and $|[a]- [a]^*|\equiv 0\pmod{p}$, and set $\widetilde{[a]}=C^a\cup[a]^*$. By Lemma~\ref{lem:permutation-of-R-class} $(\sH^\const)^{ \ell}\rightarrow^*_p\widetilde{(\sH^\const)^{\ell}}$, as for every $\RT$-class $[a]$ we can choose a permutation of order $p$ such that $\widetilde{[a]}$ is its set of fixed points. Let $\widetilde H=\bigcup_{a\in (\sH^\const)^{\ell}}\widetilde{[a]}$. Then clearly $\widetilde{(\sH^\const)^{\ell}}$ is the substructure of $(\sH^\const)^{ \ell}$ induced by $\widetilde H$.

\begin{lemma}\label{lem:equal-R}
Let $\sH$ be a $p$-rigid $\graph$-structure. Then\\[2mm]
(1) for every $x\in\sH^\ell$, $|[x]|\not\equiv0\pmod p$ and $|\widetilde{[x]}|\not\equiv0\pmod p$;\\[2mm] 
(2) $\widetilde{(\sH^\const)^{\ell}} /_\RT$ is isomorphic to $(\sH^\const)^{ \ell} /_\RT$.
\end{lemma}

\begin{proof}
(1) Let $x=(\vc x\ell)$. Then, as is easily seen, $y=(\vc y\ell)\in[x]$ if and only if $x_i\RT y_i$ for every $i\in[\ell]$. Therefore, $|[x]|=\prod_{i\in[\ell]}|[x_i]|$. Since $\sH$ is $p$-rigid, by Lemma~\ref{lem:permutation-of-R-class} $|[x_i]|\not\equiv0\pmod p$, implying the result. Also, since $|[x]|-|\widetilde{[x]}|\equiv0\pmod p$, we have $|\widetilde{[x]}|\not\equiv0\pmod p$.

(2) We prove the lemma through the following claims.

{\sc Claim 1.} The $\RT$-classes of $(\widetilde{\sH^\const)^{ \ell}}$ are the sets $\widetilde{[x]}$.

\renewcommand{\qedsymbol}{$\blacksquare$}
\begin{proof}[Proof of Claim 1] 
If $N_{\widetilde{(\sH^\const)^{ \ell}}}(a)=N_{\widetilde{(\sH^\const)^{ \ell}}}(b)$ then $a \RT b$ and $a,b$ are in a same $\RT$-class of $\sHClT$. 
Hence, it suffices to prove that if $a,b \in \widetilde{[x]}$, then $N_{\widetilde{(\sH^\const)^{ \ell}}}(a)=N_{\widetilde{(\sH^\const)^{ \ell}}}(b)$.

By the definition, if $a,b \in \widetilde{[x]}$, then $a,b \in [x]$, Thus $[a]=[b]$, Hence
\begin{align*}
c \in N_{\widetilde{(\sH^\const)^{ \ell}}}(a) \Leftrightarrow \nedge{a}{c} \in E^{{\widetilde{(\sH^\const)^{ \ell}}}} &\Leftrightarrow \nedge{[a]}{[c]} \in E^{{{(\sH^\const)^{ \ell}}}/_\RT} \\
&\Leftrightarrow \nedge{[b]}{[c]} \in E^{{{(\sH^\const)^{ \ell}}}/_\RT}
\Leftrightarrow \nedge{b}{c} \in E^{{\widetilde{(\sH^\const)^{ \ell}}}} \Leftrightarrow  c \in N_{\widetilde{(\sH^\const)^{ \ell}}}(b).
\end{align*}
\end{proof}
\renewcommand{\qedsymbol}{$\Box$}

{\sc Claim 2.} The function $\FUNC{\vf}{(\sH^\const)^{ \ell}/_\RT}{\widetilde{(\sH^\const)^{ \ell}}/_\RT}$ where $\vf([x])=\widetilde{[x]}$ is an isomorphism.

\renewcommand{\qedsymbol}{$\blacksquare$}
\begin{proof}[Proof of Claim 2]
Clearly, the function $\vf$ is bijective. We just need to prove that the function $\vf$ is edge preserving. 

Assume $\nedge{[a]}{[b]} \in E^{{{(\sH^\const)^{ \ell}}}/_\RT}$. Then,  $\vf([a])=\widetilde{[a]} \sse [a]$ and $\vf([b])=\widetilde{[b]} \sse [b]$.
So, there are $a\in \widetilde{[a]}$ and $b\in \widetilde{[b]}$ such that $\nedge{a}{b}\in E^{(\sH^\const)^{ \ell}}$, also,  $\nedge{a}{b}\in E^{\widetilde{(\sH^\const)^{ \ell}}}$. 
Hence there are $a\in \widetilde{[a]}, b\in \widetilde{[b]}$ such that $\nedge{a}{b}\in E^{\widetilde{(\sH^\const)^{ \ell}}}$, therefore, $\nedge{[a]}{[b]}\in E^{\widetilde{(\sH^\const)^{ \ell}}/_\RT}$.
\end{proof}
\renewcommand{\qedsymbol}{$\Box$}
We have an isomorphism from $\widetilde{(\sH^\const)^{ \ell}}/_\RT$ to $(\sH^\const)^{ \ell}/_\RT$.
The result follows.
\end{proof}

\begin{theorem}\label{the:FullReductionRorPower}
Let $\sH$ be a $p$-rigid $\graph$-structure and $A\sse (H^ \ell)$ such that $A = \bigcup_{i \in I} A_i$ where the $A_i$'s are $\RT$-classes of $(\sH^\const)^{ \ell}$. Let $\widetilde{A} = \bigcup_{i \in I} \widetilde{A}_i$. Then
\begin{itemize}
    \item[(1)] 
    $\widetilde A/_\RT = A/_\RT$.
    \item[(2)] 
    For any $\graph$-structure $(\sG,x)$, $\hom( (\sG,x), ((\sH^\const)^{ \ell},A))\equiv \hom ((\sG,x) , (\widetilde{(\sH^\const)^{ \ell}}, \widetilde A))\pmod p$.
\end{itemize}
\end{theorem}

\begin{proof}\noindent
(1)  It easily follows from definitions that $\widetilde A/_\RT\sse A/_\RT$, and  
we just need to prove that $A/_\RT \sse \widetilde{A}/_\RT$. Since $\sH$ is $p$-rigid, the size of all its $\RT$-classes are nonzero modulo $p$. Therefore, by Lemma~\ref{lem:factoring_R-thin} the size of each $\RT$-class of $\sH^\ell$ is also nonzero modulo $p$. Therefore, $\widetilde{A}_i \not = \emptyset$ for all $i \in I$. 

(2) Let $B$ be an $\RT$-class of $(\sH^\const)^{ \ell}$.
Then $|B-\widetilde B|$ is a multiple of $p$. By Lemma~\ref{lem:aut-reduction-structures} 
\begin{align*}
    \hom((\sG,x), &((\sH^\const)^{ \ell}, B)) \\
    &= \sum_{\ba\in B}\hom((\sG,x), ((\sH^\const)^{ \ell},\ba))\\
    &\equiv \hom((\sG,x), (\widetilde{((\sH^\const)^{ \ell}}, \widetilde B))+\sum_{\ba\in B-\widetilde B}\hom((\sG,x), ((\sH^\const)^{ \ell},\ba)) \pmod p\\ 
    &\equiv \hom((\sG,x), \widetilde{((\sH^\const)^{ \ell}}, \widetilde B)) \pmod p.
    \end{align*}
\end{proof}


\section{Expanding constraint languages}\label{sec:algebraic}

In this section we show that the standard ways of expanding relational structures used in the study of the CSP also apply in modular counting. We consider the following types of expansion: expanding by the equality relation, conjunctive formulas, constant relations, and by primitive-positive definitions. In each case we prove that modular counting over the expanded structure is reducible to that over the original structure. For the first three types of expansion the result is proved for general relational structures. However, the reduction for pp-definable structures requires Theorem~\ref{the:splitting_automorphisms}, and therefore we can only prove it in the case of graph expansions. 

\subsection{The two views on the CSP}
We defined the CSP as the problem of deciding the existence of a homomorphism between two relational structures. There is, however, another view on the CSP that is widely used in the literature and that will be very useful from the technical perspective. Firstly, note that for a $\sg$-structure $\cH$ the collection of interpretations $\rel^\cH$, $\rel\in\sg$, is just a set of relations. We call a set of relations over some set $H$ a \emph{constraint language} over $H$. Thus, for every relational structure $\cH$ there is an associated constraint language $\Gm_\cH$. Conversely, every (finite) constraint language $\Gm$ can be converted into a relational structure $\cH_\Gm$ such that $\Gm_{\cH_\Gm}=\Gm$ in a straightforward way, although in this case there is much room for the choice of a signature.

The definition of the CSP using constraint languages is as follows. Let $\Gm$ be a constraint language on a set $H$, called the \emph{domain}. The Constraint Satisfaction Problem $\CSP(\Gm)$ is the combinatorial problem with:\\
\textbf{\emph{Instance:}} a pair $\cP=(V,\cC)$ where $V$ is a finite set of variables and $\cC$ is a finite set of \emph{constraints}. Each constraint $C \in \cC$ is a pair $\ang{\bs,\rel}$ where
\begin{itemize}
    \item $\bs = (v_1, v_2,... , v_m )$  is a tuple of variables from $V$ of length $m$ for some $m$, called the \emph{constraint scope};
    \item $\rel\in\Gm$ is an $m$-ary relation, called the \emph{constraint relation}.
\end{itemize}
\textbf{\emph{Objective:}} Decide whether there is a solution of $\cP$, that is, a mapping $\vf:V\to H$ such that for each
constraint $\ang{\bs,\rel}\in \cC$ with $\bs = (\vc vm)$ the tuple $(\vf(v_1)\zd\vf(v_m))$ belongs to $\rel$.

\smallskip

In the (modular) counting version of $\CSP(\Gm)$ denoted $\NCSP(\Gm)$ ($\NpCSP(\Gm)$) the objective is to find the number of solutions of instance $\cP$ (modulo $p$).

We will refer to this definition as the \emph{standard} definition of the CSP.

It is well known, see e.g.\ \cite{Feder98:computational} and \cite{ref:POlymorphismAndUsethem_barto2017polymorphisms} that problems $\CSP(\cH)$ and $\CSP(\Gm_\cH)$ can be easily translated into each other. The same is true for $\NCSP(\cH)$ and $\NCSP(\Gm_\cH)$. The conversion procedure goes as follows. Let $\cG$ be an instance of $\CSP(\cH)$ and $\sg$ is the signature of $\cH$. Create an instance $\cP=(V,\cC)$ of $\CSP(\Gm_\cH)$ by setting $V=G$, the base set of $\cG$, and for every $\rel\in\sg$ and every $\bs\in\rel^\cG$, including the constraint $\ang{\bs,\rel^\cH}$ into $\cC$. An instance of $\CSP(\Gm)$ can be transformed to an instance of $\CSP(\cH_\Gm)$ by  reversing this process. Clearly, This transformation can be extended to counting CSPs, as well. In this paper we mainly use the standard definition of the CSP inside proofs by assuming that an instance of $\NpCSP(\cH)$ is given by variables and constraints.

\subsection{Expanding by the equality relation}\label{sec:expanding}

Let $\cH$ be a relational structure with signature $\sg$ and $\cH^=$ its expansion by adding a binary relational symbol $=$ interpreted as $=_H$, the equality relation on $H$. The following reduction is straightforward.

\begin{lemma}\label{lem:adding-equality}
For any relational structure $\cH$ and any prime $p$, $\NpCSP(\cH^=)\le_T\NpCSP(\cH)$.
\end{lemma}

\begin{proof}
Here we use the standard definition of the CSP. Let $\cP=(V,\cC)$ be an instance of $\NpCSP(\cH^=)$. Clearly, if $\cC$ contains a constraint $x=y$ for some $x,y\in V$ then for any solution $\vf:V\to H$ of $\cP$, it holds that $\vf(x)=\vf(y)$. Therefore, transform $\cP$ as follows: If $x=y$ for some $x,y\in V$, then replace every occurrence of $y$ in constraints from $\cC$ with $x$. Repeat this transformation until we obtain an instance $\cP'=(V',\cC')$ that does not contain equality constraints. We consider $\cP'$ to be an instance of $\NpCSP(\cH)$. Finally observe that if a mapping $\vf:V\to H$ is a solution of $\cP$ then $\vf_{|V'}$ is a solution of $\cP'$. On the other hand, as is easily seen, every solution of $\psi$ of $\cP'$ can be extended to a solution of $\cP$ in a unique way. The result follows.
\end{proof}

\subsection{Expanding by conjunctive definitions}

Conjunctive definitions are a special case of primitive positive definitions that do not use quantifiers. Let $\cH$ be a structure with signature $\sg$. A \emph{conjunctive formula} $\Phi$ over variables $\{\vc xk\}$ is a conjunction of atomic formulas of the form $\rel(\vc y\ell)$, where $\rel\in\sg$ is an ($\ell$-ary) symbol and $\vc y\ell\in\{\vc xk\}$. A $k$-ary predicate $\relo$ is \emph{conjunctive definable} in $\cH$ by $\Phi$ if $(\vc ak)\in\relo$ if and only if $\Phi(\vc ak)$ is true.

\begin{lemma}\label{lem:conjunctive}
Let $\cH$ be a relational structure with signature $\sg$, $\rel$ be conjunctive definable in $\cH$, and $\cH+\rel$ denotes the expansion of $\cH$ by the predicate symbol $\rel$ that is interpreted as the relation $\rel$ in $\cH$. Then $\NpCSP(\cH+\rel)\le_T\NpCSP(\cH)$.
\end{lemma}

\begin{proof}
Let $\rel$ be defined by a conjunctive formula 
\[
\relo_1(y_{11}\zd y_{1\ell_1})\wedge\dots\wedge \relo_s(y_{s1}\zd y_{s\ell_s}),
\] 
where $\vc\relo s\in\sg$ and $y_{ij}\in\{\vc xk\}$ for all $i,j$. We use the standard definition of the CSP. Let $\cP=(V,\cC)$ be an instance of $\NpCSP(\cH+\rel)$. If $\cC$ contains a constraint of the form $\ang{(\vc xk),\rel}$, replace it with $\ang{(y_{11}\zd y_{1\ell_1}),\relo_1},\zd\ang{(y_{s1}\zd y_{s\ell_s}),\relo_s}$. As is easily seen, the resulting instance has exactly the same solutions as $\cP$. Repeat this procedure while constraints containing $\rel$ remain. The resulting instance $\cP'$ has the same solutions as $\cP$, and is an instance of $\NpCSP(\cH)$.
\end{proof}

A good example of a conjunctive definable relation is the \emph{indicator problem} introduced in \cite{Jeavons97:closure,Jeavons99:expressive}. Here we will only need a simple case of the indicator problem. Slightly rephrasing the construction from \cite{Jeavons97:closure,Jeavons99:expressive}, let $\cH$ be a relational structure, $\sg$ its signature, and $H=\{\vc an\}$. Construct a conjunctive formula $\cI(\cH)$ as follows. Let $v_{a_1}\zd v_{a_n}$ be the variables of $\cI(\cH)$. For every $\rel\in\sg$ (say, $\ell$-ary), and any $(a_{i,1}\zd a_{i,\ell})\in\rel^\cH$, add the conjunct $\rel(v_{a_{i,1}}\zd v_{a_{i,\ell}})$. Since endomorphisms of $\cH$ are exactly its unary polymorphisms in the terminology of \cite{Jeavons97:closure,Jeavons99:expressive}, by Theorem~3.5 of \cite{Jeavons99:expressive} we obtain the following result.

\begin{lemma}[\cite{Jeavons97:closure,Jeavons99:expressive}]\label{exa:indicator}
Let $\cH$ be a relational structure, $|H|=n$, and let $\relo$ be defined by $\cI(\cH)$. Then $\relo=\Hom(\cH,\cH)$. In other words, $(\vc bn)\in\relo$ if and only if there is an endomorphism $\vf$ of $\cH$ such that $\vf(a_i)=b_i$ for $i\in[n]$.
\end{lemma} 

\subsection{Expanding by constant relations}\label{sec:adding-constants}

Recall that for a relational structure $\cH$ by $\cH^\const$ we denote the expansion of $\cH$ by constant relations. Theorem~\ref{the:ConstantCSP} was proved for exact counting in \cite{ref:BULATOV_TowardDichotomy} and for modular counting of graph homomorphisms in \cite{ref:CountingMod2Ini}. We use a proof similar to that in \cite{ref:BULATOV_TowardDichotomy}.

\begin{custom_num_theorem}{\ref{the:adding-constants-intro}}\label{the:ConstantCSP}
Let $\cH$ be a $p$-rigid $\sg$-structure. Then $\NpCSP(\cH^\const)$ is polynomial time reducible to $\NpCSP(\cH)$.
\end{custom_num_theorem}

\begin{proof}
We follow the same line of argument as the proof of a similar result in \cite{ref:BULATOV_TowardDichotomy} for exact counting. Let $H = \{a_1, ... , a_n\}$, and let $\relo = \{(\vf(a_1)\zd \vf(a_n))  \mid \vf \in \Hom(\cH, \cH) \}$ be the relation conjunctive definable by Lemma~\ref{exa:indicator} through the indicator problem. By Lemmas~\ref{lem:adding-equality} and~\ref{lem:conjunctive} we may assume that $\cH$ has $=_H$ and $\relo$ as its predicates.

Let $\cP = (V;\cC)$ be an instance of $\#_p \CSP(\cH^\const)$. We construct an instance $\cP'=(V',\cC')$ of $\#_p\CSP(\cH)$ as follows. 
\begin{itemize}
    \item 
    $V'= V \cup \{v_a | a \in H \}$;
    \item
    $\cC'$ consists of three parts: $\{C=\ang{\bx,\rel}\in\cC\mid \rel\in\sg\}$, $\{\ang{(v_{a_1}\zd v_{a_n}),\relo}\}$, and\linebreak  $\{\ang{(x,v_a),=_H}\mid \ang{(x),C_{H,a}}\in\cC\}$.
\end{itemize}

The number of solutions of $\cP$ equals the number of solutions $\vf$ of $\cP'$ such that $\vf(v_a) = a$ for all $a \in H$. Let $U$ be the set of all such solutions of $\cP'$ and $T=|U|$. Then $T$ can be computed in two stages.

Let again $\Part(H)$ be the poset of partitions of $H$. For every partition $\th\in\Part(H)$ we define $\cP'_\th$ as the instance $(V',\cC_\th)$, where $\cC_\th=\cC'\cup\{ \constCSP{ (v_a, v_{a'})}{=_H}\}\mid a,a'$ belong to the same class of $\theta \}$). Note that if $\vf$ is a solution of $\cP'$, then $\vf$ is a solution of $\cP'_{\theta}$ if and only if $\vf(v_a) = \vf(v_{a'})$ for every $a,a'$ from the same class of $\theta$. Let us denote by $M(\th)$ the number of solutions of $\cP'_\th$. The number $M(\th)$ can be computed using the oracle $\#_p\CSP(\cH)$, since we assume that $=_H$ and $\relo$ are predicates of $\cH$.

Next we find the number of solutions $\vf$ of $\cP'$ that assign $v_a$, $a\in A$, pairwise different values. Let $W$ be the set of all such solutions. Let us denote by $N(\th)$ the number of all solutions $\vf$ of $\cP'_\th$ such that $\vf(v_a)=\vf(v_b)$ if and only if $a,b$ belong to the same class of $\th$. In particular, $N(\zz)=|W|$. The number $N(\zz)$ can be obtained using the M\"obius inversion formula for the poset $\mathsf{Part}(H)$. Let $\FUNC{w}{\mathsf{Part(H)}}{\mathbb{Z}}$ be defined as in Section~\ref{sec:mobius-inversion}. Also, observe that for any $\th\in\Part(H)$
\[
M(\th)=\sum_{\eta\ge\th}N(\eta).
\]
Therefore, 
\[
N(\zz) = \sum_{\th\in\Part(H)} w(\th) M(\th).
\]
Thus $N(\zz)$ can be found through a constant number of calls to $\#_p\CSP(\cH)$.

Now, we express $T$ via $N(\zz)$. Let $G=\Aut(\cH)$ be the automorphism group of $\cH$. We show that $W= \{ g\circ\vf | g \in G, \vf \in U \}$. For every solution $\vf$ in $U$ and every $g \in G$, $g\circ\vf$ is also a solution of $\cP'$. Moreover, since $g$ is one-to-one, $g\circ\vf$ is in $W$. Conversely, for every $\psi \in W$, there exists some $g \in G$ such that $g(a) =\psi(v_a)$, $a \in H$. Note that $g^{-1} \in G$ implies $\vf =g^{-1}\circ\psi \in U$, which witnesses that  $\psi= g\circ\vf$ is of the required form.
Finally, for every $\vf,\vf' \in U$ and every $g,g'\in  G$, if $\vf|_{\{v_a | a \in H \}}\ne\vf'|_{\{v_a | a \in H \}}$ or $g\ne g'$ then $g\circ\vf\ne g'\circ \vf'$. Thus, $N(\zz)=|G|\cdot T$. Since $\cH$ is $p$-rigid,  $|G|\not \equiv 0\pmod p$. Therefore  $T \equiv |G|^{-1}\cdot N(\zz)\pmod{p}$.
\end{proof}

\subsection{Homomorphisms, pp-definitions and factorizations}

In this section we discuss several issues related to the use of pp-definitions and the results of Section~\ref{sec:auto}.

Primitive positive definitions have close connections to homomorphisms, see \cite{Feder98:computational,ref:kolaitis2004constraint}.

\begin{lemma}\label{lem:pp-gadget}
A predicate $\rel(\vc xk)$ is pp-definable in a $\sg$-structure $\cH$ containing the equality predicate if and only if there exists a $\sg$-structure $(\cG_R,(\vc xk))$ such that for all $(\vc{a}{k})\in R$
\[
\hom((\cG_\rel,(\vc xk)) ,(\cH,(\vc ak) ) )\ne0 \qquad\text{if and only if}\qquad (\vc ak)\in\rel.
\]
\end{lemma}

\begin{proof}
Let $\rel$ be defined by a pp-formula $\exists \vc ys\Phi(\vc xk,\vc ys)$ and let $\Phi(\vc xk,\vc ys)$ be its quantifier free part.  As $\cH$ has the equality predicate, $\Phi$ is a conjunction of atomic formulas of the form $\relo(\vc z\ell)$, where $\relo\in\sg$ and $\vc z\ell\in\{\vc xk,\vc ys\}$. The corresponding $\sg$-structure $\cG_\rel$ is constructed as follows. Its base set is $G_\rel=\{\vc xk,\vc ys\}$. Then for any $\relo\in\sg$, say, $\ell$-ary, $(\vc z\ell)\in\relo^{\cG_\rel}$ if and only if $\relo(\vc z\ell)$ is an atom in $\Phi$. The result now follows from the observation that every satisfying assignment of $\Phi$ is also a homomorphism from $\cG_\rel$ to $\cH$.

Conversely, suppose that there exists a structure $\cG_\rel$ such that 
\[
\hom((\cG_\rel,(\vc xk)),(\cH,(\vc ak)))\ne0
\] 
if and only if $(\vc ak)\in\rel$, and $\{\vc ys\}=G_\rel-\{\vc xk\}$. Then the transition to a pp-formula can be carried out as above only in the reverse direction.
\end{proof}

The results of Section~\ref{sec:auto} will be intensively used in this section. So, before we embark on studying the connections between pp-definable relations and modular counting we need to make two important remarks.

First, let $\cH$ be a graph expansion and $\sH$ its underlying $\graph$-structure. We define the $\RT$-relation on $\cH$ as that on $\sH$. Note however that this does not translate into the same concept of $p$-reduction and local $p$-automorphisms. Specifically, while for $\sH$ whenever $a\RT b$ there is an automorphism that maps $a$ to $b$, it is no longer true for $\cH$, as it may have additional relations preventing such an automorphism. This means that $\RT$-classes $\widetilde{[x]}$ defined in $\sH$ and in $\cH$ are different, although $\widetilde{[x]}$ for $\sH$ is always a subset of that for $\cH$. This does not affect the argument in this and subsequent sections, in particular, Lemma~\ref{lem:equal-R} and Theorem~\ref{the:FullReductionRorPower} remain true for $\cH$, but should be noted.

Second, we are going to use prime factorizations of $\sH$. Unfortunately, such prime factorizations do not necessarily give rise to factorizations of $\cH$. This means that we have to only use structures that avoid such phenomenon. A graph expansion $\cH$ will be called \emph{factorization-regular} if there exists a factorization $\cH=\cH_1\tm\dots\tm\cH_r$ such that $\sH=\sH_1\tm\dots\tm\sH_r$, where $\sH,\sH_1\zd\sH_r$ are the underlying $\graph$-structures of $\cH,\vc\cH r$, respectively, is a prime factorization of $\sH$. In other words $\cH$ is factorization-regular if there is a prime factorization $\sH=\sH_1\tm\dots\tm\sH_r$ of $\sH$ such that every relation $\rel$ of $\cH$ can be represented as $\rel=\rel_1\tm\dots\tm\rel_r$ for some relations $\vc\rel r$ on $\vc Hr$, respectively.

Fortunately, all the structures we use in this paper are factorization-regular.

\begin{lemma}\label{lem:f-regularity}
Let $\cH$ be a factorization-regular graph expansion. Then
\begin{itemize}
    \item[(a)]
        if $\cH$ is a $\graph$-structure, it is factorization-regular;
    \item[(b)]
        $\cH^\const$ is factorization-regular;
    \item[(c)]
        $\cH$ expanded by the equality relation $=_H$ is factorization-regular;
    \item[(d)]
        if $\rel$ is pp-definable in $\cH$ then $\cH+\rel$ is factorization-regular.
\end{itemize}
\end{lemma}

\begin{proof}
Let $\cH=\cH_1\tm\dots\tm\cH_r$ be a factorization of $\cH$ that is also a prime factorization of $\sH$.

(a) is obvious. 

(b),(c) It suffices to observe that any constant relation $C_{H,a}$ can be represented as $C_{H_1,a_1}\tm\dots\tm C_{H_r,a_r}$ for some $a_s\in H_s$, $s\in[r]$. In the same way, the equality relation $=_H$ equals $=_{H_1}\tm\dots\tm =_{H_r}$.

(d) By Lemma~\ref{lem:pp-gadget} if $\rel$ (say, $k$-ary) is pp-definable in $\cH$, then for some structure $(\cG,(\vc xk))$ we have 
\[
\rel=\{(\vc ak)\mid \Hom((\cG,(\vc xk)),(\cH,(\vc ak)))\ne\emptyset\}.
\]
Set
\[
\rel_s=\{(\vc{a^{(s)}}k)\mid \Hom((\cG,(\vc xk)),(\cH_s,(\vc{a^{(s)}}k)))\ne\emptyset\}.
\]
By Proposition~\ref{pro:ProductHom} $\rel=\rel_1\tm\dots\tm\rel_r$.
\end{proof}

\subsection{Adding Primitive-Positive definitions}\label{sec:pp-def}
The main goal of this section, Proposition~\ref{the:CloneTheorem}, is to prove that for a $p$-rigid relational structure $\cH$ that is a graph extension, $\NpCSP(\cH+\rel)$ for a pp-definable relation $\rel$ reduces to $\NpCSP(\cH)$. By Lemma~\ref{lem:adding-equality} we may assume that $\cH$ has the equality predicate.

We start with an auxiliary claim that shows that if a predicate $\rel$ is pp-definable in $\cH$, it is always possible to find a pp-definition that somewhat uniformizes the number of extensions of tuples from $\rel$. Let $\rel$ be pp-definable in $\cH$ by a pp-formula 
\begin{equation*}
\rel(\vc xk)=\exists\vc ys \Phi(\vc xk,\vc ys).   
\end{equation*}
For $\ba\in\rel$ by $\ext_\Phi(\ba)$ we denote the number of assignments $\bb\in H^s$ to $\vc ys$ such that $\Phi(\ba,\bb)$ is true.

\begin{proposition}\label{pro:GadgetExists}
Let $\cH$ be a $p$-rigid structure with equality that is a factorization-regular graph expansion and $p$ a prime. Let $\rel$ be a relation that is pp-definable in $\cH^\const$. Then there exists a pp-definition 
\[
R(\vc xk)=\exists\vc ys \Phi(\vc xk,\vc ys) 
\]
of $\rel$ such that for any $\ba\in\rel$, $\ext_\Phi(\ba)\equiv1\pmod p$.
\end{proposition}
\begin{proof}
\def\relQH{Q^{\widehat \cH}}
To simplify the notation we assume that $\cH$ contains all the constant relations, that is, $\cH=\cH^\const$.

We use the equivalence between pp-definitions and homomorphisms from Lemma~\ref{lem:pp-gadget}. Let $\rel=\{\vc \ba\ell\}\sse H^k$ where $\ba_i=(a_{i,1}\zd a_{i,k})$, be a relation pp-definable in $\cH$. Our goal is to find a structure $\cG$ such that for some $(\vc xk)\in G^k$, 
\[
\hom((\cG,(\vc xk)),(\cH,(\vc{{a_i}}{k})))\equiv1\pmod p \text{ for } 1 \leq i \leq \ell,
\]
and 
\[
\hom((\cG,(\vc xk)),(\cH,(\vc bk)))=0 \text{ for all } (\vc bk) \not \in\rel
\]
Note that by Proposition~\ref{pro:ProductHom} and Fermat's Little Theorem it suffices to prove that such a structure exists satisfying $\hom((\cG,(\vc xk)),(\cH^\const,\ba))\not\equiv0\pmod p$.

Firstly, observe that if the pp-definition of $\rel$ uses no existential quantifiers, then $\ext_\Phi(\ba)=1\pmod p$ for any $\ba\in\rel$, as the empty extension is the only option. This pp-definition also satisfies the requirements of the proposition. Thus, it suffices to consider the case when $\rel$ is given by $\exists y Q(\vc xk,y)$, where $Q$ is a pp-definable relation in $\cH$, and there exists a structure $(\cK,(\vc xk,y))$ such that $(\vc dk,e)\in Q$ if and only if,
\begin{equation}\label{eq:Q-def}
\hom( (\cK, (\vc xk,y)), (\cH, (\vc dk,e)) )\equiv 1 \pmod p, 
\end{equation}
and 
\[
\hom( (\cK, (\vc xk,y)), (\cH, (\vc dk,e)) )=0
\]
otherwise. Note that the structure $\cK$ also satisfies the condition
\[
\hom( (\cK, (\vc xk)), (\cH, (\vc dk)) )\ne0\text{ if and only if } (\vc dk)\in\rel.
\]

Consider $\cH^\ell=\cH\tm\dots\tm\cH$ with distinguished vertices $\ba^1\zd\ba^k$, where $\ba^j=\vvecad{a_{1,j}}{a_{\ell, j}}$, $j\in[k]$. By Proposition~\ref{pro:ProductHom} for any $(\cG,(\vc xk))$
\[
\hom((\cG, (\vc xk)),(\cH^\ell,(\ba^1, \dots , \ba^k))) =\prod_{i\in[\ell]}\hom((\cG,(\vc xk)),(\cH,\ba_i)),
\]
where $\ba_i=(a_{i,1}\zd a_{i,k})$.
\def\relQ{Q^{(\cH^{\const})^{\ell}}}
\def\relQR{Q^{(\cH^{\const})^{\ell}/_\RT}}
\def\compH{({\cH^{\const})}^{\ell}}
If $(\cH^\ell,(\ba^1\zd\ba^k))$ was $p$-rigid, we could apply M\"obius inversion formula in a way similar to Lemma~\ref{lem:mobius-point}(2) to infer the existence of a required $\cG$. However, there is no guarantee this is the case, and we need to make one more step.

Note that $\ba^1\zd\ba^k$ as well as each of the constant tuples is a fixed point of any automorphism of $(\cH^\ell,(\ba^1\zd\ba^k))$. This means that the $p$-reduced form of the structure $(\cH^\ell,(\ba^1\zd\ba^k))$ contains $\ba^1\zd\ba^k$ and all the constant tuples. Let $\widehat{\cH}=((\cH^{\ell})^{* p},(\ba^1\zd\ba^k))$. Thus
\begin{align*}
\hom((\cG,(\vc xk)),&(\cH^\ell,(\ba^1\zd\ba^k)))\\
&\equiv \hom((\cG,(\vc xk)),((\widehat \cH,(\ba^1\zd\ba^k))) \pmod p.
\end{align*}

\smallskip
{\sc Claim 1.} 
There exists $\vv c=\vvecad{c_{1}}{c_{\ell}}\in \widehat\cH$ such that $(a_{i,1}\zd a_{i,k},c_i) \in Q$ for every $i\in[\ell]$.
\renewcommand{\qedsymbol}{$\blacksquare$}
\begin{proof}[Proof of Claim 1] 
First, observe that it suffices to show that the required tuples exist in $\widehat \cH/_\RT$. Indeed, suppose that there is $\vvecad{[c_1]}{[c_\ell]}$ in the $p$-reduced form of $(\cH/_\RT)^\ell$ such that $([a_{i,1}]\zd [a_{i,k}],[c_i]) \in Q/_\RT$. By Lemma~\ref{lem:R-factor-auto} every $p$-automorphism of $\cH^{\ell}$ is either local or gives rise to an automorphism of $\cH^{\ell}/_\RT$. By Lemma~\ref{lem:equal-R}, as $\cH$ is $p$-rigid, reducing by local automorphisms does not eliminate tuples from $\vvecad{[c_1]}{[c_\ell]}$. By the assumption this tuple also withstands reducing by automorphisms of the former kind. Therefore,  there are $c'_i\in[c_i]$, $i\in[\ell]$, such that $(\vc{c'}\ell)\in\widehat\cH$. Moreover, by the definition of the relation $\RT$ we have $(a_{i,1}\zd a_{i,k},c'_i) \in Q$ for $i\in[\ell]$.

Let $\cH=\cH_1 \times\dots\times \cH_r$ be a factorization of $\cH$ into its prime factors. By Lemma~\ref{lem:factoring_R-thin} we have $\cH/_\RT= \cH_1/_\RT \times\dots \times \cH_r/_\RT$. Let $[a_{i,j}]=( [a_{i,j}^{(1)}]\zd [a_{i,j}^{(r)}] )$, where $i\in[l],j\in[k]$. Observe also that there exist relations $Q^{(s)}$ on $\cH_s$, $s\in[r]$, satisfying the following condition. For a tuple $(\vc b{k+1})\in\cH^{k+1}$, where $b_i=(b_i^{(1)}\zd b_i^{(r)})$ is the representation of $b_i$ in the factorization of $\cH$, it holds that $(\vc b{k+1})\in Q$ if and only if $(b_1^{(s)}\zd b_{k+1}^{(s)})\in Q^{(s)}$. Indeed, by Proposition~\ref{pro:ProductHom} a mapping $\vf$ from $(\cK,(\vc xk,y))$ to $(\cH,(\vc b{k+1}))$ is a homomorphism if and only if there are homomorphisms $\vc\vf r$ such that $\vf_s:(\cK,(\vc xk,y))\to(\cH_s,(b_1^{(s)}\zd b_{k+1}^{(s)})$ for all $s\in[r]$. Thus, $Q^{(s)}$ can be chosen to be the set of all $(b_1^{(s)}\zd b_{k+1}^{(s)})\in\cH_s^{k+1}$ such that there is a homomomorphism from $(\cK,(\vc xk,y))$ to $(\cH_s,(b_1^{(s)}\zd b_{k+1}^{(s)}))$.

Since $\ba_i\in\rel$ for $i\in[\ell]$, there are $c_i^{(s)}\in\cH_s$ such that $(a_{i,1}^{(s)}\zd a_{i,k}^{(s)},c_i^{(s)})\in Q^{(s)}$. By the observation above we can choose $c_i^{(s)}$ in such a way that $c_i^{(s)}=c_{i'}^{(s)}$ for any $s\in[r]$ and $i,i'\in[\ell]$ whenever $a_{i,j}^{(s)}=a_{i',j}^{(s)}$ for all $j\in[k]$. In other words, if instead of the matrix $(\ba^1\zd\ba^k)$ consisting of elements of $\cH$ we consider a similar matrix 
\[
A^{(s)}=\left(\begin{array}{ccc}
a_{1,1}^{(s)}&\dots & a_{1,k}^{(s)}\\
\vdots & & \vdots\\
a_{\ell,1}^{(s)}&\dots& a_{\ell,k}^{(s)}
\end{array}\right)
\]
for each $s\in[r]$, then we choose $c_i^{(s)}=c_{i'}^{(s)}$ whenever the rows $i$ and $i'$ of $A^{(s)}$ are equal. In all other cases the choice of $c_j^{(s)}$ is unrestricted. 

We prove that the tuple $([c_1]\zd[c_\ell])$ constructed this way is a fixed point of any automorphism of $(\cH/_\RT)^\ell$.

Let $\vf$ be an automorphism of $(\cH/_\RT)^\ell$. Since all the constant tuples are fixed points of $\vf$, by Proposition~\ref{pro:power-rigid} there are permutations $\pi_s$ on $[\ell]$, $s\in[r]$, such that for any $(\vc d\ell)\in\cH^\ell$
\[
\vf\vvecad{([d_{1}^{(1)}]\zd [d_{1}^{(r)}])}{ ([d_{\ell}^{(1)}]\zd [d_{\ell }^{(r)}])}=\vvecad{( [d_{\pi_1(1)}^{(1)}]\zd [d_{\pi_r(1)}^{(r)}] )}{ ([d_{\pi_1(\ell)}^{(1)}]\zd [d_{\pi_r(\ell) }^{(r)}]) }
\]
In particular, $\vf$ acts on every factor $\cH^{(s)}/_\RT$ separately. Let us consider one of the factors of $\cH$, $\cH^{(s)}$. The action of $\vf$ on $[A^{(s)}]$ can be represented as follows
\[
\vf([A^{(s)}])=\left(\begin{array}{ccc}
[a_{\pi_s(1),1}^{(s)}]&\dots & [a_{\pi_s(1),k}^{(s)}]\\
\vdots & & \vdots\\ \mbox{}
[a_{\pi_s(\ell),1}^{(s)}]&\dots& [a_{\pi_s(\ell),k}^{(s)}]
\end{array}\right)
\]
As $[\ba^1]\zd[\ba^k]$ are fixed points of $\vf$, it holds that $\vf([A^{(s)}])=[A^{(s)}]$ for every $s\in[r]$. Therefore, if $\pi_s(i)=i'$ it must be that $[a_{i,j}^{(s)}]=[a_{i',j}^{(s)}]$ for all $j\in[k]$. That is, the rows of $[A^{(s)}]$ indexed by an orbit of $\pi_s$ are equal. By the choice of $\bc$, if $\pi_s(i)=i'$, we also have $[c_i^{(s)}]=[c_{i'}^{(s)}]$, and the entries $[c^{(s)}_1]\zd[c^{(s)}_\ell]$ are also equal on orbits of $\pi_s$. This means that $[c_i^{(s)}]=[c_{\pi_s(i)}^{(s)}]$ for every $i\in[\ell]$ and every $s\in[r]$, that is, $\vf([\bc])=[\bc]$. Thus, $[\bc]\in\widehat\cH/_\RT$ and the result follows.
\end{proof}
\renewcommand{\qedsymbol}{$\Box$}

Now, we can continue with the rest of the proof of Proposition~\ref{pro:GadgetExists}. We apply M\"obius inversion to $(\widehat{\cH} ,(\ba^1\zd\ba^k))$. For $\th\in\Part(\widehat{\cH})$ we set
\begin{align*}
    M(\th) &=\hom((\widehat{\cH}/_\th,(\ba^1/_\th\zd\ba^k/_\th)),(\widehat{\cH},(\ba^1\zd\ba^k)),\\
    N(\th) &=\inj((\widehat{\cH}/_\th,(\ba^1/_\th\zd\ba^k_\th)),(\widehat{\cH},\ba^1\zd\ba^k)).
\end{align*}
Then we proceed as in the proof of Lemma~\ref{lem:mobius-point} to conclude that if $M(\th)\equiv0\pmod p$ for all $\th\in\Part(\widehat{H})$, then $(\widehat{\cH},(\ba^1\zd\ba^k))$ has a $p$-automorphism, leading to a contradiction with the construction of $(\widehat{\cH},(\ba^1\zd\ba^k))$ which is the p-reduced form of $\cH^\ell$ and is $p$-rigid.
Therefore, there exists $\th\in\Part(\widehat{H})$ such that
\[
\hom((\widehat{\cH}/_\th,(\ba^1/_\th\zd\ba^k/_\th)),(\widehat{\cH},(\ba^1\zd\ba^k))\not\equiv0\pmod p.
\]

\def\relR{R^{\widehat \cH}}
\def\relQH{Q^{\widehat \cH}}

It remains to show that for every $\bb=(\vc bk)\not\in\rel$ 
\[
\hom((\widehat{\cH}/_\th,(\ba^1/_\th\zd\ba^k/_\th)),(\cH,\bb)=0.
\]
Suppose the contrary, that is, there is a homomorphism $\psi$ such that
\[
\FUNC{\psi}{(\widehat{\cH}/_\th,(\ba^1/_\th\zd\ba^k/_\th))}{(\cH,\bb)}.
\]
By Claim 1 there exists $\vv c = (\vc c\ell)\in\widehat\cH$ 
such that $(a_{i,1}\zd a_{i,k},c_i)\in Q$. Let $\vv q=(\vv a ^1,...,\vv a^k, \vv c)$. We make use of the $\graph$-structure $\cK$ defined in \eqref{eq:Q-def}. By Proposition~\ref{pro:ProductHom} and Lemma~\ref{lem:aut-reduction-structures},
\[
\hom( (\cK, (\vc xk,y)), (\widehat{\cH}, \vv q ))\equiv\hom( (\cK, (\vc xk,y)), (\cH^\ell, \vv q )\equiv 1 \pmod p.
\]
So, there is a homomorphism $\FUNC{\vf_1}{(\cK,(\vc xk))}{(\widehat \cH,(\ba^1,...,\ba^k))}$.

By Lemma~\ref{lem:hom-for-factors} we also have the following homomorphism
\[
\FUNC{\vf_2}{(\widehat{\cH},(\ba^1\zd\ba^k))}{(\widehat{\cH}/_\th,(\ba^1/_\th\zd\ba^k/_\th))},
\]
However, the homomorphism 
\[
\FUNC{\psi \circ\vf_2\circ\vf_1}{(\cK, (\vc xk))}{(\cH,\bb)},
\]
witnesses that for some $d\in H$
\[
\Hom((\cK, (\vc x{k+1})),(\cH,(\bb,d))\ne\eps.
\]
By the choice of $(\cK, (\vc x{k+1}))$ this means that $(\bb,d)\in Q^\cH$, which contradicts the assumption that $\bb\not\in\rel$.
\end{proof}

We are now in a position to prove the main result of this subsection.

\begin{theorem}\label{the:CloneTheorem}
Let $\cH$ be a $p$-rigid relational structure that is a factorization-regular graph expansion. If $\rel$ is pp-definable in $\cH$, then $\#_p\CSP(\cH+\rel)$ is polynomial time reducible to $\#_p\CSP(\cH)$.
\end{theorem}

\begin{proof}
By Lemma~\ref{lem:adding-equality} we may assume that $\cH$ is with equality. By Proposition~\ref{pro:GadgetExists} there exists a pp-definition of $\rel(\vc xk)=\exists\vc ys\Phi(\vc xk,\vc ys)$ of $\rel$ such that for any $\ba\in\rel$, $\ext_\Phi(\ba)\equiv1\pmod p$. Let $\cP=(X,\cC)$ be a $\NpCSP(\cH+\rel)$ instance. Without loss of generality assume that $\vc Cm$ are the constraints containing $\rel$. Construct a $\#_p\CSP(\cH)$ instance $\cP'=(X',\cC')$ as follows:
\begin{itemize}
    \item 
    For each $i\in[m]$ introduce new variables $y_{i1}\zd y_{is}$ and set $X'=X\cup\bigcup_{i\in[m]}\{y_{i1}\zd y_{is}\}$.
    \item
    The set $\cC'$ contains every constraint from $\cC-\{\vc Cm\}$. Also, we replace each $C_i=\ang{(x_{i1}\zd x_{ik}),\rel}$ with the definition $\Phi(x_{i1}\zd x_{ik},y_{i1}\zd y_{is})$ of $\rel$. 
\end{itemize}
Next, we find the number of solutions of $\cP'$. As is easily seen, for every solution $\vf$ of $\cP$ there are 
\[
\prod_{i\in[m]}\ext_\Phi(\vf(x_{i_1})\zd \vf(x_{ik}))\equiv1\pmod p
\]
solutions of $\cP'$. Therefore if $N,N'$ denote the number of solutions of $\cP,\cP'$, respectively, we have $N\equiv N'\pmod p$.
\end{proof}

One important special case of pp-definable relations that we will use all the time is unary pp-definable relations or \emph{subalgebras}. Recall that if a structure $\cH$ is a graph expansion, we define the neighbourhood $N_\cH(A)$ of a set $A\sse H$ in terms of the underlying graph. 

\begin{lemma}\label{lem:pp-neighborhood}
Let $\cH$ be a graph expansion.\\[2mm]
1. Let $a\in H$. Then $N_\cH(a)$ is a subalgebra of $\cH^\const$.\\[2mm]
2. Let $A\sse H$ be a subalgebra of $\cH^\const$. Then $N_\cH(A)$ is a subalgebra of $\cH^\const$.
\end{lemma}

\begin{proof}
Let $E$ be the edge relation of $\cH$.\\[2mm]
1. Since $\cH^\const$ has the constant relation $C_{H,a}$, $N_\cH(x)=\exists y (E(x,y)\wedge C_{H,a}(y))$ is pp-definable.\\[2mm] 
2. In this case the argument is similar. Let the subalgebra $A$ be pp-defined by $A(x)=\exists\vc ys\Phi(x,\vc ys)$. Then $\relo(x)=\exists z\exists\vc ys(E(x,z)\wedge \Phi(z,\vc ys))$ defines $N_\cH(A)$.
\end{proof}

\subsection{Factor structures}\label{sec:factor}

An important tool for both decision CSPs and exact counting is a reduction from factor structures modulo a pp-definable equivalence relation. Here we prove that a similar reduction works in the case of modular counting. We start with an auxiliary statement.

\begin{lemma}\label{lem:gadget-exists-homimage}
    Let $\cH$ be a $\graph$-structure with vertex set $V$, and let $W\sse V$ and $\theta$ be a unary relation and an equivalence relation on $W$ with equivalence classes $\vc Ar$, both primitive positive definable in $\cH^\const$. Then, there exists a structure $\cG$ and $x\in G$ such that
    \[
    \hom((\cG,x), (\cH^\const, A_i)) \equiv 1 \pmod p
    \]
    for all $i\in[r]$. 
\end{lemma}

\begin{proof}
To simplify the notation, we assume that $\cH$ contains all the constant relations, that is, $\cH = \cH^\const$. Let $\cH=\cH_1 \times\dots\times \cH_k$ be a factorization of $\cH$ into its prime factors. By Lemma~\ref{lem:factoring_R-thin}, we have $\cH/_\RT= \cH_1/_\RT \times\dots \times \cH_k/_\RT$. Let $[\vv a]=([a^{(1)}]\zd [a^{(k)}])$. 

\smallskip

{\sc Claim 1.}
(a) There exist $\cH'_t\sse\cH_t$ for $t\in[k]$ such that $W=\cH'_1 \times\dots\times \cH'_k$.\\[2mm]
(b) There exist equivalence relations $\theta^{(t)}$ on $\cH'_t$ for $t\in[k]$, whose equivalence classes are denoted by $A^{(t)}_i$, $i\in[r_t]$, satisfying the following condition:  

- For tuples $(\vc b{s}), (\vc c{s}) \in \cH^{s}$, where $b_i=(b_i^{(1)}\zd b_i^{(k)})$ and $c_i=(c_i^{(1)}\zd c_i^{(k)})$ represent $b_i$ and $c_i$ in the factorization of $\cH$, respectively, it holds that $(\vc b{s}) \theta (\vc c{s})$ if and only if $b_j^{(t)} \; \theta^{(t)} \; c_j^{(t)}$ for all $t \in [k]$ and $j \in [s]$.

\begin{proof}[Proof of Claim 1]
We use the equivalence between pp-definitions and homomorphisms from Lemma~\ref{lem:pp-gadget}. 

(a) Let $(\cK_0,x)$ be the structure that defines the subalgebra $W$. Then $a\in W$ if and only if $\hom((\cK_0,x),(\cH,a))\not\equiv0\pmod p$. Let 
\[
H'_t=\{a_t\mid a=(\vc ak)\in W\}.
\]
Since 
\[
\hom((\cK_0,x),(\cH,a))=\prod_{t=1}^k \hom((\cK_0,x),(\cH_t,a_t)),
\]
it holds that $\hom((\cK_0,x),(\cH_t,a_t))\not\equiv0\pmod p$ whenever $a\in W$. For the same reason if $a_t\in H'_t$ for $t\in[k]$, then $\hom((\cK_0,x),(\cH,a))\not\equiv0\pmod p$, $a=(\vc at)$ and $a\in W$.

(b) Let $(\cK,x,y)$ be the structure that defines the relation $\th$, that is, 
\[
\hom((\cK,x,y),(\cH,a,b))\not\equiv0\pmod p
\]
if and only if $(a,b)\in\th$. Define $\th^{(i)}\sse\cH_i^2$, $i\in[k]$, by the condition 
\[
\text{$(a_i,b_i)\in\th^{(i)}$ if and only if }\hom((\cK,x,y),(\cH_i,a_i,b_i))\not\equiv0\pmod p.
\]
By Proposition~\ref{pro:ProductHom} for $a=(\vc ak),b=(\vc bk)\in\cH$, $(\vc ak)\th(\vc bk)$ if and only if $a_i\th^{(i)}b_i$ for all $i\in[k]$. It remains to show that every $\th^{(i)}$ is an equivalence relation. This however is straightforward from the property above and the reflexivity, symmetricity, and transitivity of $\th$.
\end{proof}

Claim~1 and Proposition~\ref{pro:ProductHom} imply that it suffices to find a structure $(\cG,x)$ such that  
\[
    \hom((\cG,x), (\cH_t, A^{(t)}_i)) \equiv 1 \pmod p
\]
for every $t\in[k]$ and every $i\in[r_t]$. In order to simplify notation, in the sequel we will assume that $(\cH_s,A^{(s)}_i),(\cH_t,A^{(t)}_j)$ are not isomorphic for any $s,t\in[k]$ and any $i\in[r_s],j\in[r_t]$. Indeed, otherwise the congruence above for one of them implies that for the other, and only one of the isomorphic structures is included in the construction below. We construct a relation $\cJ$ as follows:
\[
\cJ=\cH^{r_1}_1\tm\dots\tm\cH^{r_k}_k,
\]
and a set $M\sse\cJ$ as 
\[
M=A^{(1)}_1\tm\dots\tm A^{(r_1)}_1\tm A^{(1)}_2\tm\dots\tm A^{(1)}_k\tm\dots\tm A^{(r_k)}_k.
\]
By Proposition~\ref{pro:ProductHom}, and Fermat's Little Theorem it suffices to find $(\cG,x)$ with  
\[
\hom((\cG,x), (\cJ, M) \not\equiv 0 \pmod p.
\]
It will be convenient to denote the components of tuples $\ba\in\cJ$ using two parameters, say, $\ba[t,i]$, that is, $\ba=(\ba[1,1]\zd\ba[1,r_1]\zd\ba[k,1]\zd\ba[k,r_k])$.

If $(\cJ, M)$ were $p$-rigid, and $M$ were automorphism-stable, we could apply the M\"obius inversion formula, similar to Lemma~\ref{lem:mobius-point}(1), to conclude the existence of the required $\cG$. However, there is no guarantee this is the case, and a few more steps are needed. Let $\th_\cJ$ denote a binary relation on $\cJ$ given by $(\ba,\bb)\in\th_\cJ$ if and only if $(\ba[s,i],\bb[s,i])\in\th^{(s)}$ for all $s\in[r]$ and $i\in[r_s]$. As is easily seen $\th_\cJ$ is an equivalence relation on a subset of $\cJ$ and $M$ is a $\th_\cJ$-block. 

\smallskip 

{\sc Claim 2.} 
Let $\vf$ be an automorphism of $\cJ$ and $\ba,\bb\in\cJ$ such that $(\ba,\bb)\in\th_\cJ$. Then $(\vf(\ba),\vf(\bb))\in\th_\cJ$.

\begin{proof}[Proof of Claim~2]
The claim follows from the fact that $\th_\cJ$ is pp-definable in $\cJ$. Indeed, let $(\cK,x,y)$ be the structure that defines $\th$ and all the $\th^{(s)}$, $s\in[k]$. Then, as $(a,b)\in\th^{(s)}$ if and only if $\hom((\cK,x,y),(\cH_s,a,b)\ne0$, we also have $(\ba,\bb)\in\th_\cJ$ if and only if $\hom((\cK,x,y),(\cJ,\ba,\bb))\ne0$. Let $(\ba,\bb)\in\th_\cJ$ and $\psi:(\cK,x,y)\to(\cJ,\ba,\bb)$ a homomorphism mapping $x$ to $\ba$ and $y$ to $\bb$. Then $\vf\circ\psi$ is also a homomorphism from $(\cK,x,y)$ to $(\cJ,\vf(\ba),\vf(\bb))$ witnessing that $(\vf(\ba),\vf(\bb))\in\th_\cJ$.
\end{proof}

Next, we show that $M$ is automorphism-stable for $\cJ$. In fact, if this holds, then the reduced form of $M$ will be automorphism-stable in $\cJ^{*p}$. 

\smallskip
\noindent
{\sc Claim 3.}  
The set $M$ is automorphism-stable for $\cJ$ and $M/_\RT$ is automorphism-stable for $\cJ/_\RT$.

\begin{proof}[Proof of Claim 3]
Pick an arbitrary $\ba\in M$ and let $G$ be the set of automorphisms $\vf$ of $\cJ$ such that $\vf(\ba)\in M$. It suffices to show that $G$ is a subgroup of $\Aut(\cJ)$. As $G$ contains the identity automorphism, we only need to verify that $G$ is closed under composition. Let $\vf_1,\vf_2\in G$ and $\vf_1(\ba)=\bb\in M$. As $\vf_2(\ba)\in M$, by Claim~2 $\vf_2(\bb)\in M$, as well, implying $\vf_2\circ\vf_1(\ba)\in M$ and $\vf_2\circ\vf_1\in G$. For $M/_\RT$ and $\cJ/_\RT$ a proof is similar.
\end{proof}  

By Theorem~\ref{the:FullReductionRorPower}, $\cJ$ and $M$ can be replaced by $\wJ$ and $\widetilde{M}$, respectively, obtained by reducing $\cJ$ using local $p$-automorphisms, such that $\cJ/_{\RT}=\wJ/_{\RT}$ and $M/_{\RT}=\widetilde{M}/_{\RT}$. Moreover, for any $(\cG,x)$  
\[
\hom((\cG,x),(\wJ,\wM))\equiv\hom((\cG,x),(\cJ,M))\pmod p.
\]  

Note that $\wJ$ is not $p$-rigid yet; in fact, it is possible to reduce it even further. The crucial point here is that $\widetilde{M}$ will survive after $\wJ$ is completely $p$-reduced. To conclude this, we need to prove the following claims.  

\smallskip
\noindent
{\sc Claim 4.}
$\wM$ is a union of $\RT$-classes and $\wM$ is automorphism-stable in $\wJ$.

\begin{proof}[Proof of Claim 4]
Note that, as $\cJ/_{\RT}=\wJ/_{\RT}$, $M/_{\RT}=\widetilde{M}/_{\RT}$, and $\wJ$ is obtained from $\cJ$ by local automorphisms, it suffices to show that $M$ is a union of $\RT$-classes. Let $\ba\in M$ and $\bb\in\cJ$ such that $\ba\RT\bb$. As $\th_\cJ$ is an equivalence relation $\hom((\cG,x,y),(\cJ,\ba,\ba))\ne0$, let $\vf:\cG\to\cJ$ with $\vf(x)=\vf(y)=\ba$. Since $\ba$ and $\bb$ have the same neighbors, the mapping $\vf'$ that is equal to $\vf$ except $\vf'(y)=\bb$ is a homomorphism from $\cG$ to $\cJ$, witnessing that $(\ba,\bb)\in\th_\cJ$ and therefore $\bb\in M$.

Similar to Claim~3 pick an arbitrary $\ba\in\wM$ and let $G$ be the set of automorphisms $\vf$ of $\wJ$ such that $\vf(\ba)\in\wM$. It suffices to show that $G$ is a subgroup of $\Aut(\wJ)$. As $G$ contains the identity automorphism, we only need to verify that $G$ is closed under composition. Let $\vf_1,\vf_2\in G$. By Lemma~\ref{lem:R-factor-auto} $\vf_1,\vf_2$ induce automorphisms $\vf_1/_\RT,\vf_2/_\RT$ of $\wJ/_\RT$. By Claim~3 $\vf_2/_\RT\circ\vf_1/_\RT$ maps $\wM/_\RT$ to itself. Therefore, since $(\vf_2\circ\vf_1)/_\RT=\vf_2/_\RT\circ\vf_1/_\RT$, and by what is proved above, $\vf_2\circ\vf_1$ maps $\wM$ to itself.
\end{proof}

Finally, we show that $(\wJ,\wM)$ has no non-local automorphisms.

\smallskip
\noindent
{\sc Claim 5}
$(\wJ,\wM)$ has no non-local $p$-automorphisms.

\begin{proof}[Proof of Claim 5]  
Suppose that $\vf$ is a non-local $p$-automorphism of $(\wJ,\wM)$. Then $\vf/_\RT$ is a nontrivial $p$-automorphism of $(\wJ/_\RT,\wM/_\RT)$. By Proposition~\ref{pro:power-rigid} we have
\begin{align*}  
\vf([\vv a]) &= \vf(([\ba[1,1]]\zd[\ba[1,r_1]]\zd[\ba[k,1]]\zd [\ba[k,r_k]])) \\  
&= (([\ba[1,\pi_1(1)]]\zd[\ba[1,\pi_1(r_1)]]\zd[\ba[k,\pi_1(1)]]\zd [\ba[k,\pi_k(r_k)]]))
\end{align*}  
for $\ba\in\wJ$. If for some $i\in[k]$ the mapping $\pi_i$ is not identity, say, $\pi_i(j)\ne j$ for $j\in[r_i]$ then for any $\ba\in\wM$ it holds that $\ba[i,\pi_i(j)]\in A^{(i)}_j\cap A^{(i)}_{\pi_i(j)}$, and this cannot be true for any $\ba\in\wM$, because $A^{(i)}_j\ne A^{(i)}_{\pi_i(j)}$. This implies that every $\pi_i$ is the identity mapping, $\vf/_\RT$ is the identity mapping, a contradiction with the choice of $\vf$.
\end{proof}  

Now, we apply Lemma~\ref{lem:mobius-point}(1) to $(\wJ^{*p}, \wM)$ and obtain the desired $(\cG, x)$. 
\end{proof}

\begin{theorem}\label{the:homomoprhic-image-graph}
Let $\cH$ be a $p$-rigid relational structure that is a factorization-regular graph expansion. If $W$ is a subalgebra of $\cH^\const$ and $\theta$ is an equivalence relation on $W$ pp-definable in $\cH$, then $\#_p\CSP(\cH'/_\theta)$ is polynomial time reducible to $\#_p\CSP(\cH)$, where $\cH'$ is the substructure of $\cH$ induced by $W$.
\end{theorem}

\begin{proof}
We use the homomorphism formulation of the CSP. Also, for every $\rel$ (say, $k$-ary) from the vocabulary of $\cH$, let $\rel_\th$ be defined by 
\[
\rel_\th(\vc xk)=\exists \vc yk\ \  \rel(\vc xk)\wedge\bigwedge_{i=1}^k\th(y_i,x_i).
\] 
By Theorem~\ref{the:CloneTheorem} we may assume that $W,\th$ and all the $\rel_\th$ are in the vocabulary of $\cH,\cH/_\th$, and $\th$ is the equality relation in $\cH/_\th$. Then $\rel_\th^{\cH/_\th}$ is the same as $\rel^{\cH/_\th}$, and 
\[
\rel_\th^\cH=\bigcup_{\ba\in\rel^\cH}\ba[1]/_\th\tm\dots\tm\ba[k]/_\th,
\]
where $a/_\th$ denotes the $\th$-class containing $a$.

Let $\cJ$ be an instance of $\#_p \CSP(\cH/_\theta)$ and $N=\hom(\cJ,\cH/_\th)$. By the observation above we may assume that $\cJ$ only uses predicates of the form $\rel_\th$, that is, the remaining predicates are interpreted as empty relations. Our goal is to construct an instance $\cJ'$ of $\#_p \CSP(\cH)$ such that $\hom(\cJ',\cH^\const)\equiv N\pmod p$. Note that there is no guarantee that $\cJ$ itself works as $\cJ'$, because every homomorphism to $\cH/_\th$ blows up to an unpredictable number of homomorphisms to $\cH$. Therefore, $\cJ$ needs to be modified.

Let $\cG$ and $x\in G$ be as in Lemma~\ref{lem:gadget-exists-homimage}. We construct an instance $\cJ'$ of $\#_p \CSP(\cH)$ as follows:  
\begin{itemize}  
\item 
For every vertex $y\in J$ of $\cJ$ create a copy $(\cG_y,y)$ of $(\cG,x)$ (thus identifying $y$ and the distinguished vertex of $\cG$) and set $J'=J\cup\bigcup_{y\in J}G_y$.
\item 
For every $\rel_\th$ from the vocabulary of $\cH$ set $\rel_\th^{\cJ'}=\rel_\th^\cJ$.
\item 
For every $\rel$ from the vocabulary of $\cH$ that is not of the form $\rel_\th$ set $\rel^{\cJ'}=\bigcup_{y\in J}\rel^{\cG_y}$.
\end{itemize}  

Next, we analyze the number of homomorphisms from $\cJ'$ to $\cH$. Note that for every homomorphism $\vf:\cJ\to\cH$ the mapping $\vf/_\th:\cJ\to \cH/_\th$ given by $\vf/_\th(y)=\vf(y)/_\th$ is a homomorphism from $\cJ$ to $\cH/_\th$. Therefore, it suffices to fix a homomorphism $\psi:\cJ\to\cH/_\th$ and evaluate the number $N_\psi$ of homomorphisms $\vf:\cJ'\to\cH$ such that $(\vf_{|J})/_\th=\psi$. We show that $N_\psi\equiv1\pmod p$, thus proving that the reduction is parsimonious modulo~$p$. Since all the predicates involved in $\cJ$ are of the form $\rel_\th$, any mapping $\vf':\cJ\to\cH$ such that $\vf'(y)\in\psi(y)$ for $y\in\cJ$ is a homomorphism from $\cJ$ to $\cH$. Therefore
\begin{align*}
N_\psi &=\prod_{y\in\cJ}\prod_{a\in\psi(y)}\hom((\cG_y,y),(\cH,a))\\
&=\prod_{y\in\cJ}\hom((\cJ_y,y),(\cH,\psi(y))\\
&\equiv 1\pmod p,
\end{align*}
where the last congruence is by Lemma~\ref{lem:gadget-exists-homimage}. The result follows.
\end{proof}

\subsection{Nested subalgebras}\label{sec:nested}

In this section we prove an auxiliary statement that will be instrumental later on.

\begin{lemma}\label{lem:t-12-recognizable}
    Let $p$ be a prime number and let $\cH$ be a $p$-rigid structure. Suppose that $A,B\sse H$ are nonempty sets such that $A$ and $A \cup B$ are subalgebras of $\cH^\const$. Then there exists a structure $\cK$ with a distinguished vertex $x$ such that
\begin{align*}
    \begin{split}
        \hom((\cK,x),(\cH^\const,A)) &\not \equiv 0 \pmod{p}, \\
        \hom((\cK,x),(\cH^\const,B)) &\not \equiv 0 \pmod{p}, \\
        \hom((\cK,x),(\cH^\const,a)) &\equiv 0 \pmod{p} \qquad \text{for all } a \notin A \cup B.
    \end{split}
\end{align*}
\end{lemma}

\begin{proof}
For reasons that will be clear shortly we consider two cases: $p>2$ and $p=2$.

\smallskip

{\sc Case 1. $p>2$.}

\smallskip

Set $\cJ=(\cH^\const)^2$, $M=(A\tm B)\cup(B\tm A)$, and $M'=(A\cup B)\tm(A\cup B)$. As we demonstrate in Claim~4, it suffices to show that for some $(\cG,x)$ it holds that 
\[
\hom((\cG,x),(\cJ,M))\not\equiv0\pmod p.
\] 
Indeed, if this is the case, by Proposition~\ref{pro:ProductHom} we would conclude that 
\[
2\hom((\cG,x),(\cH^\const,A)),2\hom((\cG,x),(\cH^\const,C))\not\equiv0\pmod p,
\]
and $(\cG',x)$ with the required properties can be easily constructed from $(\cG,x)$. 

We will apply Lemma~\ref{lem:mobius-point}(1) to prove the existence of such a structure. However, in order to apply Lemma~\ref{lem:mobius-point}(1) the set $M$ needs to be automorphism-stable, and there must be no $p$-automorphism $\vf$ of $\cJ$ such that $\vf(a)\in M$ for some $a\in M$. Unfortunately, $\cJ$ and $M$ do not satisfy any of these conditions. We will modify them to enforce the conditions.

Firstly, by Theorem~\ref{the:FullReductionRorPower}, $\cJ,M$ can be replaced by $\wJ$ and $\widetilde M$, respectively, obtained by reducing $\cJ$ using local $p$-automorphisms, such that $\cJ/_{\RT}=\wJ/_{\RT}$, $M/_{\RT}=\widetilde M/_{\RT}$. Moreover, for any $(\cG,x)$
\[
\hom((\cG,x),(\wJ,\wM))\equiv\hom((\cG,x),(\cJ,M))\pmod p.
\]

\smallskip
\noindent
{\sc Claim 1.}
$\wJ$ is $p$-rigid.
\renewcommand{\qedsymbol}{$\blacksquare$}
\begin{proof}[Proof of Claim 1]
Since $p>2$, by Corollary~\ref{cor:bipartite-p-auto} every $p$-automorphism of $\cJ$ is local. As $\wJ/_{\RT}=\cJ/_{\RT}$ the same holds for $\wJ$. By the construction $\wJ$ does not have local $p$-automorphisms. The claim follows.
\end{proof}
\renewcommand{\qedsymbol}{$\Box$}
\smallskip
\noindent
{\sc Claim 2.}
$\wM$ is a union of $\RT$-classes.
\renewcommand{\qedsymbol}{$\blacksquare$}
\begin{proof}[Proof of Claim 2]
For any $a \in A \cup B$ and $b \in \cH$ such that $a \RT b$, the map swapping $a$ and $b$ (and fixing everything else) is an automorphism of $\cH$ and hence must preserve the subalgebra $A \cup B$. This implies that $A \cup B$ is a union of $\RT$-classes.

Since $A$ is also a subalgebra, the same argument applies to it: $A$ is automorphism-stable and hence a union of $\RT$-classes. Now, since $A \cap B = \emptyset$, and both $A$ and $A \cup B$ are unions of $\RT$-classes, it follows that $B$ is also a union of $\RT$-classes.

We now turn to the structure of $\WT M$. By construction, $\WT M = (A \times B) \cup (B \times A)$. Since both $A$ and $B$ are unions of $\RT$-classes, Lemma~\ref{lem:R-product} and Lemma~\ref{lem:factoring_R-thin} imply that both $A \times B$ and $B \times A$ are unions of $\RT$-classes. Therefore, so is their union $\WT M$.
\end{proof}
\renewcommand{\qedsymbol}{$\Box$}
\noindent
{\sc Claim 3.}
The set $\wM$ is automorphism-stable for $\wJ$.

\renewcommand{\qedsymbol}{$\blacksquare$}
\begin{proof}[Proof of Claim 3]
We need to show that for some $a\in \wM$, $\Stab(a,\wM)$ is a subgroup of $\Aut(\wJ)$. In fact we show that for any $a\in \wM$, $\Stab(a,\wM)=\Aut(\wJ)$. In order to do that, it suffices to prove that for any $\vf\in\Aut(\wJ)$ it holds that $\vf(a)\in \wM$. Also, it suffices to prove this for $\WT M /_\RT$ by Claim 2.

We first argue that no element of $\WT M/_\RT$ is isomorphic to an element of $\WT M'/_\RT - \WT M/_\RT$, that is, no automorphism of $\wJ/_\RT$ can map an element of $\WT M/_\RT$ into $\WT M'/_\RT - \WT M/_\RT$. Note that $\WT M'/_\RT - \WT M/_\RT=A^2/_\RT \cup B^2/_\RT$.

Assume for contradiction that there exists an automorphism $\vf \in \Aut(\wJ/_\RT)$ and a pair $(a/_\RT,b/_\RT) \in \WT M/_\RT$ such that
\[
\vf(a/_\RT,b/_\RT)\in A^2/_\RT \cup B^2/_\RT 
\]
Note that by Proposition~\ref{pro:power-rigid}, since $\wJ/_\RT$ is $\RT$-thin, we actually have $\vf(a/_\RT,b/_\RT) = (\vf_1(a/_\RT), \vf_2(b/_\RT))$ for some automorphisms $\vf_1,\vf_2$ of $\cH/_\RT$. Without loss of generality, suppose that $(a/_\RT,b/_\RT) \in A/_\RT \times B/_\RT$, so
\begin{equation}
    a/_\RT \in A/_\RT, \qquad b/_\RT \in B/_\RT. \label{eq:b-in-T} 
\end{equation}

Now consider two cases:

Case 1: Suppose $(\vf_1(a/_\RT), \vf_2(b/_\RT)) \in A^2/_\RT$. Then both $\vf_1(a/_\RT)$ and $\vf_2(b/_\RT)$ are in $A/_\RT$. Apply the inverse automorphism $\psi = \vf_2^{-1} \in \Aut(\wJ/_\RT)$. Since $\vf_2(b/_\RT) \in A/_\RT$, we conclude that $b/_\RT = \psi(\vf_2(b/_\RT)) \in \psi(A/_\RT) = A/_\RT$, because $A/_\RT$ is a subalgebra. But this contradicts \eqref{eq:b-in-T}, which says $b/_\RT \in B/_\RT$, and $A/_\RT \cap B/_\RT = \emptyset$ by assumption.

Case 2: Suppose $(\vf_1(a/_\RT), \vf_2(b/_\RT)) \in B^2/_\RT$. Then both $\vf_1(a/_\RT)$ and $\vf_2(b/_\RT)$ are in $B/_\RT$. However, as $A$ is a subalgebra, we should have $\vf_1(a/_\RT)\in A/_\RT$, a contradiction.

Therefore, no automorphism $\vf \in \Aut(\wJ)$ can map a pair in $\WT M$ to an element of $\WT M' - \WT M$. Hence, $\WT M$ is closed under automorphisms of $\wJ$.

Next, we show that $\WT M'$ is a subalgebra of $\wJ$. Since $A\cup B$ is a subalgebra of $\cH$, it is pp-definable. By Proposition~\ref{pro:GadgetExists}, there exists $(\cG,x)$ such that
\[
\hom((\cG,x),(\cH,a)) \not\equiv 0 \pmod{p} \text{ iff } a \in A \cup B,
\]
and $0$ otherwise. Then, by Proposition~\ref{pro:ProductHom}(2), the same $(\cG,x)$ shows that $M'$ is a subalgebra of $\cJ$. Consequently, the image $\WT M'$ under reduction is a subalgebra of $\wJ$, and thus automorphism-stable. The claim follows.
\end{proof}

\renewcommand{\qedsymbol}{$\Box$}

\noindent{\sc Claim 4.}
For any $(\cG,x)$,
\[
\hom((\cG,x),(\wJ,\widetilde M)) \equiv 2\hom((\cG,x),(\cH^\const,A))\cdot\hom((\cG,x),(\cH^\const,B))\pmod p.
\]
\renewcommand{\qedsymbol}{$\blacksquare$}
\begin{proof}[Proof of Claim 4]
By Theorem~\ref{the:FullReductionRorPower} it suffices to prove the claim for $\cJ$ and $M$. By Proposition~\ref{pro:power-rigid} every homomorphism $\vf:\cG\to \cJ$ can be represented as $\vf(y)=(\vf_1(y),\vf_2(y))$, where $\vf_1,\vf_2$ are homomorphisms $\cG\to \cH^\const$. Conversely, if $\vf_1,\vf_2:\cG\to \cH$ are homomorphisms, then the mapping given by $\vf(y)=(\vf_1(y),\vf_2(y))$ is a homomorphism from $\cG$ to $\cJ$. Finally, $\vf(x)\in M$ if and only if $\vf_1(x)\in A,\vf_2(x)\in B$ or $\vf_1(x)\in B,\vf_2(x)\in A$. Since swapping coordinates of $\cJ$ is an automorphism of $\cJ$ the number of homomorphisms of the two kinds is the same. 
\end{proof}
\renewcommand{\qedsymbol}{$\Box$}
By Lemma~\ref{lem:mobius-point}(1) there is $(\cG,x)$ such that 
\[
\hom((\cG,x),(\wJ,\wM))\not\equiv0\pmod p.
\]
Then by Claim~4 we also have 
\[
\hom((\cG,x),(\cH^\const,A)),\hom((\cG,x),(\cH^\const,C)) \not\equiv0\pmod p.
\]
To obtain a required gadget $(\cK,x)$ we need to ensure that $\hom((\cK,x),(\cH^\const,a))=0$ whenever $a\not\in A\cup B$. Since $A\cup B$ is a subalgebra, there is $(\cG',x)$ such that $\hom((\cG',x),(\cH^\const,a))\equiv1\pmod p$ whenever $a\in A\cup B$, and $\hom((\cG',x),(\cH^\const,a))\equiv 0 \pmod p$ otherwise. The structure $(\cK,x)=(\cG,x)\odot(\cG',x)$ satisfies the required conditions.

\smallskip

{\sc Case 2. $p=2$.}

Let $\sH=\sH_1\times\dots\times\sH_k$ be a prime factorization of $\sH$, the underlying $\graph$-structure of $\cH$. By Proposition~\ref{pro:power-rigid} every automorphism $\vf$ of $(\cH^\const)^2$ has the form
\[
\vf\left(\nvedge{ a_{1,1} }{ a_{2,1}  }\zd \nvedge{ a_{1,k} }{ a_{2,k}  }\right) =\left(\nvedge{ a_{\pi_1(1),1} }{ a_{\pi_1(2),1}  }\zd \nvedge{ a_{\pi_k(1),k} }{ a_{\pi_k(2),k}  }\right).
\]
where $(a_{i,1}\zd a_{i,r})$ 
is an element of the $i$th copy of $\cH$ and $\vc\pi k$ are permutations of $[2]$, that is, either identity mappings or involutions. Let $J_\vf=\{i\in[k]\mid \pi_i \text{ is an involution}\}$.

We plan to use Lemma~\ref{lem:mobius-point}(1), and for that we need to find 
\[
\ba=(a_{1,1}\zd a_{1,k},a_{2,1}\zd a_{2,k})\in A\tm B
\]
such that $\Stab(\ba,A\tm B)$ is as small as possible. Suppose that for some $\ba\in A\tm B$ and an automorphism $\vf$ of $(\cH^\const)^2$ we have $\vf(\ba)\in A\tm B$. If $\vf$ is not an identity mapping, $J_\vf\ne\eps$. Choose $\ba$ and $\vf$ such that $J_\vf$ is maximal possible. Without loss of generality, $J_\vf=[s]$, that is,
\begin{equation}
\vf\left(\nvedge{ a_{1,1} }{ a_{2,1}  }\zd \nvedge{ a_{1,k} }{ a_{2,k}  }\right) = \left( \nvedge{a_{2,1}}{a_{1,1}},..., \nvedge{a_{2,s}}{a_{1,s}}, \nvedge{a_{1,s+1}}{a_{2,s+1}},...,\nvedge{a_{1,k}}{a_{2,k}} \right).\label{equ:involutions}
\end{equation}
Note that $J_\vf\ne[k]$, because in this case $(\vc ak)\in A\cap B$, which is impossible, as $A$ and $B$ are disjoint. 

By the choice of $\ba$ and $\vf$ we have $(a_{2,1}\zd a_{2,s},a_{1,s+1}\zd a_{1,k})\in A$. 
Therefore, by (\ref{equ:involutions})
\[
\bb=\left( \nvedge{a_{2,1}}{a_{2,1}},..., \nvedge{a_{2,s}}{a_{2,s}}, \nvedge{a_{1,s+1}}{a_{2,s+1}},...,\nvedge{a_{1,k}}{a_{2,k}} \right)\in A\tm B
\] 
is a fixed point of $\vf$. Let $\bar\cH^2$ denote the structure $(\cH^\const)^2$ reduced using the automorphism $\vf$. In other words, $\bar\cH^2$ is the induced structure on the set 
\[
S=\left\{ \left( \nvedge{ a_{1,1} }{ a_{2,1}  }\zd \nvedge{ a_{1,k} }{ a_{2,k}  } \right)\in\cH^2\mid a_{1,i}=a_{2,i},\ i\in[s]\right\}.
\]
Also, let $M=(A\tm B)\cap S$. Note that $M\ne\eps$ because $\bb\in M$. 

\smallskip

{\sc Claim 5.}
For any $\bc\in M$, the only automorphism $\psi$ of $\bar\cH^2$ such that $\psi(\bc)\in M$ is the identity mapping.

\renewcommand{\qedsymbol}{$\blacksquare$}
\begin{proof}[Proof of Claim 5]
Any nontrivial homomorphism $\psi$ of $(\cH^\const)^2$ has the form
\[
\psi\left(\nvedge{ a_{1,1} }{ a_{2,1}  }\zd \nvedge{ a_{1,k} }{ a_{2,k}  }\right) = \left(\nvedge{ a_{\pi_1(1),1} }{ a_{\pi_1(2),1}  }\zd \nvedge{ a_{\pi_k(1),k} }{ a_{\pi_k(2),k}  }\right).
\]
Suppose that $\pi_i$ is an involution for some $i\in\{s+1\zd k\}$ and there is $\bc\in M$ such that $\psi(\bc)\in M$. But then for the automorphism $\psi'$ of $(\cH^\const)^2$ given by 
\[
\psi'\left(\nvedge{ a_{1,1} }{ a_{2,1}  }\zd \nvedge{ a_{1,k} }{ a_{2,k}  }\right) = \left(\nvedge{ a_{\pi'_1(1),1} }{ a_{\pi'_1(2),1}  }\zd \nvedge{ a_{\pi'_k(1),k} }{ a_{\pi'_k(2),k}  }\right).
\]
where $\vc{\pi'}s$ are involutions and $\pi'_i=\pi_i$ for $i\in\{s+1\zd k\}$ we have $\psi'(\bc)\in A\tm B$. Since $I_\vf\subset I_{\psi'}$ we get a contradiction with the choice of $\ba$ and $\vf$.
\end{proof}
\renewcommand{\qedsymbol}{$\Box$}

Claim~5 implies that $\Stab(\bc,M)$ contains only the identity mapping, and therefore is a subgroup of $\Aut(\bar\cH^2)$. Therefore $M$ is automorphism stable. By Lemma~\ref{lem:mobius-point}(1) there is $(\cG,x)$ such that $\hom((\cG,x),(\bar\cH^2,M))\equiv1\pmod2$. Therefore 
\begin{align*}
1 &\equiv\hom((\cG,x),(\bar\cH^2,M))\equiv\hom((\cG,x),((\cH^\const)^2,A\tm C))\\
&=\hom((\cG,x),((\cH^\const),A))\cdot\hom((\cG,x),((\cH^\const),C)).
\end{align*}
Since $A\cup B$ is a subalgebra we obtain a structure $(\cK,x)$ satisfying the conditions of Lemma~\ref{lem:t-12-recognizable}.
\end{proof}




\section{Bipartite graphs}\label{sec:Z-nondegenerate}

In this and the next sections we apply the results of Sections~\ref{sec:expanding}--\ref{sec:pp-def} to advance along the chain of reductions from Fig.~\ref{eq:chain_of_reductions}. As this chain splits into two, for bipartite and nonbipartite graphs, we also consider two cases. This section tackles bipartite graphs. 

\subsection{Searching for a thick Z-graph}\label{sec:Z-reduction}

In this subsection we prove Lemma~\ref{lem:getting-Z-graph-intro} as well as a stronger result that will be needed for the hardness proof.

\begin{lemma}\label{lem:getting-Z-graph}
Let $\cH$ be a bipartite graph expansion such that its underlying $\bip$-structure $\sH$ is connected but not a complete bipartite graph. Then there are subsets $A,C\sse \Lside^\cH, B,D\sse \Rside^\cH$ such that $A\cup C, B\cup D$ are subalgebras of $\cH^\const$, and the subgraph induced by $A\cup B\cup C\cup D$ is a thick Z-graph. Moreover, $A\cup C, B\cup D$ are unions of $\RT$-classes of $H$.
\end{lemma}

\begin{proof}
Let $E$ be the binary predicate of $\cH$ defining the underlying bipartite graph. We proceed by induction on the number of vertices in $\cH$. If $|\Lside^\cH|=|\Rside^\cH|=2$, $\bip$-structure $\sH=(H,E,\Lside,\Rside)$ is either disconnected, or complete, or a Z-graph. For the inductive step it suffices to show that if $\sH$ is not complete and $\sH$ is also not a thick Z-graph, then there are $\Lside'\sse \Lside^\cH, \Rside'\sse \Rside^\cH$ such that $\Lside',\Rside'$ are subalgebras of $\cH$, one of the inclusions is strict, and the subgraph induced by $\Lside'\cup \Rside'$ is connected and not complete. If a connected $\bip$-structure $\sH$ is not complete, then either it is a thick Z-graph, or there are $v,w\in H$ (assume $v,w\in \Lside^\sH$) such that $N_\sH(v)\ne \Rside^\sH$ and $\eps\ne N_\sH(v)\cap N_\sH(w)\ne N_\sH(v)$. Then  $\Rside'=N_\sH(v)$, $\Lside'=N_\sH(N_\sH(v))$ are subalgebras of $\cH$, and the subgraph induced by $\Lside'\cup \Rside'$ is connected and is not a complete bipartite graph.
\end{proof}

Recall that by $\cH^\sZ$ we denoted the expansion of the structure $\cH^\const$ by a thick Z-graph. 

\begin{corollary}\label{cor:Z-reduction}
If $\cH$ is an expansion of a connected non-complete bipartite graph then there exists a thick Z-graph $\sZ$ such that $\NpCSP(\cH^\sZ)\leq_T\NpCSP(\cH^\const)$.
\end{corollary}

One potential difficulty with the Z-graph we have found is that the subsets involved in it may have sizes divisible by $p$, in which case it is not suitable for a hardness result. Next, we show that the Z-graph satisfies certain additional conditions.

Recall that for a bipartite graph (or a $\bip$-structure) $\sG$ by $\Lside^\sG,\Rside^\sG$ we denote the two parts of its bipartition. Consider the structure $\cH^\sZ$ constructed in above. As a part of its signature it contains the two parts of a thick Z-graph $\sZ$. Let $A,B,C,D$ be the components of $\sZ$. The graph $\sZ$ is said to be \emph{non-degenerate} if there exist structures with distinguished vertices  $(\cK_L,x),(\cK_R,x)$ such that 
\begin{align}\label{eq:soundness_gadget}
    \begin{split}
        \hom((\cK_L,x),(\cH^\sZ, A)) \equiv \alpha_1 &\not \equiv 0 \pmod p,\\
        \hom((\cK_L,x),(\cH^\sZ, C)) \equiv \alpha_2 &\not \equiv 0 \pmod p,\\
        \hom((\cK_R,x),(\cH^\sZ, D)) \equiv \beta_1  &\not \equiv 0 \pmod p,\\
        \hom((\cK_R,x),(\cH^\sZ, B)) \equiv \beta_2  &\not \equiv 0 \pmod p,
    \end{split}
\end{align}
and 
\begin{align}\label{eq:completness_gadget}
    \begin{split}
       \hom((\cK_L,x),(\cH^\sZ,v)) & =0, \qquad \text{for $v\in  \Lside^\cH-(A\cup C)$}, \\
       \hom((\cK_R,x),(\cH^\sZ,v) & = 0 \qquad \text{for $v\in \Rside^\cH - (B\cup D)$}.
    \end{split}
\end{align}
Note that the non-degeneracy of $\sZ$ is a property of the entire structure $\cH^\sZ$, and not only of the graph $\sZ$ itself.

\begin{proposition}\label{pro:non-degenerate}
$\cH^\const$ contains a non-degenerate thick Z-subgraph.
\end{proposition}

\begin{proof}
Let $\sZ$ be the thick Z-subgraph found in Lemma~\ref{lem:getting-Z-graph}. In particular, $A,B,C,D$ are such that $A\cup C, B\cup D$ are subalgebras of $\cH$ and are unions of $\RT$-classes. We need to prove the existence of structures $(\cK_L,x),(\cK_R,x)$ from the definition of non-degeneracy. Applying Lemma~\ref{lem:t-12-recognizable} it suffices to show that $B$ and $C$ are also subalgebras of $\cH^\const$. This however is witnessed by the following pp-definitions (inside the Z-graph) where $a\in A,b\in B,c\in C,d\in D$
\begin{align*}
B(x) &=\exists y,z\ (E(x,y)\wedge E(x,z)\wedge C_{H,a}(y)\wedge C_{H,c}(z)),\\
C(x) &=\exists y,z\ (E(x,y)\wedge E(x,z)\wedge C_{H,b}(y)\wedge C_{H,d}(z)).
\end{align*}
\end{proof}


\subsection{Weighted $\#\BIS$ and thick Z-graphs}

To prove the hardness of $\NpCSP(\cH^\sZ)$ we reduce the Weighted $\#_p\BIS$ problem to $\NpCSP(\cH^\sZ)$. For a graph $G$ let $\IS(G)$ denote the set of its independent sets. Let $\al,\beta\not\equiv0\pmod p$. Then $\#_p\BIS(\al,\beta)$ is the problem defined as follows: given a $\bip$-structure $\sG$, find the value
\[
\mathcal{Z}_{\alpha, \beta}(\sG)=\sum_{I\in\IS(\sG)}\al^{|I\cap \Lside^\sG|}\cdot\beta^{|I\cap \Rside^\sG|}
\]
modulo $p$. The problem of computing the function $\mathcal{Z}_{\alpha, \beta}(\sG)$  is proven to be $\#_p P$-hard in \cite{ref:CountingModPToTrees_gbel_et_al_LIPIcs}.

We show that if $\sZ$ is non-degenerate, then $\#_p\BIS(\al,\beta)$ for some $\al,\beta\not\equiv0\pmod p$ is polynomial time reducible to $\NpCSP(\cH^\sZ)$. Together with Proposition~\ref{pro:non-degenerate} this gives the hardness result for bipartite graphs.

\begin{theorem}\label{the:bis-reduction}
Let there exist an induced subgraph of $\cH$ that is a non-degenerate thick Z-graph with parameters $\al_1,\al_2,\beta_1,\beta_2$, then $\#_p\BIS (\alpha_1 \alpha_2^{-1}, \beta_1 \beta_2^{-1})$ (multiplication and inversion here are modulo $p$) is polynomial time reducible to $\NpCSP(\cH)$.
\end{theorem}

\begin{proof}
Let $\sG=(\Lside^\sG \cup \Rside^\sG, E)$ be a $\bip$-structure, an instance of $\#_p\BIS$.  We construct a structure $\cG'$, an instance of $\NpCSP(\cH^\sZ)$ as follows:
\begin{itemize}
    \item 
    For each vertex $x \in \Lside^\sG$ and each vertex $y \in \Rside^\sG$, we introduce vertices $l^x$ and $r^y$, respectively. If $\nedge{x}{y}\in E$, then we add an edge between $l^x$ and $r^y$ in $\cG'$.
    \item
    For $x\in \Lside^\sG$ let $\cK_L(x)$ denote a copy of $\cK_L$ attached to $l^x$, that is, $x$ in $(\cK_L,x)$ is identified with $l^x$. For $y\in \Rside^\sG$ a copy $\cK_R(y)$ is attached in the same way.
\end{itemize}

\begin{figure}[h]
    \centering
    \includegraphics[height=4cm]{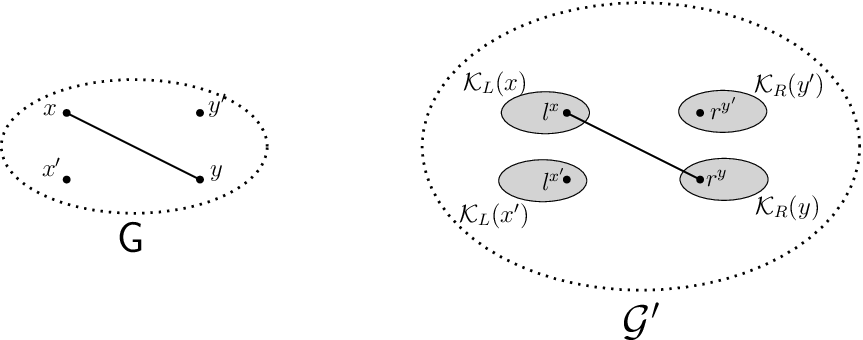}
    \caption{The structure of $\mathcal{G}'$.}
    \label{fig:reduction}
\end{figure}

The vertices from $A$ and $D$ will help to encode independent sets of $\sG$. Specifically, with every independent set $I$ of $\cG$ we associate the set of homomorphisms $\vf:\cG'\to \cH^\sZ$ such that for every vertex $x\in \Lside^\sG$, $x\in I$ if and only if $\vf(l^x)\in A$ (recall that $x$ is also a vertex of $\cG'$ identified with $l^x$ in $(\cK_L(x),l^x)$); and similarly, for every $y\in \Rside^\sG$, $y\in I$ if and only if $\vf(y)\in D$. Finally, the structure of $\cH^\sZ$ makes sure that every homomorphism from $\cG'$ to $\cH$ is associated with an independent set. Note that just an association of independent sets with collections of homomorphisms is not enough, the number of homomorphisms in those collections have to allow one to compute the weight of an independent set in $\#_p\BIS$.

For each $\vf\in \Hom(\cG',\cH)$, define  
\begin{equation*}
    I_{\vf}=\{ x \in \Lside^\sG : \vf(l^x)\in A \} \cup\{ y \in \Rside^\sG : \vf(r^y)\in D \}.
\end{equation*}
\renewcommand{\qedsymbol}{$\blacksquare$}
{\sc Claim 1.} $I_{\vf}$ is an independent set of $\sG$.
\begin{proof}[Proof of Claim 1]
Assume that for some $\vf\in\Hom(\cG',\cH)$ the set $I_{\vf}$ is not an independent set in $\cG$, i.e.\  there are two vertices $a,b\in I_{\vf}$ such that $\nedge{a}{b}\in E^\sG$.

Without loss of generality, let $a\in \Lside^\sG$ and $b \in \Rside^\sG$. 
By the construction of $I_{\vf}$, $\vf(a)\in A$ and $\vf(b)\in D$ and by definition of $\cH^\sZ$ there is no edge between $A$ and $D$, that is $\vf$ is not a homomorphism, a contradiction.
\end{proof}
\renewcommand{\qedsymbol}{$\Box$}
Let $\bumpeq$ be the relation on $\Hom(\cG',\cH)$ given by $\vf\bumpeq \vf'$ if and only if $I_{\vf} = I_{\vf'}$.
Obviously $\bumpeq$ is an equivalence relation on $\Hom(\cG',\cH)$. We denote the class of $\Hom(\cG',H) \mathbin{/}_{\bumpeq}$ containing $\vf$ by $\vf/_{\bumpeq}$. Clearly, the $\bumpeq$-classes correspond to the  independent sets of $\sG$. We will need the corresponding mapping
\begin{equation*}
\mathfrak{F}:\Hom(\cG',\cH) \mathbin{/}_{\bumpeq} \to \IS(\sG),
\quad\text{where}\quad \mathfrak{F}(\vf/_\bumpeq)=I_{\vf}
\end{equation*}

First, we prove that $\mathfrak{F}$ is bijective, then compute the cardinalities of classes $\vf/_{\bumpeq}$.

\smallskip

\noindent
{\sc Claim 2.}
The mapping $\mathfrak{F}$ is bijective.
\renewcommand{\qedsymbol}{$\blacksquare$}
\begin{proof}[Proof of Claim 2]
By the definition of $\mathfrak{F}$, it is injective. To show surjectivity let $I\in \IS(\sG)$ be an independent set. We construct a homomorphism $\vf\in\Hom(\cG',\cH)$ such that $I_{\vf}=I$. Pick arbitrary $a\in A, b\in B, c\in C, d\in D$ such that $\hom((\cK_L,x),(\cH^\sZ, a)), \hom((\cK_L,x),(\cH^\sZ, c)),\hom((\cK_R,x),(\cH^\sZ, b))$,\linebreak $\hom((\cK_R,x),(\cH^\sZ, d))\ne 0$, and 
\begin{itemize}
    \item For every vertex $x \in I \cap \Lside^\sG$ set $\vf(l^x)=a$, 
    \item For every vertex $\bar{x} \in \Lside^\sG - I$ set $\vf(l^{\bar x})=c$, %
    \item For every vertex $y \in I \cap \Rside^\sG$ set $\vf(r^y)=d$, 
    \item For every vertex $\bar{y} \in \Rside^\sG - I$ set $\vf(r^{\bar y})=b$. 
\end{itemize}

By the construction of $\vf$, if $\nedge{x}{y}\in E^\sG$ and $\vf(l^x)\in A$, then $\vf(r^y)\in B$. Similarly, if $\vf(r^y)\in D$, then $\vf(l^x)=C$. If none of the endpoints of an edge $\nedge{x}{y}$ belongs to $I$ then $\vf(l^x)\in C $ and $\vf(r^y)\in B$. By the assumption on $(\cK_L,x),(\cK_R,x)$ and the choice of $a,b,c,d$, the mapping $\vf$ can be extended to a homomorphism from $\cG'$ to $\cH$. Hence $\mathfrak{F}$ is surjective.
\end{proof}
\renewcommand{\qedsymbol}{$\Box$}
\smallskip
\noindent
{\sc Claim 3.} 
$|\vf/_{\bumpeq}|\equiv 
\alpha_1^{|\Lside^\sG \cap I_{\vf}|}\;\; \alpha_2^{|\Lside^\sG - I|} \;\;
\beta_1^{|\Rside^\sG \cap I_{\vf}|}\;\; \beta_2^{|\Rside^\sG - I|}\pmod{p}$.

\renewcommand{\qedsymbol}{$\blacksquare$}
\begin{proof}[Proof of Claim 3]
We find the number of homomorphisms $\vf' \in \vf/_{\bumpeq}$. 
The structure $\cG'$ consists of copies of $(\cK_L(x),l^x)$ and $(\cK_R(y),r^y)$. By Claim~2, and by \eqref{eq:completness_gadget} and \eqref{eq:soundness_gadget}
\begin{align*}
    |\vf/_{\bumpeq}|&= 
    \Big( \prod_{x \in \Lside^\sG \cap I_{\vf}} \hom((\cK_L(x),l^x),(\cH, A))\Big)\cdot
    \Big( \prod_{x \in \Lside^\sG - I_{\vf}} \hom((\cK_L(x),l^x),(\cH, C))\Big)\\
    &\tm \Big( \prod_{y \in \Rside^\sG \cap I_{\vf}} \hom((\cK_R(y),r^y),(\cH, D))\Big)\cdot
    \Big( \prod_{y \in \Rside^\sG - I_{\vf}} \hom((\cK_R(y),r^y),(\cH, B))\Big)\\
    &\equiv \alpha_1^{|\Lside^\sG \cap I_{\vf}|}\;\; \alpha_2^{|\Lside^\sG - I_\vf|} \;\;
\beta_1^{|\Rside^\sG \cap I_{\vf}|}\;\; \beta_2^{|\Rside^\sG - I_\vf|}\pmod{p}.
\end{align*}
\end{proof}
\renewcommand{\qedsymbol}{$\Box$}
Assume that $\bumpeq$ has $M$ classes and $\vf_i$ is a representative of the $i$-th class. Note that there is no homomorphism $\vf$ such that $\vf(l^x)\not\in A\cup C$ or $\vf(r^y)\not\in B\cup D$.
\begin{flalign*}
    |\Hom(\cG',H)| &=\sum_{i=1}^M |\vf_i/_{\bumpeq_I}| \equiv \sum_{i=1}^M \alpha_1^{|\Lside^\sG \cap I_{\vf_i}|}\;\; \alpha_2^{|\Lside^\sG - I_{\vf_i}|} \;\;
\beta_1^{|\Rside^\sG \cap I_{\vf_i}|}\;\; \beta_2^{|\Rside^\sG - I_{\vf_i}|}\\
                &\equiv \sum_{I \in \IS(\sG)} \alpha_1^{|\Lside^\sG \cap I|}\;\; \alpha_2^{|\Lside^\sG - I|} \;\;
\beta_1^{|\Rside^\sG \cap I|}\;\; \beta_2^{|\Rside^\sG - I|} \\ 
&\equiv \alpha_2^{|\Lside^\sG|}\beta_2^{|\Rside^\sG|} 
\sum_{I \in \IS(\sG)} \alpha_1^{|\Lside^\sG \cap I|}\;\; \alpha_2^{-|\Lside^\sG|}\alpha_2^{|\Lside^\sG - I|} \;\;
\beta_1^{|\Rside^\sG \cap I|}\;\; \beta_2^{-|\Rside^\sG|} \beta_2^{|\Rside^\sG - I|}\\
&\equiv \alpha_2^{|\Lside^\sG|}\beta_2^{|\Rside^\sG|} 
\sum_{I \in \IS(\sG)} (\alpha_1 \alpha_2^{-1})^{|\Lside^\sG \cap I|} \;\;
(\beta_1\beta_2^{-1})^{|\Rside^\sG \cap I|} \\
&\equiv \alpha_2^{|\Lside^\sG|}\beta_2^{|\Rside^\sG|} \mathcal{Z}_{(\alpha_1 \alpha_2^{-1}), (\beta_1 \beta_2^{-1})}(\sG) \pmod p,
\end{flalign*}
as required.
\end{proof}

\section{Non-bipartite graphs}

We now consider nonbipartite graphs. As we are going to use different methods, this case is split into two subcases. The first one concerns irreflexive, i.e.\ simple graphs, in which case we reduce the problem of counting satisfying assignments to a 3-CNF to $\#_p\CSP(\cH^\const)$, and the second one concerns graph with loops, in which case we reduce either $\#_p\BIS$ or the similar problem of counting independent sets in arbitrary graphs.

\subsection{Irreflexive nonbipartite graphs} 



In this subsection we prove Lemma~\ref{lem:thick-3SAT}, that is, we show that if $H$ is a nonbipartite graph then $\cH^\const$ pp-defines a 3-SAT structure (as introduced in Section~\ref{sec:H3SAT}), and obtain the reduction $\#_p\CSP(\cH^{3SAT})\le_T\#_p\CSP(\cH^\const)$ as an immediate consequence. This result is (almost) readily available from \cite{ref:BULATOV_HColoring}, and we only need to combine those results in the right way. What  \cite{ref:BULATOV_HColoring} proves can be stated as follows. Recall that a relational structure $\cG$ is a \emph{core} if every homomorphism of $\cG$ into itself is an automorphism.

\begin{theorem}[\cite{ref:BULATOV_HColoring}]\label{the:triangle-exists}  
If $H$ is simple non-bipartite graph that is a core, then there exists a subalgebra $K$ of $\cH^\const$, a binary symmetric relation $E'$ on $K$ pp-definable in $H^\const$, and an equivalence relation $\theta$ on $K$ pp-definable in $\cH^\const$ such that $K/_\theta$ is $K^\const_3$, that is, a complete 3-element graph with all the constant relations, where $\cK$ is the relational structure with the base set $K$, edge set $E'$, and unary relations $C_0,C_1,C_2$ that are the $\th$-classes (these relations become the constant relations of $K^\const_3$).  
\end{theorem} 

This theorem does not immediately prove Lemma~\ref{lem:thick-3SAT}. Firstly, it requires that $H$ is a core. However, a thorough analysis of proofs in \cite{ref:BULATOV_HColoring} shows that the only reason to require that $H$ is a core is to be able to add constant relations to the signature of $H$. Since we are seeking pp-definitions in $\cH^\const$, the requirement of being a core is not needed if in Theorem~\ref{the:triangle-exists} $H$ is replaced with $\cH^\const$. Secondly, as $\cH^\const$ contains all the constant relations and $\th$ is pp-definable in $\cH^\const$, the relations $C_0,C_1,C_2$ are also pp-definable in $\cH^\const$. Indeed, to pp-define $C_i$, $i=0,1,2$, pick $a\in C_i$, then $C_i(x)=\exists y\,(E(x,y)\wedge C_{H,a}(y))$. Thirdly, Theorem~\ref{the:triangle-exists} allows us to define $K^\const_3$ rather than a 3-SAT structure. Overcoming this requires a bit more work.

\begin{corollary}[Lemma~\ref{lem:thick-3SAT}]\label{cor:thick-3SAT}
If $H$ is a simple nonbipartite graph that is not a complete graph or a single vertex, then $\cH^\const$ contains a pp-definable 3-SAT structure.
\end{corollary}

\begin{proof}
Using the standard algebraic techniques, see, e.g.\ \cite{Bulatov05:classifying,ref:POlymorphismAndUsethem_barto2017polymorphisms}, it can be shown that $K^\const_3$ pp-defines any relation on its set of vertices. More specifically, we employ the notion of polymorphisms. The following statements can be found e.g.\ in \cite{ref:POlymorphismAndUsethem_barto2017polymorphisms}: \\
(1) If $H$ is a complete graph, then every polymorphism of $\cH^c$ is a projection.\\[2mm]
(2) If every polymorphism of a relational structure $\cH$ is a projection, any relation on $H$ is pp-definable in $\cH$.\\[2mm]
\indent
Rename the vertices of $K^\const_3$ so that $0,1$ are among them; suppose that these vertices correspond to the sets $C_0,C_1$ in $\cH^\const$. Then the relations $R^*_{ijk}=\{0,1\}^3-\{(i,j,k)\}$, $i,j,k\in\{0,1\}$, are pp-definable in $K^\const_3$ (see Section~\ref{sec:H3SAT}). Let $\Psi_{i,j,k}$ be a pp-definition of $R^*_{ijk}$. Let also $\Phi$ be a pp-definition of $E$ in $\cH^\const$ and use $O,I$ to denote $C_0,C_1$. Then, as is easily seen, replacing in $\Psi_{ijk}$, $i,j,k\in\{0,1\}$, every occurrence of $E(x,y)$ with  $\Phi(x,y)$ and every occurrence of $C_0(x),C_1(x)$ (constant relations) with $C_0(x),C_1(x)$ (relations of $H^\const$), we obtain a pp-definition of $R_{ijk}$.
\end{proof}

By Theorem~\ref{the:CloneTheorem} Lemma~\ref{lem:thick-3SAT} follows. 

\begin{corollary}\label{cor:thick-3SAT-reduction}
Let $\cH^{3SAT}$ be the expansion of $\cH^\const$ by a new unary predicate $O\cup I$ and the relations $R_{ijk},i,j,k\in\{O,I\}$, as in Corollary~\ref{cor:thick-3SAT}. Then $\#_p\CSP(\cH^{3SAT})\le_T\#_p\CSP(\cH^\const)$.
\end{corollary}

We are now in a position to prove Theorem~\ref{the:main-intro} in the case of simple non-bipartite graphs.

\begin{proof}[Proof of Theorem~\ref{the:main-intro} for non-bipartite graphs]  
By Corollary~\ref{cor:thick-3SAT}, there exists a subalgebra $K=O\cup I$ and pp-definable relations $R_{ijk}$, $i,j,k\in\{O,I\}$, such that $K$ with these relations is a 3-SAT structure. Also, there exists a pp-definable equivalence relation $\theta$ on $K$ such that $K/_\theta$ is isomorphic to $\cS=(\{0,1\},R^*_{ijk},i,j,k\in\{0,1\})$ and $R^*_{ijk}=\{0,1\}^3-\{(i,j,k)\}$. By Theorems~\ref{the:CloneTheorem} and~\ref{the:homomoprhic-image-graph}  
\[
\#_p \CSP(\cS) \leq\#_p \CSP(K/_\theta) \leq \#_p \CSP(\sH).  
\]  
It remains to observe that $\CSP(\cS)$ is the 3-SAT problem.
\end{proof}

\subsection{Nonbipartite graphs with loops}\label{sec:reflexive}

In this section we consider the case when $H$ has loops. First, similar to the bipartite case and thick Z-graphs we find a subalgebra of $\cH^\const$ that induces a subgraph with certain properties. This construction in part is similar to the argument from \cite{Dyer00:complexity} only using Theorem~\ref{the:CloneTheorem}. Then we show that this special induced subgraph is non-degenerate in the same sense as thick Z-graphs from Section~\ref{sec:Z-nondegenerate} and show how to reduce $\#_p\BIS$ or $\#_p\IS$ to $\#_p\CSP(\cH^\const)$.

We start with introducing two special types of graphs. A reflexive graph $G=(V,E)$ is called a \emph{thick star} if there is a partition of $V$ into $V_0,V_1\zd V_k$, $k\ge2$, such that $G|_{V_i}$ is a clique with all loops present, every vertex from $V_0$ is connected with an edge with every other vertex, and there are no other edges. A graph $G=(V,E)$ is said to be an \emph{independent 3-path} if there is a partition of $V$ into $V_0,V_1,V_2$ such that $G|_{V_2}$ is an independent set and contains no loops, $G|_{V_1}$ is a disjoint union of complete graphs $V_1^1\zd V_1^k$ with all loops present, $G|_{V_0\cup V_1^i}$ is a complete graph with all loops present for $i\in[k]$, every vertex of $V_2$ is connected with all the vertices of $V_0$, but to no vertex of $V_1$, see Fig.~\ref{fig:thick-star}.

\begin{lemma}\label{lem:reflexive-reduction}
Let $H$ be a connected graph that is not a clique and contains a loop. Then there exists a subalgebra $W$ of $\cH^\const$ such that $H|_W$ is a thick star or an independent 3-path.
\end{lemma}

\begin{proof}
Suppose first that $H=(V,E)$ is reflexive. In this case we prove that if $H$ is neither a clique nor a thick star, then there is a subalgebra $W\subset V$ of $\cH^\const$ such that $H|_W$ is not a clique. Then the result follows, as $H$ is finite.

Let $V_0\sse V$ be the set of all vertices $v$ such that $N(v)=V$. As $H$ is connected but not a complete graph, there are $u,v\in V$, $(u,v)\not\in E$, such that $(w,u),(w,v)\in E$ for some $w\in V$. If $w\not\in V_0$, then $N(w)$ is a subalgebra, $N(w)\subset V$, and $H|_{N(w)}$ is not a complete graph. Therefore, $H|_{N(w)}$ is as required.

Let $V_1\zd V_k$ be the connected components of the graph $H'$ obtained by removing $V_0$ from $H$. If all of them are complete graphs, then $k\ge2$ (as $H$ is not complete), and $H$ is a thick star. Suppose $H|_{V_1}$ is not a complete graph. As before, $H'$ contains $u,v\in V_1$, $uv\not\in E$ and $w\in V_1$ such that $(w,u),(w,v)\in E$. Since $w\not\in V_0$, the graph $H|_{N(w)}$ satisfies the conditions of the lemma.

Next, suppose that $H$ contains both reflexive (with a loop attached) and irreflexive vertices. As $H$ is a connected graph, there is $uv\in E$ such that $u$ is reflexive and $v$ is irreflexive. If $u$ can be chosen such that $N(u)\ne V$ then the graph $H|_{N(u)}$ still satisfies the conditions of the lemma. Therefore, we can assume that $N(u)=V$ for every reflexive $u\in V$ such that $uv\in E$ for some irreflexive $v\in V$, and that a reflexive $u\in V$ with $N(u)=V$ exists. If there is $uv\in E$ such that both $u,v$ are irreflexive, then $N(u)\subset V$ and $H|_{N(u)}$ still satisfies the conditions of the lemma. Therefore, we can assume that the set $V_2$ of all irreflexive vertices is an independent set. As shown before, vertices from the set $V_0$ of neighbors of irreflexive vertices are connected to every vertex of $V$. Finally, let $V_1=V-V_0-V_2$. This set only contains reflexive vertices. Suppose that $H|_{V_1}$ does not satisfy the conditions of independent 3-paths. Then there are $u,v\in V_1$ with $(u,v)\not\in E$ and $w\in V_1$ such that $(w,u),(w,v)\in E$. The graph $H|_{N(w)}$ is reflexive but not complete.
\end{proof}

Next, we show that the problem of counting the number of independent sets either in bipartite or general graphs is reducible to $\#_p\CSP(\cH^\const)$ whenever $H$ has a loop but is not a complete graph. Recall that $\#_p\IS$ is the problem of computing the number of independent sets in a given (not necessarily bipartite) graph. This problem is $\#_p P$-hard as it follows from \cite{faben2008complexity}. 

\begin{proposition}\label{pro:thick-star-reduction}
Let $H$ be graph that has a loop and is not a complete graph with all loops present. Then $\#_p\BIS (1, 1)$ or $\#_p\IS$ is polynomial time reducible to $\NpCSP(\cH^\const)$.
\end{proposition}

\begin{proof}
By Lemma~\ref{lem:reflexive-reduction} $\cH^\const$ has a subalgebra $W$ such that $H'=H|_W$ is either a thick star or an independent 3-path. Let $H'=(W,E')$. In both cases there is a partition of $W$ into several sets: $V_0,V_1\zd V_k$ in the case of a thick star and $V_0,V_1,V_2$, where $V_1=V_1^1\cup\dots\cup V_1^k$, in the case of an independent 3-path. We unify the notation by setting $A=V_0$ in both cases, $B=V_1$, $C=\bigcup_{i=2}^{k}V_i$ if $H'$ is a thick star, and $B=V_1^1$, $C=V_2\cup\bigcup_{i=2}^{k}V_1^i$ if $H'$ is an independent 3-path. Observe that both $A$ and $A\cup B$ are subalgebras of $\cH^\const$. Indeed, pick $u\in B$ and $v\in C$, then $A,A\cup B$ are defined by the following formulas
\[
(A\cup B)(x)=\exists y\ (W(x)\wedge C_u(y)\wedge E(x,y)),\ 
A(x)=\exists y\ ((A\cup B)(x)\wedge C_v(y)\wedge E(x,y)). 
\]
Let $\cH^\dg$ denote the expansion of $\cH^\const$ with subalgebras $W,A,A\cup B$. Since both $A$ and $A \cup B$ are subalgebras, by Lemma~\ref{lem:t-12-recognizable} there exists a structure $\cK'_L(x)$ that satisfies the conditions of Lemma~\ref{lem:t-12-recognizable} for $A,B$. Similarly, since both $A \cup B$ and $W=A\cup B\cup C$ are subalgebras, by Lemma~\ref{lem:t-12-recognizable} there exists a structure $\cK'_R(x)$ that satisfies the conditions of Lemma~\ref{lem:t-12-recognizable} for $A \cup B$ and $C$. 
\smallskip

First, we consider the case when $H'$ is a thick star or an independent 3-path with $V_1\ne\emptyset$. In this case we reduce $\#_p\BIS(1,1)$. Let $\sG = (\Lside^\sG \cup \Rside^\sG, F)$ be a bipartite graph, an instance of $\#_p\BIS$. We construct a new graph $\cG'$, actually, a graph expansion, which is an instance of $\NpCSP(\cH^\dg)$, as follows:

\begin{itemize}
\item 
Include all the vertices and edges of $\sG$.
\item 
For each $x \in \Lside^\sG$, attach $p-1$ copies of $\cK'_L(x)$ to $x$, denoting them by $\cK_L(x)$ and identifying $x$ in the gadget with the vertex $x$.
\item 
For each $y \in \Rside^\sG$, attach $p-1$ copies of $\cK'_R(y)$ to $y$, denoting them by $\cK_R(y)$ and identifying $y$ in the gadget with the vertex $y$.
\end{itemize}
Let $F'$ denote the edge set of $\cG'$.

Now note that for any homomorphism $\varphi: \cG' \to \cH^\dg$ and any edge $xy\in F'$ we have 
\begin{itemize}
    \item 
    if $\varphi(x) \in B$, then $\varphi(y)$ is any vertex in $A\cup B$, because there are no edges between $B$ and $C$;
    \item 
    if $\varphi(x) \in A$, then there are no restrictions on $\varphi(y)$ apart from $\vf(y)\in W$.     
\end{itemize}
With this observation the following table summarizes the possible combinations of $\varphi(x)$ and $\varphi(y)$, along with whether a homomorphism exists in each case (indicated by $1$) or not (indicated by $0$)
\begin{align*}
\kbordermatrix{
     & y \to A & y \to B & y \to C \\
x \to A & 1 & 1 & 1 \\
x \to B & 1 & 1 & 0 \\
x \to C & 0 & 0 & 0 \\
}
\end{align*}
Note that due to the properties of $\cK_L$, no homomorphism maps $x\in\Lside^\sG$ to $C$. The sets $B$ and $C$ are used to encode independent sets of $\sG$. 
More precisely, for each $\varphi \in \Hom(\cG', \cH^\dg)$, define the set
\[
I_{\varphi} = \{ x \in \Lside^\sG : \varphi(x) \in B \} \cup \{ y \in \Rside^\sG : \varphi(y) \in C \}.
\]
We now show that this encoding gives rise to a reduction from $\#_p\BIS(1,1)$.

\renewcommand{\qedsymbol}{$\blacksquare$}
{\sc Claim 1.} $I_{\vf}$ is an independent set of $\sG$.
\begin{proof}[Proof of Claim 1]
Assume that for some $\vf\in\Hom(\cG',\cH^\dg)$ the set $I_{\vf}$ is not an independent set in $\cG$, i.e.\  there are two vertices $a,b\in I_{\vf}$ such that $\nedge{a}{b}\in E^\sG$. Without loss of generality, let $a\in \Lside^\sG$ and $b \in \Rside^\sG$. By the construction of $I_{\vf}$, $\vf(a)\in B$ and $\vf(b)\in C$ and by definition of $\cH^\dg$ there is no edge between $B$ and $C$, that is $\vf$ is not a homomorphism, a contradiction.
\end{proof}

\renewcommand{\qedsymbol}{$\Box$}
Let $\bumpeq$ be the relation on $\Hom(\cG',\cH^\dg)$ given by $\vf\bumpeq \vf'$ if and only if $I_{\vf} = I_{\vf'}$. Obviously $\bumpeq$ is an equivalence relation on $\Hom(\cG',\cH^\dg)$. We denote the class of $\Hom(\cG',\cH^\dg) \mathbin{/}_{\bumpeq}$ containing $\vf$ by $\vf/_{\bumpeq}$. Clearly, the $\bumpeq$-classes correspond to the independent sets of $\sG$. We will need the corresponding mapping
\begin{equation*}
\mathfrak{F}:\Hom(\cG',\cH^\dg) \mathbin{/}_{\bumpeq} \to \IS(\sG),
\quad\text{where}\quad \mathfrak{F}(\vf/_\bumpeq)=I_{\vf}.
\end{equation*}

First, we prove that $\mathfrak{F}$ is bijective, then compute the cardinalities of classes $\vf/_{\bumpeq}$.

\smallskip

\noindent
{\sc Claim 2.}
The mapping $\mathfrak{F}$ is bijective.
\renewcommand{\qedsymbol}{$\blacksquare$}
\begin{proof}[Proof of Claim 2]
By the definition of $\mathfrak{F}$, it is injective. To show surjectivity let $I\in \IS(\sG)$ be an independent set. We construct a homomorphism $\vf\in\Hom(\cG',\cH)$ such that $I_{\vf}=I$. Since $\cK_L$ and $\cK_R$ consist of $p-1$ copies of $\cK'_L$ and $\cK'_R$ respectively, by Fermat's Little Theorem
\begin{align*}
\hom((\cK_L(x),x),(\cH^\dg,A)), \hom((\cK_L(x),x),(\cH^\dg,B)) &\equiv 1\pmod p,\\ 
\hom((\cK_R(x),x),(\cH^\dg,A\cup B)),\hom((\cK_R(x),x),(\cH^\dg,C)) &\equiv 1\pmod p,
\end{align*}
there are $a\in A, b\in B, c\in C, d\in A\cup B$ such that 
\begin{align*}
    \hom((\cK_L(x),x),(\cH^\dg, a)), \hom((\cK_L(x),x),(\cH^\dg, b)) & \not\equiv 0 \pmod p, \\
    \hom((\cK_R(x),x),(\cH^\dg, d)), \hom((\cK_R(x),x),(\cH^\dg, c)) & \not\equiv 0 \pmod p. 
\end{align*}
Then
\begin{itemize}
    \item for every  $x \in I \cap \Lside^\sG$ set $\vf(x)=b$, 
    \item for every  $x \in \Lside^\sG - I$ set $\vf(x)=a$, 
    \item for every  $y \in I \cap \Rside^\sG$ set $\vf(y)=c$, 
    \item for every  $y \in \Rside^\sG - I$ set $\vf(y)=d$. 
\end{itemize}

By the construction of $\vf$, if $\nedge{x}{y}\in F$ and $\vf(x)\in B$, then $\vf(y)\in A \cup B$. 
By the assumption on $(\cK_L(x),x),(\cK_R(x),x)$ and the choice of $a,b,c,d$, the mapping $\vf$ can be extended to a homomorphism from $\cG'$ to $\cH^\dg$. Hence $\mathfrak{F}$ is surjective.
\end{proof}

\renewcommand{\qedsymbol}{$\Box$}
\smallskip
\noindent
{\sc Claim 3.} 
$|\vf/_{\bumpeq}|\equiv 1 \pmod{p}$.

\renewcommand{\qedsymbol}{$\blacksquare$}
\begin{proof}[Proof of Claim 3]
We find the number of homomorphisms $\vf' \in \vf/_{\bumpeq}$. 
The structure $\cG'$ consists of copies of $(\cK_L(x),x)$ and $(\cK_R(y),y)$. By Claim~2, and by the choice of $\cK_L,\cK_R$ 
\begin{align*}
    |\vf/_{\bumpeq}|&= 
    \Big( \prod_{x \in \Lside^\sG \cap I_{\vf}} \hom((\cK_L(x),x),(\cH, B))\Big)\cdot
    \Big( \prod_{x \in \Lside^\sG - I_{\vf}} \hom((\cK_L(x),x),(\cH, A))\Big)\\
    &\times \Big( \prod_{y \in \Rside^\sG \cap I_{\vf}} \hom((\cK_R(y),y),(\cH,C))\Big)\cdot
    \Big( \prod_{y \in \Rside^\sG - I_{\vf}} \hom((\cK_R(y),y),(\cH,  A \cup B))\Big)\\
    &\equiv 1\pmod{p}.
\end{align*}
\end{proof}
\renewcommand{\qedsymbol}{$\Box$}
Assume that $\bumpeq$ has $M$ classes where each class represents an independent set in $\sG$, hence
\begin{flalign*}
    |\Hom(\cG',\cH^\dg)|=\sum_{i=1}^M |\vf_i/_{\bumpeq_I}| \equiv M \equiv \#_p\BIS(1,1)(\sG) \pmod p.
\end{flalign*}
as required.

\medskip

Now suppose that $H'$ is an independent 3-path and $V_1=\emptyset$. We reduce the problem $\#_p\IS$ to $\#_p\CSP(\cH^\const)$. Let $\sG = (U,F)$ be an input for $\#_p\IS$, and construct a structure $\cG'$ as follows:
\begin{itemize}
\item 
Include the vertices and edges of $\sG$ into $\cG'$.
\item 
Since both $A=A\cup B$ and $A\cup B\cup C$ are subalgebras, by Lemma~\ref{lem:t-12-recognizable} with parameters $A\cup B=A$ and $C$ a gadget $(\cK'(x),x)$ satisfying the conditions of Lemma~\ref{lem:t-12-recognizable} exists. Attach $p-1$ copies of $(\cK'(x),x)$ to each $x\in U$ and denote it by $(\cK(x),x)$.
\end{itemize}

Suppose $\varphi: \cG' \to\cH^\dg$ is a homomorphism. For any edge $(x,y) \in F$, the possible combinations of the images of $x,y$ are
\begin{itemize}
    \item If $\varphi(x) \in A$, then $\varphi(y)$ can be in either $A$ or $C$.
    \item If $\varphi(x) \in C$, then $\varphi(y)$ must be in $A$.
\end{itemize}

This results in the following table
\begin{align*}
\kbordermatrix{
 &  y \to A & y \to C \\
x \to A & 1 & 1  \\
x \to C & 1 & 0  \\
}
\end{align*}

In this construction, the set $C$ encodes independent sets of $\sG$. For a homomorphism $\varphi: \cG' \to \cH^\dg$, we interpret $x \in I$ if and only if $\varphi(x) \in C$.

This construction ensures that adjacent vertices cannot both map to $C$, and thus, each homomorphism corresponds to an independent set in $\sG$. 

Moreover, due to the design of the gadgets, each such assignment contributes $1 \bmod p$ to the total homomorphism count. Therefore, we can recover the value of $\#_p\IS$ on the input graph $\sG$, completing the reduction.
\end{proof}

\section{Omitted proofs}\label{sec:appendix}
 

In this section, we give a full proof of Theorem~\ref{the:splitting_automorphisms}. To avoid confusion and utilize the existing results, we demote  $\bip$-structures to merely bipartite graphs with fixed left and right sides and use the appropriate terminology. 

We denote the disjoint union of graphs $G$ and $H$ by $G+H$. It will also be convenient to use the standard graph theory notation, and denote graphs in regular font and use $V(H),E(H)$ for the sets of vertices and edges. The neighborhood of a vertex $v\in V(G)$ or a set $A\sse V(G)$ in $G$ is denoted $N_G(v),N_G(A)$, respectively. However, we will use $\edge ab$ to denote edges, because sometimes vertices will have a complicated structure.

\subsection{Graph Products}

We will need three types of graph products. For Cartesian product we use the results of Chapter~6 of \cite{ref:ProductOfGraphs}, which are valid for arbitrary graphs.

The \emph{Cartesian product} of two simple graphs $A, B$ is the graph $A \carts B$ with the vertex set $V(A)\times V(B)$ and the edge set $E(A \carts B) =\{\edge{(a, b)}{(a', b')} | \edge{a}{a'} \in E(A) \text{ and } b = b', \text{ or } a = a' \text{ and } \edge b{b'} \in E(B) \}$.

A graph $G$ is called \emph{prime} with respect to Cartesian product if whenever $G=G_1\carts G_2$, one of $G_1,G_2$ contains only one vertex.
 
\begin{theorem}[Lemma 6.7, Theorem 6.8, \cite{ref:ProductOfGraphs}]\label{the:FactorizationOfCartesian}
Let $G, H$ be isomorphic connected graphs $G = G_1 \carts \dots \carts G_k$ and
$H = H_1 \carts \dots \carts H_l$, where each factor $G_i$ and $H_i$ is prime with respect to Cartesian product. Then $k = l$, and for any isomorphism $\varphi : G \rightarrow H$,
there is a permutation $\pi$ of $[k]$ and isomorphisms $\varphi_i : G_{\pi(i)} \rightarrow H_i$ for which
\begin{equation*}
    \varphi(\vc xk)=(\varphi_1(x_{\pi(1)}),\zd\varphi_k(x_{\pi(k)})). 
\end{equation*}
\end{theorem}

The direct product of two graphs $A,B$ is defined in the regular way, as the graph $A \times B$ with the vertex set $V(A)\times V(B)$ and the edge set $E(A \times B) =\{\edge{(a, b)}{(a', b')} | \edge{a}{a'} \in E(A) \text{ and } \edge{b}{b'} \in E(B) \}$.

The diamond product is defined with bipartite graphs in mind. For a bipartite graph $G=(L\cup R , E)$, we denote the two parts of the bipartition by $L^G=L$ and $R^G=R$.

\begin{definition}\label{def:FullDiamonProduct}
The \emph{diamond product} of two simple bipartite graphs $G=(L^G\cup R^G, E(G))$ and $H=(L_H\cup R_H, E(H))$ is defined as follows: 

\begin{align*}
     V(G \diamond H) &= (L^G\times L^H) \cup (R^G\times R^H) \\
     E(G \diamond H) &= \{\edge{(a, b)}{(a', b')} | \edge{a}{a'} \in E(G) \text{ and } \edge{b}{b'} \in E(H) \}
\end{align*} 
\end{definition}

\begin{remark}\label{rem:diamondToDirect}
An alternative way to view the diamond product is to consider it as the product of two relational $\bip$-structures.
\end{remark}

\begin{remark}
The diamond product for simple graphs $G$ and $H$ is commutative and associative up to an isomorphism, meaning $G \diamond H \cong H \diamond G$, and $(G \diamond H) \diamond \sK \cong G \diamond (H \diamond \sK) $
\end{remark}

The following statement is straightforward by Remark~\ref{rem:diamondToDirect}

\begin{proposition}
    Let $G,H,K$ be simple bipartite graphs with $r$ distinguished vertices $\vv x, \vv y, \vv z$ respectively. Then
    \begin{align*}
    \hom((K,G \diamond H) &= \hom(K,G)\cdot\hom(K,H),\\  
    \hom((K,\vv z),((G,\vv x)\diamond (H,\vv y)) &= \hom((K,\vv z),(G,\vv x))\cdot\hom((K,\vv z),(H,\vv y)).   
    \end{align*}
\end{proposition}

We denote the $k$-th diamond power of $H$ by $H^{\diamond k}$.

\subsection{$\RT$-thin Graphs}

We recall the definition of $\RT$-thinness for graphs here.
Relation $\RT_G$ on $V(G)$ for a graph $G$ is defined as follows. Vertices $x$ and $x'$ are in $\RT_G$, written $x\RT x'$, if and only if  $N_G(x) = N_G(x')$. 
Clearly, $\RT_G$ is an equivalence relation. Normally we will omit the subscript $G$.

We denote the quotient of $G$ modulo $\RT$ by $G/_\RT$. Given $x \in V(G)$, let $[x] = \{x'\in V (G) | x'\RT x\}$ denote the $\RT$-class containing $x$. Then $V(G/_\RT)=\{[x]\mid x\in V(G)\}$ and $E(G/_\RT)=\{\edge{[x]}{[y]}\mid \edge xy\in E(G)\}$. Since $\RT_G$ is defined entirely in terms of $E(G)$, it is easy to see that for an isomorphism $\vf : G \rightarrow H$, we have $x\RT_G y$ if and only if $\vf(x)\RT_H\vf(y)$.
Thus, $\vf$ maps equivalence classes of $\RT_G$ to equivalence classes of $\RT_H$, and can be defined to act on $V(G/_\RT)$ by $\vf([x]) = [\vf(x)]$.

We need the following lemmas from \cite{ref:ProductOfGraphs}.

\begin{lemma}[Proposition 8.4, \cite{ref:ProductOfGraphs}]\label{lem:cong_R-thin}
Suppose $G,H$ are simple graphs. Then $G \cong H$ if and only if $G/_\RT \cong H/_\RT$ and there is an isomorphism $\FUNC{\psi}{G/_\RT}{H/_\RT}$ with $|[x]| = |\psi([x])|$ for each $[x] \in V (G/_\RT)$. In fact, given an isomorphism $\FUNC{\psi}{G/_\RT}{H/_\RT}$, any map $\FUNC{\vf}{V(G)}{V(H)}$ that restricts to a bijection $\FUNC{\vf}{[x]}{\psi([x])}$ for every $[x] \in G/_\RT$ is an isomorphism from $G$ to $H$.
\end{lemma}

\begin{lemma}[Proposition 8.5, \cite{ref:ProductOfGraphs}]\label{lem:R-product}
If graphs $G$ and $H$ are simple graphs and have no isolated vertices, then $V((G\times H)/_\RT) =\{(x,y)  | x \in V (G/_\RT), y \in V (H/_\RT) \}$. 
In particular, $[(x, y)] = [x] \times [y]$. Furthermore,
$(G \times H)/_\RT \cong G/_\RT \times H/_\RT$, and $\FUNC{\psi}{(G \times H)/_\RT}{ G/_\RT \times H/_\RT}$ with $\psi ([x,y])=([x],[y])$ is an isomorphism.
\end{lemma}

\begin{remark}\label{rem:diamond_prop}
If $G=H\diamond K$, then $N_G((h, k)) = N_H(h) \times N_K(k)$ for any $(h,k)\in V(H\diamond K)$. Recall however that if $G$ and $H$ are bipartite, then for any $(h,k)\in V(H\diamond K)$ either $h\in L_G, k\in L_K$ or $h\in R_G, k\in R_K$.
\end{remark}

\begin{lemma}[Lemma~\ref{lem:factoring_R-thin} rephrased for bipartite graphs]
If graphs G and H are simple bipartite graphs and have no isolated vertices, then $(G \diamond H)/_\RT \cong G/_\RT \diamond H/_\RT$, and $\FUNC{\psi}{(G \diamond H)/_\RT}{ G/_\RT \diamond H/_\RT}$ with $\psi ([x,y])=([x],[y])$ is an isomorphism. Also $[(x, y)] = [x] \times [y]$.
\end{lemma}

\begin{proof}
Consider a vertex $[(x, y)]$ of $(G \diamond H)/_\RT$, By Remark~\ref{rem:diamond_prop}, we have
\begin{align*}
(x', y') \in [(x, y)] &\Leftrightarrow N_{G\diamond H}((x', y')) = N_{G\diamond H}((x, y))\\
&\Leftrightarrow N_G(x') \times N_H(y') = N_G(x) \times N_H(y) \\
&\Leftrightarrow N_G(x') = N_G(x) \mbox{ and } N_H(y') = N_H(y) \\
&\Leftrightarrow x' \in [x] \mbox{ and } y' \in [y] \\
&\Leftrightarrow (x', y') \in [x] \times [y] .
\end{align*}

Thus $[(x, y)] =
[x] \times [y]$. To complete the proof, we show that $\psi([(x, y)])= ([x], [y])$ is an isomorphism. 
\begin{align*}
    \edge{[(x, y)]}{[(x', y')]} \in E((G\diamond H)/_\RT) &\Leftrightarrow \edge{(x, y)}{(x' , y')} \in E(G \diamond H) \\
    &\Leftrightarrow \edge{x}{x'} \in E(G) \mbox{ and } \edge{y}{y'} \in E(H) \\
    &\Leftrightarrow \edge{[x]}{[x']} \in E(G/_\RT) \mbox{ and } \edge{[y]}{[y']} \in E(H/_\RT) \\
    &\Leftrightarrow \edge{([x],[x'])]}{[([y],[y'])} \in E(G)/_\RT \diamond E(H)/_\RT).
\end{align*}
\end{proof}

\subsection{Cartesian Skeleton}
In this subsection we prove a result similar to Theorem~\ref{the:FactorizationOfCartesian} only for direct products of bipartite graphs. First, we introduce several definitions and results similar to those in Chapter 8 of \cite{ref:ProductOfGraphs}.
The \emph{Boolean square} of a graph $G$ is the graph $G^s$ with $V(G^s) = V(G)$ and $E(G^s) = \{\edge xy | N_G(x) \cap N_G(y) \not = \emptyset \}$.

\begin{lemma}\label{lem:skeleton}
If $\vc Gk$ are  bipartite graphs, then $(G_1\diamond\dots\diamond G_k)^s = G^s_1\diamond\dots\diamond G^s_k$. 
\end{lemma}

\begin{proof}
Observe that $\edge{(\vc xk)}{(\vc yk)} \in E ((G_1 \diamond \dots\diamond G_k)^s)$ if and only if there is a walk
of length 2 joining $(\vc xk)$ and $(\vc yk)$ in $G_1 \diamond \dots\diamond G_1$. By definition of $\diamond$ such a walk exists if and only if each $G_i$ has a walk of length 2 between $x_i$ and $y_i$. The latter is equivalent to $\edge{x_i}{y_i}\in  E((G_i)^s)$ for each $i\in [k]$, which happens if and only if $\edge{(\vc xk)}{(\vc yk)}$
is an edge of $G^s_1 \diamond\dots\diamond G^s_k$.
\end{proof}

Next, following \cite{ref:ProductOfGraphs} we introduce \emph{Cartesian skeleton} of a bipartite graph. While this concept is not very intuitive, if the graph $G$ under consideration is a direct product of two bipartite graphs, the construction aims to eliminate `diagonal' edges like $\edge{(h, k)}{(h', k')}$ from the Boolean square of $G$, thereby making it similar to a Cartesian product. 

Given a factorization $G = H \diamond K$, we say that an edge $\edge{(h, k)}{(h', k')}$ of the Boolean square $G^s$ is \emph{Cartesian} relative to the factorization if either $h = h'$ and $k \not = k'$, or $h \not = h'$ and $k = k'$.
An edge $\edge xy$ of the Boolean square $G^s$ is \emph{dispensable} if it is a loop, or if there exists some $z \in V (G)$ for which both of the following statements hold:
\begin{enumerate}
    \item $N_G(x) \cap N_G(y) \subset N_G(x) \cap N_G(z)$ or $N_G(x) \subset N_G(z)\subset N_G(y)$,
    \item $N_G(y) \cap N_G(x) \subset N_G(y) \cap N_G(z)$ or $N_G(y) \subset N_G(z)\subset N_G(x)$.
\end{enumerate}

\begin{definition}
The \emph{Cartesian skeleton} $S(G)$ of a graph $G$ is the spanning subgraph of the Boolean square $G^s$ obtained by removing all dispensable edges from $G^s$.
\end{definition}

The following statements are from \cite{ref:ProductOfGraphs}.

\begin{lemma}[Proposition 8.11, \cite{ref:ProductOfGraphs}]\label{lem:IsoOnSkeleton}
Any isomorphism $\vf : G \rightarrow H$, as a map $\vf: V(G) \rightarrow V(H)$, is also an isomorphism $\vf : S(G) \rightarrow S(H)$.
\end{lemma}

\begin{lemma}[Proposition 8.13, \cite{ref:ProductOfGraphs}]\label{lem:BipOfS}
Suppose a graph $G$ is connected.
\begin{enumerate}
    \item If $G$ has an odd cycle, then $S(G)$ is connected. 
    \item If $G$ is bipartite, then $S(G)$ has two connected components whose respective vertex sets are the two parts of the bipartition of $G$.
\end{enumerate}
\end{lemma}

If $G$ is bipartite, we denote the connected component of the $S(G)$ corresponding to the left part of $G$ by $S_L(G)$ and the one corresponding to the right part of $G$ by $S_R(G)$. Note that if $G=H\diamond K$, by Remark~\ref{rem:diamond_prop} $N_G((h, k)) = N_H(h) \times N_K(k)$. This implies the following:

\begin{lemma}[\cite{ref:ProductOfGraphs}]\label{lem:CartesEdge}
If $G$ is an R-thin bipartite graph with an arbitrary factorization $G = H\diamond K$, then every edge of $S(G)$ is Cartesian relative to this factorization.
\end{lemma}

Observe that in the case of bipartite graphs the Cartesian skeleton is restricted to the left and right parts of the bipartition of the graph. Hence, we can modify Proposition~8.10 of \cite{ref:ProductOfGraphs} for diamond product. 

\begin{proposition}\label{pro:FactorSkeleton}
If $A,B$ are $\RT$-thin bipartite graphs without isolated vertices, then 
\begin{equation*}
    S(A\diamond B) =  S_L(A)\carts S_L(B) + S_R(A)\carts S_R(B).
\end{equation*}
\end{proposition}

\begin{proof}
As is easily seen, $S(A \diamond B)$ has two connected components $S_L$ and $S_R$. We prove that the $S_L(A\diamond B)$ is equal to $S_L(A)\carts S_L(B)$. The proof for the right part is similar.

First, we show $E(S_L(A)\carts S_L(B)) \subseteq E(S_L(A\diamond B))$. Take an edge in $E(S_L(A)\carts S(B)_L)$, say $\edge{(a, b)}{(a', b)}$ with $\edge{a}{a'} \in E(S_L(A))$. We must show that $\edge{(a, b)}{(a', b)}$ is not dispensable in $(A \diamond B)^s$. Suppose it is. Then there is a vertex $z = (c', c'')$ in $G=S_L(A\diamond B)$ such that the dispensability conditions (1) and (2) hold for $x = (a, b), y = (a', b)$, and $z = (c', c'')$. The various cases are considered below. Each leads to a contradiction.

Suppose $N_G(x) \subset N_G(z) \subset N_G(y)$. This means $N_A(a) \times N_B(b) \subset N_A(c') \times N_B(c'') \subset N_A(a') \times N_B(b)$, so $N_B(c'') = N_B(b)$. Then the fact that $N_B(b)\neq \emptyset$ permits cancellation of the common factor $N_B(b)$, so $N_A(a) \subset N_A(c') \subset N_A(a')$, and $\edge{a}{a'}$ in ${A^s}$ is dispensable. We will reach the same
contradiction if $N_G(y) \subset N_G(z)\subset N_G(x)$.

Finally, suppose there is a $z = (c', c'')$ for which both $N_G(x) \cap N_G(y) \subset N_G(x) \cap N_G(z)$ and $N_G(y) \cap N_G(x) \subset N_G(y) \cap N_G(z)$. Rewrite this as
\begin{align*}
N_G((a, b)) \cap N_G((a', b)) &\subset N_G((a, b)) \cap N_G((c', c'')),\\
N_G((a', b)) \cap N_G((a, b)) &\subset N_G((a', b)) \cap N_G((c', c'')),
\end{align*}
which is the same as
\begin{align*}
(N_A(a) \cap N_A(a'))\times N_B(b) &\subset (N_A(a) \cap N_A(c'))\times (N_B(b) \cap N_B(c'')),\\
(N_A(a') \cap N_A(a))\times N_B(b) &\subset (N_A(a') \cap N_A(c'))\times (N_B(b) \cap N_B(c'')).
\end{align*}
Thus $N_B(b) \subset N_B(b) \cap N_B(c'')$, so $N_B(B) = N_B(b) \cap N_B(c'')$, whence
\begin{align*}
N_A(a) \cap N_A(a') &\subset N_A(a) \cap N_A(c'),\\
N_A(a') \cap N_A(a') &\subset N_A(a') \cap N_A(c').
\end{align*}
Thus $\edge{a}{a'}$ in ${A^s}$ is dispensable, a contradiction.

We now show that $E(L_{S(A \diamond B)})\subseteq E(S_L(A)\carts S_L(B))$.
By Lemma~\ref{lem:CartesEdge}, all edges of $L_{S(A \diamond B)}$ are Cartesian, so we just need to show that $\edge{(a, b)}{(a', b)} \in E(L_{S(A \diamond B)})$ implies $\edge{a}{a'} \in E(S(A))$ (The same argument will work for edges of the form $\edge{(a, b)}{(a, b')}$).

Suppose for the sake of contradiction that$\edge{(a, b)}{(a', b)} \in E(L_{S(A \diamond B)})$, but $\edge{a}{a'} \not \in E(S(A))$, Thus $\edge{a}{a'}$ is dispensable in $L_{A^s}$, so there is a $c\in V(A)$ for which both of the following conditions hold:
\begin{itemize}
    \item [1. ] $N_A(a) \cap N_A(a') \subset N_A(a) \cap N_A(c)$ or $N_A(a) \subset N_A(c)\subset N_A(a')$,
    \item [2. ] $N_A(a') \cap N_A(a) \subset N_A(a') \cap N_A(c)$ or $N_A(a') \subset N_A(c)\subset N_A(a)$.
\end{itemize}
There are no isolated vertices, so $N_B(d)\neq \emptyset$. Now,  we can multiply each neighborhood
$N_A(x)$ in Condition 1 and 2 by $N_B(d)$ on the right and still preserve the proper inclusions. Then the fact that $N_{(A\diamond B)}((a, b)) = N_A(a) \times N_B(b)$ ($a\in L_A$ and $b \in L_B$) yields the dispensability conditions (1) and (2), where
$x = (a, b)$ and $y = (a', b)$ and $z = (c, b)$. Thus $\edge{(a, b)}{(a', b)} \not\in  E(L_{S(A \diamond B)})$, a contradiction.
\end{proof}

\subsection{Diamond product factorization}

A bipartite graph $G$ is said to be \emph{prime with respect to diamond product} if for any $G\cong H\diamond K$ one of $H,K$ is an edge. The following lemma is the core technical statement of this section.

\begin{lemma}\label{lem:auto_dependency_r-thin}
Consider any isomorphism $\FUNC{\vf}{G_1\diamond\dots\diamond G_k}{H_1\diamond\dots \diamond H_\ell}$, where $\vf(\vc xk)=(\vf_1(\vc xk)\zd\vf_\ell(\vc xk))$ (recall that $\vf$ preserves the left and right parts of the graphs) and  all the factors are connected and $\RT$-thin. If a factor $G_i$ is prime, then exactly one of the functions $\vf_1, \vf_2, ... , \vf_\ell$ depends on $x_i$.
\end{lemma}

\begin{proof}
By grouping and permuting the factors we may assume that $k=\ell=2$ and $G_1$ is prime. We prove the lemma by showing that if it is not the case that exactly one of $\vf_1$ and $\vf_2$ depends on $x_1$, then $G_1$ is not prime. More specifically, we represent the Cartesian skeletons of $G_1,G_2,H_1,H_2$ as Cartesian products of prime graphs and show that if both $\vf_1$ and $\vf_2$ depend on $x_1$, than those Cartesian products can be rearranged in such a way that $G_1$ can be further decomposed into a direct product.

We denote the connected components of $S(G_1 \diamond G_2)$ by $S_L$ and $S_R$, and the connected components of $S(H_1 \diamond H_2)$ by $C_L$ and $C_R$ which correspond to the left and right parts of the bipartition.

By Lemma~\ref{lem:IsoOnSkeleton} $\vf$ is also an isomorphism for Cartesian skeletons:
\begin{equation}\label{eq:isooncarts}
    \FUNC{\vf}{S_L + S_R }{C_L+C_R},
\end{equation}
which gives us isomorphisms on each connected component:
\begin{equation*}
    \FUNC{\vf_L}{S_L }{C_L}, \; \; \FUNC{\vf_R}{S_R}{C_R}.   
\end{equation*}
Note that although $G_1$ is prime with respect to diamond product, $S(G_1)$ as well as $S_L,S_R$ are not necessarily prime with respect to Cartesian product. Take prime factorizations of each component with respect to Cartesian product, when both $G$ and $H$ are bipartite: 
\begin{align*}
    S_L(G_1)= A_1 \carts \dots \carts A_t \; &, \; \; S_R(G_1)= B_1 \carts\dots\carts B_s,\\
     S_L(G_2)=D_1 \carts \dots \carts D_w \; &, \; \; S_R(G_2)= E_1 \carts \dots \carts E_r,\\
     C_L= C_1 \carts \dots \carts C_m \; &, \; \; C_R = C'_1 \carts \dots \carts C'_n.
\end{align*}
As $\vf$ is an isomorphism, we  have
\begin{align*}
    \FUNC{\vf_L}{(A_1 \carts \dots \carts A_t) \carts (D_1 \carts \dots \carts D_w) &}{ (C_1 \carts \dots \carts C_m)},\\ 
    \FUNC{\vf_R}{(B_1 \carts \dots \carts B_s) \carts (E_1 \carts \dots \carts E_r) &}{(C'_1 \carts \dots \carts C'_n)}.
\end{align*}
We relabel the vertices of $G_1$ with 
$V((A_1 \carts \dots \carts A_t) + (B_1 \carts \dots \carts B_s))$, and those of $G_2$ with
$V(D_1 \carts \dots\carts D_w + E_1 \carts \dots \carts E_r)$. We relabel vertices of $H_1\diamond H_2 $ with $V(C_1 \carts \dots \carts C_m+ C'_1 \carts \dots \carts C'_n)$. By Proposition~\ref{pro:FactorSkeleton}  the isomorphism $\vf$ can be represented as follows:
\begin{align*}
    \vf_L:(A_1 \carts \dots \carts A_t) \carts (D_1 \carts \dots \carts D_w) &\rightarrow \\
    (A_1 \carts \dots \carts A_{t'} \carts D_1 \carts \dots \carts D_{w'}) & \carts (A_{t'+1} \carts \dots \carts A_{t} \carts D_{w'+1} \carts \dots \carts D_{w}), \\ 
    \vf_R:
    (B_1 \carts \dots \carts B_s) \carts (E_1 \carts \dots \carts E_r) &\rightarrow \\
    (B_1 \carts \dots \carts B_{s'} \carts E_1 \carts \dots \carts E_{r'}) & \carts (B_{s'+1} \carts \dots \carts B_{s} \carts E_{r'+1} \carts \dots \carts E_{r}).
\end{align*}

To simplify the notation we will denote a left vertex $(a_1\zd a_{t'}, a_{t'+1}\zd a_t)$ of $G_1$ by $(x_L,y_L)$, where $x_L=(a_1\zd a_{t'})$ and $y_L=(a_{t'+1}\zd a_t)$, and right vertex $(b_1\zd b_{s'}, b_{s'+1}\zd b_s)$ of $G_1$ by $(x_R,y_R)$, where $x_R=(b_1,\zd b_{s'})$ and $y_R=(b_{s'+1}\zd b_s)$. Similarly, we
denote a left vertex $(d_1\zd d_{w'}, d_{w'+1}\zd d_w)$ of $G_2$ by $(u_L,v_L)$, where $u_L=(d_1\zd d_{w'})$ and $v_L=(d_{w'+1}\zd d_w)$,
and a right vertex $(e_1\zd e_{r'}, e_{r'+1}\zd e_r)$ of $G_2$ by $(u_R,v_R)$, where $u_R=(e_1\zd e_{r'})$ and $v_R=(e_{r'+1}\zd e_r)$. By Theorem~\ref{the:FactorizationOfCartesian}, isomorphism $\vf$ is represented as follows
\begin{align*}
    \vf_L((x_L,y_L),(u_L,v_L)) &=((x_L,u_L),(y_L,v_L)),\\
    \vf_R((x_R,y_R),(u_R,v_R)) &=((x_R,u_R),(y_R,v_R)).
\end{align*}
Since this holds for both left and right parts $G_1 \diamond G_2$, we can represent the isomorphism like $\vf((x,y),(u,v))=((x,u),(y,v)) $.

Now we find a nontrivial factorization $G_1\cong S \diamond S'$. Define $S$ and $S'$ as follows:
\begin{align*}
    V(S)&=\left\{x_L |\; \Big((x_L, y_L), (u_L, v_L) \Big) \in V(G_1 \diamond G_2) \right\}\\
    &\hspace*{5cm}\cup
    \left\{ x_R |\; \Big((x_R, y_R), (u_R, v_R) \Big) \in V(G_1 \diamond G_2) \right\}, \\
    E(S)&= \left\{ \edge{x_L}{x_R} |\; \edge{ \Big((x_L, y_L), (u_L, v_L)\Big)}{\Big((x_R, y_R), (u_R, v_R) \Big)}\in E(G_1 \diamond G_2)  \right\},
    \end{align*}
and
\begin{align*}
    V(S')&=\left\{y_L |\; \Big((x_L, y_L), (u_L, v_L)\Big) \in V(G_1 \diamond G_2) \right\}\\
    &\hspace*{5cm}\cup
    \left\{ y_R |\; \Big((x_R, y_R), (u_R, v_R)\Big) \in V(G_1 \diamond G_2) \right\}, \\
    E(S')&= \left\{ \edge{y_L}{y_R} |\; \edge{ \Big((x_L, y_L), (u_L, v_L)\Big)}{\Big((x_R, y_R), (u_R, v_R)\Big)}\in E(G_1 \diamond G_2)  \right\}.
\end{align*}

We need to prove that $\edge{(x,y)}{(x',y')} \in E(G_1)$ if and only of $\edge{(x,y)}{(x',y')} \in E(S \diamond S')$.
If $\edge{(x,y)}{(x',y')} \in E(G_1)$, then for some $u,v,u',v'$ there is an edge $\edge{ \Big((x, y), (u, v)\Big)}{\Big((x', y'), (u', v')\Big)}\in E(G_1 \diamond G_2)$, and hence there is an edge $S \diamond S'$ which is $\edge{ (x, y)}{ (x', y')}\in E(S \diamond S')$.
Next, suppose that $\edge{(x,y)}{(x',y')} \in E(S \diamond S')$. Then, $\edge{x}{x'} \in E(S)$ and $\edge{y}{y'}\in E(S')$. By the construction of $S$ and $S'$ we have that there exist $a,b, u, v,u',v'$ such that
\begin{align*}
\edge{\Big((x, a), (u, v)\Big)}{\Big((x', b), (u', v')\Big)}\in E(G_1 \diamond G_2),
\end{align*}
and there exists $c,d,u'',v'',u''',v'''$, such that 
\begin{align*}
\edge{\Big((c, y), (u'', v''')\Big)}{\Big((d , y'), (u''', v''')\Big)}\in E(G_1 \diamond G_2).
\end{align*}
Now, apply the isomorphism $\vf$
\begin{align*}
\edge{\Big((x, u), (a, v)\Big)} {\Big((x', u'), (b, v')\Big)}&\in E(H_1 \diamond H_2),\\
\edge{\Big((c, u''), (y, v''')\Big)}{\Big((d ,u''' ), (y' , v''')\Big)}&\in E(H_1 \diamond H_2).
\end{align*}
Hence, $\edge{(x, u)}{(x', u')} \in E(H_1)$ and also  $\edge{(y, v''')}{(y' , v''')}\in E(H_2)$. Therefore, we have \linebreak $\edge{\Big((x,u),(y,v''')\Big)}{ \Big((x',u'),(y' , v''')\Big)} \in E(H_1 \diamond H_2)$. Applying $\vf^{-1}$ we get
\[
    \edge{\Big((x,y),(u,v''')\Big)}{\Big((x',y'),(u' , v''')\Big)}\in E(G_1 \diamond G_2).
\] 
Thus, $\edge{(x,y)}{(x',y')}\in E(G_1)$.
\end{proof}
Let $G^{\diamond \ell}$ denote the diamond product of $G$ to itself for $\ell$\ times, i.e. $G^{\diamond \ell} = G \diamond ... \diamond G$. 

\begin{corollary}\label{cor:splitting_automorphisms-r-thin}
Let $G$ be a prime $\RT$-thin graph and let $\vf$ be an automorphism of $G^{\diamond \ell}$. Then there is a permutation $\pi$ such that 
\begin{equation*}
   \vf(\vc x\ell) =(\vf_1(x_{\pi(1)})\zd \vf_\ell(x_{\pi(\ell)}) )
\end{equation*}
where each $\vf_i$ is an automorphism of $G$. 
\end{corollary}

\subsection{Proof of Theorem~\ref{the:splitting_automorphisms}}
\begin{theorem}[Theorem~\ref{the:splitting_automorphisms}, the bipartite form]
Suppose $\vf$ is an automorphism of $G^{\diamond \ell}$.  Let $\FUNC{\psi}{G^{\diamond \ell}/_\RT}{ G^{\diamond \ell}/_\RT}$ be the  automorphism induced by $\vf$, and let $G=G_1 \diamond \dots \diamond G_k$ be a prime factorization of $G$.
Then there is a permutation $\pi$ of $[\ell]\times [k]$ such that every automorphism of $G^{\diamond \ell}/_\RT$ can be split into $k\ell$ automorphisms:
\begin{align*}
    \psi([v_1]\zd [v_\ell])
    &=\Big(\big(\psi_{1,1}([v_{\pi(1,1)}])\zd\psi_{1,k}([v_{\pi(1,k)}])\big),\\
    &\;\;\;\;\;\;\vdots\\
    &\;\;\;\;\;\; \big(\psi_{\ell,1}([v_{\pi(\ell,1)}])\zd\psi_{\ell,k}([v_{\pi(\ell,k)}])
    \big)\Big).
\end{align*}
\end{theorem}

We start with two auxiliary lemmas.

\begin{lemma}\label{lem:LiftingLemma}
Let $G$ be a simple connected bipartite graph such that $G/_\RT \cong H_1 \diamond H_2$. Then every vertex $(x, y) \in V (H_1 \diamond H_2)$ corresponds to a unique $\RT$-class of $G$, denote its cardinality by $|(x, y)|$. If there are functions $\FUNC{\omega_1}{V(H_1)}{\mathbb{N}}, \FUNC{\omega_2}{V(H_2)}{\mathbb{N}}$ such that $|(x, y)| = \omega_1(x)\cdot \omega_2(y)$, then there are bipartite graphs $H'_1$ and $H'_2$ such that $G \cong H'_1 \diamond H'_2$, and $H'_1/_\RT=H_1$ and $H'_2/_\RT=H_2$.
\end{lemma}

\begin{proof}
By Lemma~\ref{lem:factoring_R-thin} $H_1,H_2$ are $\RT$-thin. Define $H'_1$ as follows. Take a family $\{A_x \mid x \in  V (H_1)\}$ of disjoint sets such that $|A_x| = \omega_1(x)$ for each $x \in V (H_1)$. Set $V (H'_1) = \bigcup_{x \in V(H_1)} A_x$, and $E(H'_1)= \{ \edge ab | a\in A_x, b\in A_y \mbox{ and } \edge{x}{y} \in E(H_1) \}$. A graph $H'_2$ is constructed in the same way by choosing sets $B_x$ with $|B_x| = \omega_2(x)$ for each $x \in V(H_2)$. Set $V(H'_2) = \bigcup_{y \in V(H_1)} B_x$, and $E(H'_2)= \{ \edge ab | a\in B_x, b\in B_y \mbox{ and } \edge{x}{y} \in E(B) \}$.

\smallskip
\renewcommand{\qedsymbol}{$\blacksquare$}
{\sc Claim 1. } The $\RT$-classes of $H'_1,H'_2$ are the sets $A_x,B_x$, respectively.
\begin{proof}[Proof of Claim 1]
It suffices to prove that if $a,b\in A_x$, then $N_{H'_1}(a)=N_{H'_2}(b)$. If there is a $c$ such that $\edge{a}{c} \in E(H'_1)$, then there is $z$ such that $c\in A_z$ and $\edge{x}{z} \in E(H_1)$, then by the definition of $E(H'_1)$ we have $\edge{b}{c}\in E(H'_1)$. For $H'_2$ the proof is analogous.
\end{proof}

\renewcommand{\qedsymbol}{$\Box$}
By Claim~1 there are isomorphisms $\FUNC{\psi_1}{H_1}{H'_1/_\RT}$ and $\FUNC{\psi_2}{H_2}{H'_2/_\RT}$, hence, there is a isomorphism $\FUNC{\psi}{H_1\diamond H_2}{H'_1/_\RT \diamond H'_2/_\RT}$. Therefore, $(H'_1 \diamond H'_2)/_\RT \cong H'_1/_\RT \diamond H'_2/_\RT \cong H_1\diamond H_2 \cong G/_\RT $. By Lemma~\ref{lem:cong_R-thin} we have $H'_1\diamond H'_2 \cong G$. 
\end{proof}

\begin{lemma}\label{lem:splitting_isomprphism}
Suppose there is an isomorphism $\FUNC{\vf}{G_1 \diamond \dots \diamond G_k}{H_1\diamond \dots \diamond H_\ell}$ where $\vf(\vc xk) =(\vf_1(\vc xk)\zd \vf_\ell(\vc xk))$ and all factors are connected and such that $G_i/_\RT,H_i/_\RT$ are nontrivial. Let $\FUNC{\psi}{G_1/_\RT\diamond\dots\diamond G_k/_\RT}{ H_1/_\RT\diamond\dots \diamond  H_\ell/_\RT}$ be the induced isomorphism
\begin{equation*}
   \psi([x_1]\zd [x_k]) =(\psi_1([x_1]\zd [x_k]), ... , \psi_l([x_1]\zd [x_k]) ).
\end{equation*}
If some $G_i$ is prime, then at most one of the functions $\psi_j$ depends on $[x_i]$.
\end{lemma}

\begin{proof}
Grouping and permuting factors, it suffices only to prove the lemma in the case when $G_1$ is prime and $k = l= 2$. We rewrite $\FUNC{\vf}{G_1 \diamond G_2}{ H_1 \diamond H_2}$ as $\vf(x, y) =(\vf_1(x, y), \vf_2(x, y))$. As a consequence we have $\FUNC{\psi}{G_1/_\RT \diamond G_2/_\RT}{H_1/_\RT \diamond H_2/_\RT}$, where $\psi([x], [y]) =(\psi_1([x], [y]), \psi_2([x], [y]))$. As $G_1,G_2,H_1,H_2$ are connected, each of $G_1/_\RT,G_2/_\RT,H_1/_\RT,H_2/_\RT$ is connected and $\RT$-thin.
We prove that if $\psi_1$ and $\psi_2$ depend on $[x]$, then $G_1$ is not prime, a contradiction with the assumption made.

Lemma~\ref{lem:auto_dependency_r-thin} implies that if both $\psi_1$ and $\psi_2$ depend on $[x]$, then  $G_1/_\RT$ is
not prime. Take a prime factorization $G_1/_\RT = A_1 \diamond \dots \diamond  A_n$. This gives a labeling $[x] =
(\vc an)$ of $\RT$-classes of $G_1$ with vertices of $A_1 \diamond \dots \diamond A_n$. Then $\psi$ can be viewed as an isomorphism $\FUNC{\psi}{A_1 \diamond \dots \diamond A_n \diamond G_2/_\RT}{H_1/_\RT \diamond H_2/_\RT}$. Note that when we switch from $G_1/_\RT$ to its prime factorization, the connection between individual vertices of $G_1$ and the vertices of the prime factors disappear. This means that $[x]$ has nothing to do with $([a_1]\zd [a_n])$.
By Lemma~\ref{lem:auto_dependency_r-thin} for each $i \in [n] $, exactly one of $\psi_1$ and $\psi_2$ depends on $a_i$. We order the factors of $G_1/_\RT$ = $A_1 \diamond \dots \diamond A_n$ so that $\psi_1$ depends on $\vc as$, but not on $a_{s+1}\zd a_n$, and $\psi_2$ depends on $a_{s+1}\zd a_n$ but not on $\vc as$. Then we have
\[
\psi(\vc as, a_{s+1},\zd a_n, [y]) =(\psi_1(\vc as, [y]), \psi_2(a_{s+1}\zd a_n, [y]).
\]
We know that $(\vc an) = [x] \in V (G_1/_\RT)$ is an $\RT$-class of $G_1$, so it makes sense to speak of its cardinality $|(\vc an)|$. By Lemma~\ref{lem:LiftingLemma} graph $G_1$ has a nontrivial factorization (i.e., is non-prime) if we can define functions $\FUNC{\omega_1}{A_1\diamond \dots \diamond A_s}{\mathbb{N}}$ and  $\FUNC{\omega_2}{A_{s+1}\diamond \dots \diamond A_n}{\mathbb{N}}$, for which $|(\vc as, a_{s+1}\zd a_n)| = \omega_1(\vc as)\cdot \omega_2(a_{s+1}\zd a_n)$. 
Let $[y] \in V (G_2/_\RT)$ be an $\RT$-class of $G_2$. Observe that the isomorphism $\FUNC{\vf}{G_1 \diamond G_2}{H_1 \diamond H_2}$ maps each $\RT$-class $( (\vc an , [y])$ of $G_1 \diamond G_2$ bijectively to the $\RT$-class $(\psi_1(\vc as, [y]) , \psi_2(a_{s+1}\zd a_n, [y]))$ of $H_1 \diamond H_2$. Therefore
\begin{equation*}\label{eq:fraction_of_omega}
|(\vc as, a_{s+1}\zd a_n)| =  \frac{|\psi_1(\vc as, [y])|\cdot|\psi_2(a_{s+1}\zd a_n, [y])|}{|[y]|}. 
\end{equation*}
Since $|(\vc as, a_{s+1}\zd a_n)|$ is an integer, $d=|[y]|$ divides $|\psi_1(\vc as, [y])|\cdot|\psi_2(a_{s+1}\zd a_n, [y])|$. So there are $d_1$ and $d_2$ such that $d_1\cdot d_2 =d $ and $d_1$ divides $|\psi_1(\vc as, [y])|$ and $d_2$ divides $|\psi_2(a_{s+1}\zd a_n, [y])|$. Set $\omega_1(\vc as)=\frac{|\psi_1(\vc as, [y])|}{d_1}$ and $\omega_2(a_{s+1}\zd a_n)= \frac{|\psi_2(a_{s+1}\zd a_n, [y])|}{d_2}$. It is easy to see that functions $\omega_1,\omega_2$ are as required. 
\end{proof}

\begin{proof}[Proof of Theorem~\ref{the:splitting_automorphisms}]
Let $G=G_1 \diamond \dots \diamond G_k$ be a factorization of $G$ into prime factors. It will be convenient to represent $G^{\diamond \ell}/_\RT$ as 
\begin{equation}\label{eq:factorization}
   G^{\diamond \ell}/_\RT = (G_{1,1}/_\RT \diamond \dots \diamond G_{1,k}/_\RT)\diamond \dots \diamond (G_{\ell,1}/_\RT \diamond \dots \diamond G_{\ell,k}/_\RT),
\end{equation}
where $G_{i,t}=G_{j,t}$ for all $i,j\in [\ell], t\in[k]$.
Relabel each vertex $(\vc v\ell) \in V(G^{\diamond \ell})$ as
\[
(\vc v\ell)=\big((v_{1,1}\zd v_{1,k})\zd (v_{\ell,1}\zd v_{\ell,k})\big).
\]
Also, we can write the isomorphism $\psi$ as follows:
\begin{align*}
\psi([v_1]\zd [v_\ell])&=\psi\big(([v_{1,1}]\zd[v_{1,k}])\zd ([v_{\ell,1}]\zd[v_{\ell,k}])\big)\\
    &=\Big(\psi_{1,1}\big(([v_{1,1}]\zd[v_{1,k}])\zd ([v_{\ell,1}]\zd[v_{\ell,k}])\big),\\
    &\;\;\;\;\;\;\vdots\\
    &\;\;\;\;\;\; \psi_{\ell, k}\big(([v_{1,1}]\zd[v_{1,k}])\zd ([v_{\ell,1}]\zd[v_{\ell,k}])\big)\Big),
\end{align*}
By Lemma~\ref{lem:splitting_isomprphism}, since all $G_{i,t}$ for all $i \in [\ell]$, $t \in [k]$ are prime, each $\psi_{i,t}$ depends only on one of $G_{i,t}$'s. Also, note that $G_{i,t} = G_{j,t}$ for all $i, j \in [\ell]$, $t \in [k]$. Thus, there exists a permutation $\pi$ of $[\ell] \times [k]$ such that $\psi$ can be split into $k\ell$ automorphisms:
\begin{align*}
    \psi([v_1]\zd [v_\ell])&=\psi(([v_{1,1}]\zd[v_{1,k}])\zd ([v_{\ell,1}]\zd[v_{\ell,k}]))\\
    &=\Big(\big(\psi_{1,1}([v_{\pi(1,1)}])\zd\psi_{1,k}([v_{\pi(1,k)}])\big),\\
    &\;\;\;\;\;\;\vdots\\
    &\;\;\;\;\;\; \big(\psi_{\ell,1}([v_{\pi(\ell,1)}])\zd\psi_{\ell,k}([v_{\pi(\ell,k)}])
    \big)\Big),
\end{align*}
where the equality follows by putting together all $\psi_{i,j}$ for $j\in [k]$ to obtain the automorphism $\psi_i$ on $G_i/_\RT$. 
\end{proof}


\section{Acknowledgments}
We would like to thank Gregor Lagodzinski for his very useful comments and recommendations that allowed to improve the presentation.

\bibliographystyle{plain}
\bibliography{refrences.bib}

\begin{thebibliography}{10}

\bibitem{ref:POlymorphismAndUsethem_barto2017polymorphisms}
Libor Barto, Andrei Krokhin, and Ross Willard.
\newblock Polymorphisms, and how to use them.
\newblock In {\em Dagstuhl Follow-Ups}, volume~7. Schloss
  Dagstuhl-Leibniz-Zentrum fuer Informatik, 2017.

\bibitem{Barvinok16:combinatorics}
Alexander~I. Barvinok.
\newblock {\em Combinatorics and Complexity of Partition Functions}, volume~30
  of {\em Algorithms and combinatorics}.
\newblock Springer, 2016.

\bibitem{ref:BULATOV_HColoring}
Andrei~A. Bulatov.
\newblock H-coloring dichotomy revisited.
\newblock {\em Theoretical Computer Science}, 349(1):31--39, 2005.
\newblock Graph Colorings.

\bibitem{Bulatov13:counting}
Andrei~A. Bulatov.
\newblock The complexity of the counting constraint satisfaction problem.
\newblock {\em J. {ACM}}, 60(5):34:1--34:41, 2013.

\bibitem{ref:BULATOV_TowardDichotomy}
Andrei~A. Bulatov and Víctor Dalmau.
\newblock Towards a dichotomy theorem for the counting constraint satisfaction
  problem.
\newblock {\em Information and Computation}, 205(5):651--678, 2007.

\bibitem{Bulatov05:partition}
Andrei~A. Bulatov and Martin Grohe.
\newblock The complexity of partition functions.
\newblock {\em Theor. Comput. Sci.}, 348(2-3):148--186, 2005.

\bibitem{Bulatov05:classifying}
Andrei~A. Bulatov, Peter Jeavons, and Andrei~A. Krokhin.
\newblock Classifying the complexity of constraints using finite algebras.
\newblock {\em SIAM J. Comput.}, 34(3):720--742, 2005.

\bibitem{Cai89:power}
Jin{-}yi Cai and Lane~A. Hemachandra.
\newblock On the power of parity polynomial time.
\newblock In Burkhard Monien and Robert Cori, editors, {\em {STACS} 89, 6th
  Annual Symposium on Theoretical Aspects of Computer Science, Paderborn, FRG,
  February 16-18, 1989, Proceedings}, volume 349 of {\em Lecture Notes in
  Computer Science}, pages 229--239. Springer, 1989.

\bibitem{Creignou96:complexity}
Nadia Creignou and Miki Hermann.
\newblock Complexity of generalized satisfiability counting problems.
\newblock {\em Inf. Comput.}, 125(1):1--12, 1996.

\bibitem{Dalmau04:complexity}
V{\'{\i}}ctor Dalmau and Peter Jonsson.
\newblock The complexity of counting homomorphisms seen from the other side.
\newblock {\em Theor. Comput. Sci.}, 329(1-3):315--323, 2004.

\bibitem{Dyer00:complexity}
M.~Dyer and C.~Greenhill.
\newblock The complexity of counting graph homomorphisms.
\newblock {\em Random {S}tructures and {A}lgorithms}, 17:260--289, 2000.

\bibitem{Dyer07:counting}
Martin~E. Dyer, Leslie~Ann Goldberg, and Mike Paterson.
\newblock On counting homomorphisms to directed acyclic graphs.
\newblock {\em J. {ACM}}, 54(6):27, 2007.

\bibitem{Dyer13:effective}
Martin~E. Dyer and David Richerby.
\newblock An effective dichotomy for the counting constraint satisfaction
  problem.
\newblock {\em {SIAM} J. Comput.}, 42(3):1245--1274, 2013.

\bibitem{faben2008complexity}
John Faben.
\newblock The complexity of counting solutions to generalised satisfiability
  problems modulo {k}, 2008.

\bibitem{ref:CountingMod2Ini}
John Faben and Mark Jerrum.
\newblock The complexity of parity graph homomorphism: an initial
  investigation.
\newblock {\em arXiv preprint arXiv:1309.4033}, 2013.

\bibitem{Feder98:computational}
Tom{\'{a}}s Feder and Moshe~Y. Vardi.
\newblock The computational structure of monotone monadic {SNP} and constraint
  satisfaction: {A} study through datalog and group theory.
\newblock {\em {SIAM} J. Comput.}, 28(1):57--104, 1998.

\bibitem{ref:CountingMod2ToKMinor}
Jacob Focke, Leslie~Ann Goldberg, Marc Roth, and Stanislav Zivney.
\newblock Counting homomorphisms to ${K}_4$-minor-free graphs, modulo 2.
\newblock {\em CoRR}, abs/2006.16632, 2020.

\bibitem{Focke21:counting}
Jacob Focke, Leslie~Ann Goldberg, Marc Roth, and Stanislav Zivn{\'{y}}.
\newblock Counting homomorphisms to ${K}_4$-minor-free graphs, modulo 2.
\newblock In D{\'{a}}niel Marx, editor, {\em Proceedings of the 2021 {ACM-SIAM}
  Symposium on Discrete Algorithms, {SODA} 2021, Virtual Conference, January 10
  - 13, 2021}, pages 2303--2314. {SIAM}, 2021.

\bibitem{ref:CountingMod2ToSquarefree}
Andreas G{\"o}bel, Leslie~Ann Goldberg, and David Richerby.
\newblock Counting homomorphisms to square-free graphs, modulo 2.
\newblock {\em ACM Transactions on Computation Theory (TOCT)}, 8(3):1--29,
  2016.

\bibitem{ref:CountingModPToTrees_gbel_et_al_LIPIcs}
Andreas G{\"o}bel, J.~A.~Gregor Lagodzinski, and Karen Seidel.
\newblock Counting homomorphisms to trees modulo a prime.
\newblock In Igor Potapov, Paul Spirakis, and James Worrell, editors, {\em 43rd
  International Symposium on Mathematical Foundations of Computer Science (MFCS
  2018)}, volume 117 of {\em Leibniz International Proceedings in Informatics
  (LIPIcs)}, pages 49:1--49:13, Dagstuhl, Germany, 2018. Schloss
  Dagstuhl--Leibniz-Zentrum fuer Informatik.

\bibitem{guo_et_al:LIPIcs:2011:3015}
Heng Guo, Sangxia Huang, Pinyan Lu, and Mingji Xia.
\newblock {The Complexity of Weighted Boolean $\#$CSP Modulo k}.
\newblock In Thomas Schwentick and Christoph D{\"u}rr, editors, {\em 28th
  International Symposium on Theoretical Aspects of Computer Science (STACS
  2011)}, volume~9 of {\em Leibniz International Proceedings in Informatics
  (LIPIcs)}, pages 249--260, Dagstuhl, Germany, 2011. Schloss
  Dagstuhl--Leibniz-Zentrum fuer Informatik.

\bibitem{Guo13:symmetric}
Heng Guo, Pinyan Lu, and Leslie~G. Valiant.
\newblock The complexity of symmetric boolean parity holant problems.
\newblock {\em {SIAM} J. Comput.}, 42(1):324--356, 2013.

\bibitem{ref:CountingMod2ToCactus}
Andreas Göbel, Leslie~Ann Goldberg, and David Richerby.
\newblock The complexity of counting homomorphisms to cactus graphs modulo 2.
\newblock {\em ACM Transactions on Computation Theory}, 6(4):1–29, Aug 2014.

\bibitem{ref:ProductOfGraphs}
Richard Hammack, Wilfried Imrich, and Sandi Klavzar.
\newblock {\em Handbook of Product Graphs, Second Edition}.
\newblock CRC Press, Inc., USA, 2nd edition, 2011.

\bibitem{Hell04:homomorphism}
P.~Hell and Ne{\v s}et{\v r}il.
\newblock {\em Graphs and homomorphisms}, volume~28 of {\em Oxford {L}ecture
  {S}eries in {M}athematics and its {A}pplications}.
\newblock Oxford University Press, 2004.

\bibitem{Hertrampf90:relations}
Ulrich Hertrampf.
\newblock Relations among mod-classes.
\newblock {\em Theor. Comput. Sci.}, 74(3):325--328, 1990.

\bibitem{Jeavons:algebraic}
Peter Jeavons.
\newblock On the algebraic structure of combinatorial problems.
\newblock {\em Theoretical Computer Science}, 200(1-2):185--204, 1998.

\bibitem{Jeavons97:closure}
Peter Jeavons, David~A. Cohen, and Marc Gyssens.
\newblock Closure properties of constraints.
\newblock {\em J. {ACM}}, 44(4):527--548, 1997.

\bibitem{Jeavons99:expressive}
P.G. Jeavons, D.A. Cohen, and M.~Gyssens.
\newblock How to determine the expressive power of constraints.
\newblock {\em Constraints}, 4:113--131, 1999.

\bibitem{Jerrum93:polynomial}
Mark Jerrum and Alistair Sinclair.
\newblock Polynomial-time approximation algorithms for the {I}sing model.
\newblock {\em {SIAM} J. Comput.}, 22(5):1087--1116, 1993.

\bibitem{ref:CountingModPToSquarefree}
Amirhossein Kazeminia and Andrei~A Bulatov.
\newblock Counting homomorphisms modulo a prime number.
\newblock In {\em 44th International Symposium on Mathematical Foundations of
  Computer Science (MFCS 2019)}. Schloss Dagstuhl-Leibniz-Zentrum fuer
  Informatik, 2019.

\bibitem{ref:kolaitis2004constraint}
Phokion~G Kolaitis.
\newblock Constraint satisfaction, complexity, and logic.
\newblock In {\em Hellenic Conference on Artificial Intelligence}, pages 1--2.
  Springer, 2004.

\bibitem{ref:CountingModPToK33free}
J.~A.~Gregor Lagodzinski, Andreas Göbel, Katrin Casel, and Tobias Friedrich.
\newblock On counting (quantum-)graph homomorphisms in finite fields of prime
  order.
\newblock {\em CoRR}, abs/2011.04827, 2021.

\bibitem{Lieb81:general}
E.H. Lieb and A.D. Sokal.
\newblock A general {L}ee-{Y}ang theorem for one-component and multicomponent
  ferromagnets.
\newblock {\em Communications in {M}athematical {P}hysics}, 80(2):153–179,
  1981.

\bibitem{lovasz2012large}
L{\'a}szl{\'o} Lov{\'a}sz.
\newblock {\em Large networks and graph limits}, volume~60.
\newblock American Mathematical Soc., 2012.

\bibitem{Valiant79:computing}
L.~Valiant.
\newblock The complexity of computing the permanent.
\newblock {\em Theoretical {C}omputing {S}cience}, 8:189--201, 1979.

\bibitem{Valiant79:complexity}
L.~Valiant.
\newblock The complexity of enumeration and reliability problems.
\newblock {\em {SIAM} {J}ournal on {C}omputing}, 8(3):410--421, 1979.

\end{thebibliography}

\end{document}